\documentclass{elsarticle}

\PassOptionsToPackage{caption=false}{subfig}

\usepackage{graphicx}
\usepackage[caption=false,justification=centerlast]{subfig}

\usepackage{dcolumn}  
\usepackage{bm}  
\usepackage{amssymb}
\usepackage{amsmath}

\makeatother
\newcommand{\openone}{\mathbb 1}

\newcommand{\req}[1]{Eq.~(\ref{#1})}
\newcommand{\reqs}[1]{Eqs.~(\ref{#1})}
\newcommand{\rref}[1]{(\ref{#1})}

\renewcommand{\k}{\mathbf{k}}

\newcommand{\bv}{\boldsymbol{v}}

\newcommand{\beq}{\begin{equation}}
\newcommand{\eeq}{\end{equation}}
\newcommand{\be}{\begin{equation}}
\newcommand{\ee}{\end{equation}}
\newcommand{\beqa}{\begin{eqnarray}}
\newcommand{\eeqa}{\end{eqnarray}}
\newcommand{\bea}{\begin{eqnarray}}
\newcommand{\eea}{\end{eqnarray}}
\usepackage{amssymb}
\usepackage{array}
\usepackage{amsmath}
\usepackage{graphicx}\usepackage{graphics}
\usepackage{dcolumn}
\usepackage{bm}\usepackage{varioref}

\usepackage[cal=boondoxo,calscaled=1.05]{mathalfa}
\newcommand{\ldcb}{\left\{\!\!\left\{}
\newcommand{\rdcb}{\right\}\!\!\right\}} 

\newcommand{\ldb}{\left[\!\left[}
\newcommand{\rdb}{\right]\!\right]} 

\begin{document}

\title{
Bethe Ansatz Solutions for Certain Periodic Quantum  Circuits}

\date{\today}

\begin{abstract}

I  derived Bethe Ansatz equations for two model Periodic Quantum Circuits: 1) XXZ model; 2) Chiral Hubbard Model.
I obtained  explicit expressions for the spectra of the strings of any length. These analytic results may be useful for
calibration and error mitigations in modern engineered quantum platforms.

\end{abstract}


\author{Igor L. Aleiner$^{a,b}$}

\address{Google Quantum AI, Santa Barbara, CA, USA}
\address{Department of Physics, Columbia University,
  New York, NY, 10027, USA}

\maketitle
\section{Introduction}
\label{sec:intro}
One of the promising approaches to study the quantum materials is to simulate them on the engineered quantum platforms currently
under development by several groups. In order for these simulations to have predictive powers the platforms have to be properly calibrated and
errors to be mitigated. Only reliable way to do it is to compare the results of simulations with available exact analytic results.

The important step in this direction was recently 
performed \cite{Google2020} where the spectrum (Fourier transform) of the periodic circuits was studied and impressive accuracy of the
extracting the parameters of  the circuits was achieved (statistical precision below $10^{-5}$ rad).
However, those results were achieved only for the simulations in the Hilbert subspace of one particle excitations.
It is needless to say that not all the parameters of the circuits could be calibrated within study of this Hilbert subspace.

In general, the increase of the number of excitations quickly leads to the results that are untractable on such a high precision level.
The only known exceptions are exactly solvable models
(such as one dimensional $XXZ$ spin chains \cite{Bethe31} or Hubbard model \cite{Lieb-Wu}) where the number of non-linear equations describing the spectra grows
only linearly with system size. However, the
direct comparison of those analytic results with the quantum platforms simulation is not straightforward. The difficulty is
that the quantum platforms readily operate with the discrete unitary gates rather than with the continuous Hamiltonian dynamics.
The reducing the latter to the former requires the Suzuki-Trotter expansion accuracy of which requires  smaller steps and deeper circuits.
The depth of the circuits is limited due to the noise and it can not be increased in perpetuity. This motivates the interest to the models
where the discrete quantum circuits themselves remain integrable for arbitrary parameters and not only in the limit of small Trotter steps.

This paper is devoted to the analytic solutions of two such models: (i) $XXZ$ periodic quantum circuit model; (ii) Chiral Hubbard periodic
quantum circuit.
Those integrable models  in the limit of small Trotter steps match their Hamiltonian counterparts
(The integrability of quantum circuits was first pointed out before \cite{prosenXXZ} for $XXZ$ and for a non-unitary circuit somewhat
similar to our Chiral Hubbard circuit \cite{prosenH}. No expressions for spectra were given in those references).
The symmetries of the  $XXZ$ periodic quantum circuit model are the same for the Hamiltonian version of $XXZ$ spin chains.
The symmetries of the Chiral Hubbard periodic
quantum circuit is lower than that of the Hamiltonian of the fermionic Hubbard model and they are restored only in the limit of small Trotter steps.

We will obtain the complete Bethe ansatz equation and find the spectrum of the string solutions.
We will argue that the direct
excitations of the strings from the vacuum states is the most effective way to study the integrable many-body physics on the developing quantum platforms.
We will use the coordinate version of Bethe ansatz rather than algebraic Bethe ansatz \cite{KorepinBogolyubonIsergin,Review}
to keep the derivation a little bit more transparent.

\section{General properties.}
\label{sec1}

The content of this section is related to the both models considered in this paper.

\begin{figure}
 \center{ \begin{picture}(130,40)
    \put(120,17){$\hat{U}_e$}
    \put(120,35){$\hat{U}_o$}
    \put(0,0){
     \includegraphics[width=120\unitlength]{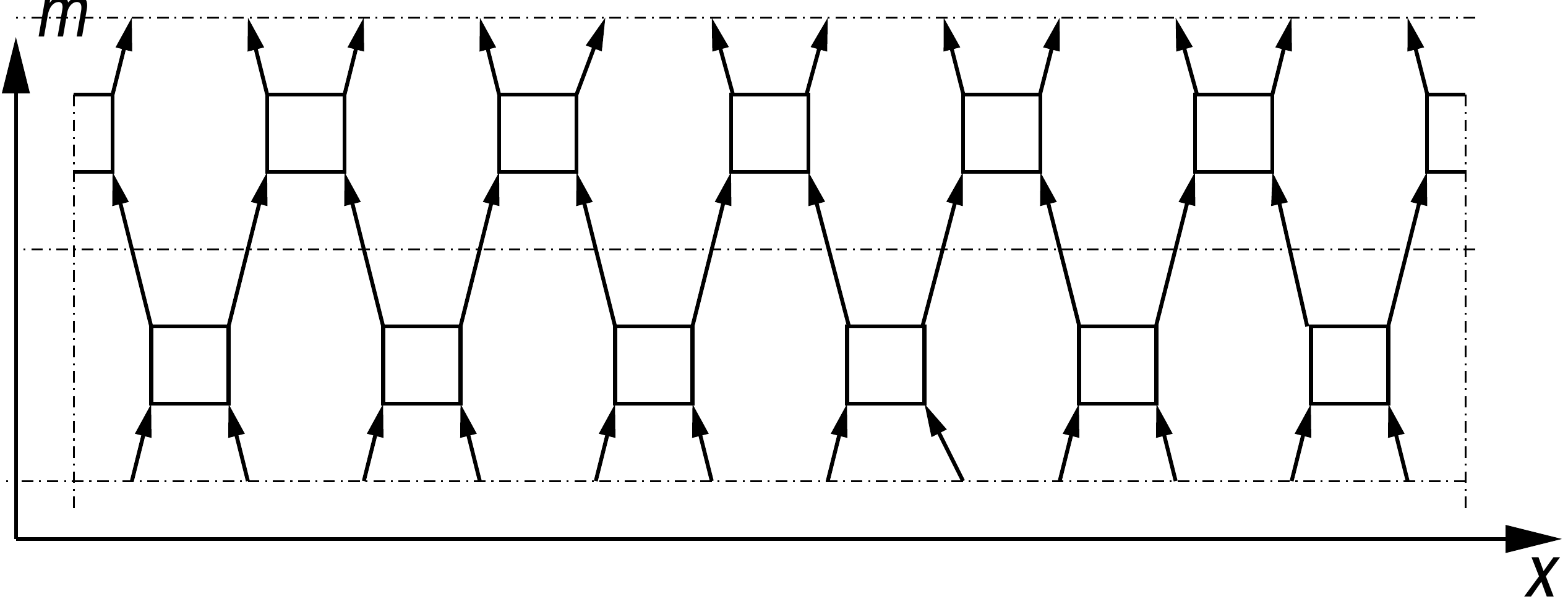}
    }
  \end{picture}
  }
  \caption{Schematic representation of the periodic quantum circuit.
    The horizontal direction is a spatial coordinate. The discrete time $m$
    is running vertically and
    the unitary operation is periodically repeated.
    Square represents an arbitrary unitary matrix operating in Hilbert space operating
    in Hilbert space of the neighboring sites. The periodic boundary
    conditions in space are assumed.
  }
\label{Fig1}
\end{figure}

A quantum circuit operating on $2L$ sites is shown schematically on Fig.~\ref{Fig1}.
Each site corresponds to either one qubit ($XXZ$) or two qubits (Chiral Hubbard).

The total Hilbert space of the system is the direct product of the Hilbert spaces for different sites:
\be
\mathcal{ H}^{(2L)}=\bigotimes^{2L}_{x=1}\mathcal{ h}_x.
\ee
The Hilbert space of one site, $\mathcal{ h}_x$, is either two-dimensional with the basis $(\left|0\right\rangle,\left|1\right\rangle)$
for the Circuit I or the direct product of the two dimensional Hilbert spaces for the different legs (spins or replicas) labeled by additional
subscript $\mathcal{ h}_x=\mathcal{ h}_{x,1}\otimes\mathcal{ h}_{x,2}$.

The unitary matrix describes the evolution of any initial wavefunction $\left|\psi(0)\right\rangle$  with the discrete time $n$
\be
\left|\psi(n)\right\rangle=\hat{\mathcal{U}}^n\left|\psi(0)\right\rangle.
\ee 
Accordingly, the quasi-energies $E$ and eigenfunctions $\left|\psi_E\right\rangle$ are defined
as
\be
\hat{\mathcal{U}}\left|\psi_E\right\rangle=e^{iE}\left|\psi_E\right\rangle.
\label{eq:quasienergy}
\ee
The quasi-energies will be chosen to lie in the first Brillouin zone
\be
E=\ldb E\rdb, \quad \ldb\dots\rdb\equiv \left[\,\left(\dots+\pi\right)\ \mathrm{mod}\  2\pi\,\right] -\pi.
\label{eq:BZ}
\ee

The reference "vacuum" state  $\left|0\right\rangle$  is assumed to be invariant with time $n$
\be
\hat{\mathcal{U}}\left|0\right\rangle=\left|0\right\rangle.
\label{eq:vacuum1}
\ee 

The evolution matrix is further decomposed,  see Fig.~\ref{Fig1},
\be
\hat{\mathcal{U}}=\hat{\mathcal{U}}_o\hat{\mathcal{U}}_e,
\label{eq:Udecomp}
\ee
with
\be
\hat{\mathcal{U}}_e=\bigotimes^{L}_{x=1}\hat{U}_{2x-1/2};
\quad \hat{\mathcal{U}}_o=\bigotimes^{L}_{x=1}\hat{U}_{2x+1/2},
\label{eq:UeUo}
\ee
where unitary $\hat{U}_{\{x+1/2\}}$ acts in  the Hilbert space of two neighboring sites $\mathcal{ h}_{\ldcb x\rdcb}\otimes\mathcal{ h}_{\ldcb x+1\rdcb}$,
and we use notation for the periodic boundary condition
\be
\ldcb x \rdcb\equiv \left[\,\left(x-1\right)\ \mathrm{mod}\  2L\,\right] + 1.
\label{eq:periodic}
\ee

To facilitate the further manipulations, let us introduce the  translation operator
$\hat{t}$ by one site. First, the vacuum state is translationally invariant
\be
\hat{t}\,\left|0\right\rangle=\left|0\right\rangle.
\label{eq:vacuum2}
\ee
Second, for any operator $\hat{\cal{O}}_{x_1,x_2,\dots,x_m}$ acting in the Hilbert space $\mathcal{ h}_{x_1}\otimes\mathcal{ h}_{x_2}
\otimes\dots \otimes\mathcal{ h}_{x_m}$
\be
\hat{t}\, \hat{\cal{O}}_{x_1,x_2,\dots,x_m}\, \hat{t}^{-1}=\hat{\cal{O}}_{\ldcb x_1+1 \rdcb,\ldcb x_2+1\rdcb,\dots,\ldcb x_m+1\rdcb}.
\label{eq:t-definition}
\ee
Then, it follows immediately from \reqs{eq:UeUo}  and \rref{eq:t-definition}
\be
\hat{t}\hat{\mathcal{U}}_e\hat{t}^{-1}=\hat{\mathcal{U}}_o,
\ \hat{t}^2\hat{\mathcal{U}}_e\hat{t}^{-2}=\hat{\mathcal{U}}_e,
\ t^{2L}=\hat{\openone}.
\label{eq:commutation-relations}
\ee

Equation \rref{eq:commutation-relations} enables one to rewrite the evolution matrix \rref{eq:Udecomp} as
(similarly to Refs.~\cite{prosenXXZ,prosenH})
\be
\hat{\mathcal{U}}=\hat{t}\hat{\mathcal{U}}_e\hat{t}^{-1}\hat{\mathcal{U}}_e=\hat{t}^2\left[\hat{\mathcal{V}}\right]^2,
\ee
where
\be
\hat{\mathcal{V}}\equiv \hat{t}^{-1}\hat{\mathcal{U}}_e; \quad \left[\hat{t}^{2};\ \hat{\mathcal{V}}\right]=0.
\label{eq:quasienergy2}
\ee
Thus, operators $\hat{t}^{2}$ and  $\hat{\mathcal{V}}$ can be diagonalized simultaneously
\be
\hat{\mathcal{V}}\left|\psi_{\mathcal{E},Q}\right\rangle=e^{i\mathcal{E}}\left|\psi_{\mathcal{E},Q}\right\rangle,
\ \hat{t}^2\left|\psi_{
    \mathcal{E},Q}\right\rangle=e^{iQ}\left|\psi_{\mathcal{E},Q}\right\rangle,
\label{eq:EQ}
\ee
with the quasi-momentum quantization condition
\be
e^{iQL}=1.
\label{eq:q-quantized}
\ee
The quasi-energy \rref{eq:quasienergy} is given by
\be
E(Q)=\ldb 2\mathcal{E}+Q\rdb.
\label{eq:quasienergies-e-q} 
\ee
We will see later that the solution of \reqs{eq:EQ} is much simpler than that of the original problem \rref{eq:quasienergy}. 

Closing this section, let us discuss the binary transformations of the circuit.

First is the spatial reflection
\begin{subequations}
\be
\begin{split}
  x\to \ldcb -x\rdcb;\ \hat{{U}}_{x+1/2} \to \hat{R}\hat{{U}}_{x+1/2}\hat{R}; \ \hat{R}^2=1,
  \ \hat{R}^\dagger=\hat{R}.
\end{split}
\ee
It is easy to see from \reqs{eq:commutation-relations}
that it leads to changes
\be
\mathcal{E}\to \mathcal{E};\ Q\to -Q.
\ee
\label{eq:reflection}
\end{subequations}

Second is the parity transformation
\begin{subequations}
\be
\begin{split}
\hat{{U}}_{x+1/2} \to \hat{t}\hat{W}\hat{t}^{-1}\left[\hat{{U}}_{x+1/2}\right]^{-1}\hat{W}^{-1};
\end{split}
\ee
where $\hat{W}$ is an arbitrary matrix.
Under this transformation $\hat{\mathcal{U}} \to \left[\hat{W} \hat{t}\right] \hat{\mathcal{U}}^{-1} \left[\hat{W} \hat{t}\right]^{-1} $
and, therefore,
\be
\mathcal{E}\to -\mathcal{E};\ Q\to Q.
\ee
\label{eq:paritty}
\end{subequations}

\section{XXZ  circuit.}
Two dimensional Hilbert space of a single qubit is  spanned by the basis $\left(\left|0_x\right\rangle,\left|1_x\right\rangle\right)$.
The relevant single qubit raising/lowering operators are
\be
\hat{\sigma}^+_x\left|0_x\right\rangle=\left|1_x\right\rangle,
\
\hat{\sigma}^-_x\left|1_x\right\rangle=\left|0_x\right\rangle,
\
\left[\hat{\sigma}^{\pm}_x\right]^2=0.
\label{eq:sigma}
\ee
We will also use the standard Pauli operators $\sigma_x^{x,y,z}$ acting in the Hilbert space of single qubit, $\mathcal{ h}_x$.
To avoid confusion, we will always use subscript to label the coordinate and the the superscript $x,y,z$ to label the corresponding Pauli matrix.

Unitary operator, $\hat{U}_{\{x+1/2\}}$, acts in the Hilbert space of
two neighboring sites $\mathcal{ h}_{\ldcb x\rdcb}\otimes\mathcal{ h}_{\ldcb x+1\rdcb}$ .
The basis for this four dimensional space is
\be
\left(
  \left|00\right\rangle,
\
\left|01\right\rangle,
\
\left|10\right\rangle,
\
\left|11\right\rangle
\right).
\label{eq:XXZbasis}
\ee

In basis \rref{eq:XXZbasis}, the elementary unitary has the form
\be
\hat{U}_{x+1/2}=
\begin{pmatrix}
1 & 0&0&0\\
0& \cos\alpha & i \sin\alpha & 0
\\
0&  i \sin\alpha & \cos\alpha  & 0\\
0&0&0&e^{2i\phi}
\end{pmatrix}
\label{eq:XXZ}
\ee
where $\alpha$ and $\phi$ are parameters of the circuit.

With the unitary \rref{eq:XXZ}, the vacuum \rref{eq:vacuum1}, \rref{eq:vacuum2} is easily found
\be
\left|0\right\rangle
=\bigotimes^{2L}_{x=1}\left|0_x\right\rangle.
\label{eq:vacuum-definition}
\ee
Moreover, both operator $\hat{\mathcal{V}}$ and $\hat{t}^2$ commute with number of the excitations operator
\be
\hat{\mathcal{N}}=\sum_{x=1}^{2L}\hat{\sigma}^+_x\hat{\sigma}^-_x,
\label{eq:N}
\ee
and it is used as one more quantum number.

It is obvious that it suffices to consider only $\phi,\alpha >0$.
Indeed change $\alpha$ to $-\alpha$ corresponds to the unitary transformation 
\[
  \hat{\mathcal{U}}
  \to \left[\bigotimes^{L}_{x=1}\hat{\sigma}_{2x}^z\right]\hat{\mathcal{U}}\left[\bigotimes^{L}_{x=1}\hat{\sigma}_{2x}^z\right],
\]  
that leaves the spectrum unchanged.
On the other hand transformation $\phi\to-\phi,\alpha\to-\alpha$ is nothing but the parity transformation
\rref{eq:paritty} changing the sign of the quasienergy $\mathcal{E} \to -\mathcal{E}$.
[The model clearly has a mirror reflection symmetry so that everything is invariant with respect to $Q\to -Q$.)

\subsection{Single excitation, $\mathcal{N}=1$}
\label{sec:xxz1}


We look for the single excitation wave function parameterized by its
rapidity $\varepsilon$ as
\be
\begin{split}
& \left|\psi_{1;\varepsilon}\right\rangle
 =\sum_{x=1}^{2L}\Upsilon_x(\varepsilon)\hat{\sigma}^+_{x}\left|0\right\rangle.
 \\
 &
 \Upsilon_x(\varepsilon)=e^{iqx/2}\tilde{\Upsilon}_x(\varepsilon)
 =\left\{
   \begin{matrix}
 e^{-iq(\varepsilon) x/2};    & x &\mathrm{even}\\
 e^{-iq(\varepsilon)(x+1)/2}\Upsilon(\varepsilon);     & x &\mathrm{odd}
   \end{matrix}
 \right.
\end{split}
\label{eq:psi1}
\ee
where the functions $q(\varepsilon),\Upsilon(\varepsilon)$ will be found later.
Periodic function $\tilde{\Upsilon}_x=\tilde{\Upsilon}_{\ldcb x+2\rdcb}$ is analogous to the Bloch amplitude
for the lattices with two sites per unit cell.
Also, because there are two sites per  unit cell,
function $\varepsilon(q)$ is two-valued.

Let us start with the translational property of the wave-function \rref{eq:psi1}.
We have
\be
\begin{split}
 \hat{t}^2\left|\psi_{1;\varepsilon}\right\rangle
 & =\sum_{x=1}^{L}\Upsilon_x(\varepsilon)\hat{t}^2\hat{\sigma}^+_{x}\hat{t}^{-2}\left|0\right\rangle
 =\sum_{x=1}^{L}\Upsilon_x(\varepsilon)\hat{\sigma}^+_{\ldcb x+2\rdcb}\left|0\right\rangle
 \\
 &=e^{iq(\varepsilon)}\left|\psi_{1;\varepsilon}\right\rangle+\left|\delta\psi\right\rangle;
 \\
 \left|\delta\psi\right\rangle=&
 \left(e^{-iq(\varepsilon)L}-1\right)\left[\Upsilon(\varepsilon)\hat{\sigma}^+_{1}+\hat{\sigma}^+_{2}\right]\left|0\right\rangle.
\end{split}
\raisetag{0.8cm}
\label{eq:psidagger-translation}
\ee
The annoying correction $\left|\delta\psi\right\rangle$ vanishes for $e^{iqL}=1$ and we find 
\be
\hat{t}^2\left|\psi_{1;\varepsilon}\right\rangle=e^{iq(\varepsilon)}\left|\psi_{1;\varepsilon}\right\rangle,
\quad e^{iq(\varepsilon)L}=1,
\label{eq:q-q}
\ee
in accordance with the general rule \rref{eq:EQ} for $Q=q$ and \req{eq:q-quantized}.

To diagonalize the first of \reqs{eq:EQ}, we use
\reqs{eq:UeUo}, \rref{eq:XXZ}, and \rref{eq:psi1}.
We require $e^{i\varepsilon}$ to be an eigenvalue of the operator $\hat{\mathcal{V}}$.
We find
\be
  \begin{split}
    &\hat{t}\left(\hat{\mathcal{V}}-e^{i\varepsilon}\right)\left|\psi_{1,\varepsilon}\right\rangle
    \\
    &
    =\sum_{x=1}^{L}e^{-iq(\varepsilon)x}
    \left[1,\Upsilon(\varepsilon)\right]\left\{\hat{\mathcal{u}}^{(1)}-
    e^{i\varepsilon}\hat{\mathbb{T}}^{(1)}(\varepsilon)\right\}
    \begin{bmatrix}\hat{\sigma}^+_{2x}\\ \hat{\sigma}^+_{2x-1}\end{bmatrix}
    \left|0\right\rangle;
    \\
     &\hat{\mathcal{u}}^{(1)}
   =\begin{pmatrix}\cos\alpha & i\sin\alpha\\i\sin\alpha & \cos\alpha
      \end{pmatrix};
     \quad
    \hat{\mathbb{T}}^{(1)}(\varepsilon)
    =\begin{pmatrix} 0 & {e^{iq(\varepsilon)}} \\
    1 & 0
    \end{pmatrix};
  \end{split}
  \label{eq:vpsi1}
  \raisetag{27mm}
  \ee
  The matrix  $\hat{\mathcal{u}}^{(1)}$ is nothing but the sub-block of gate \rref{eq:XXZ} in the subspace of
a single excitation.
The right-hand-side of \req{eq:vpsi1} vanishes if $\Upsilon(\varepsilon)$
satisfies equations
\be
\begin{bmatrix} e^{i\varepsilon}\Upsilon(\varepsilon)
  \\ e^{i\varepsilon+iq(\varepsilon)}
\end{bmatrix}=
\hat{\mathcal{u}}^{(1)}
\begin{bmatrix}
  1 \\\Upsilon(\varepsilon)
\end{bmatrix}
  .
\label{eq:eqUpsilon}
\ee
The compatibility of \reqs{eq:eqUpsilon} yields the equation for spectrum
\begin{subequations}
\be
e^{iq}=e^{-i2\varepsilon}\frac{1+i\sin\alpha\, e^{i\varepsilon}}{1-i\sin\alpha\, e^{-i\varepsilon}}.
\label{eq:q-epsilon}
\ee
Notice that $q(\varepsilon)$ is a single valued function within the first Brillouin zone, whereas its inverse $\varepsilon(q)$ is a two-valued function.
The quasi-energy of the original circuit \rref{eq:quasienergy} is than obtained using \req{eq:quasienergies-e-q} 
\be
e^{i\epsilon }=\frac{1+i\sin\alpha\, e^{i\varepsilon}}{1-i\sin\alpha\, e^{-i\varepsilon}}.
\label{eq:e-epsilon}
\ee
\label{implicit}
\end{subequations}
Equations \rref{implicit} give the parametric expression for the spectrum $\epsilon(q)$ in one particle spectrum.
Continuous parameter $\varepsilon  \in [-\pi;\pi]$ will be called rapidity.
By construction it is clear that $ e^{iq}$ winds twice around origin on the complex plane whereas $ e^{i\epsilon}$ moves along
the finite arc on the unit circle. It means that \reqs{implicit} describe two branches of spectrum $\epsilon_{\pm}(q)$.

Equations \rref{implicit} in its present form will be useful for the study of the Chiral Hubbard model.
For $XXZ$  model it can be further simplified. Straightforward calculation yields,
\be
\cos\left[ \epsilon(q)\right] =\cos^2\alpha-\sin^2\alpha\cos q,
\label{eq:cos-cos}
\ee
that describes the both branches $\epsilon_\pm(q)$ of the spectrum, $\pm \epsilon_\pm(q)>0$. 

Finally, let us show one more parameterization that is extremely handy for writing Bethe ansatz equations.
Let $\varphi$ be an arbitrary angle. Introduce new variables $\theta, \eta$ as
\be
e^{-i(\varphi+\alpha)}=-\frac{
  \sinh\left(\frac{\theta}{2}+\frac{i\eta}{2}\right)
}{\sinh\left(\frac{\theta}{2}-\frac{i\eta}{2}\right)};
  \ e^{-i(\varphi-\alpha)}=\frac{\cosh\left(\frac{\theta}{2}+\frac{i\eta}{2}\right)}{\cosh\left(\frac{\theta}{2}-\frac{i\eta}{2}\right)},
\label{eq:theta-eta}
\ee
or for the real quantities (for $\varphi < \alpha$),
\be
\tan\varphi=\frac{\sin\alpha\, \cos\eta}{\sqrt{\cos^2\alpha+\sin^2\alpha\sin^2\eta}},
\ \tan\alpha=\frac{\sinh\theta}{\sin\eta}.
\label{eq:phialpha}
\ee
Let us parameterize the rapidity $\epsilon$ as
\be
e^{i\varepsilon(v)}=e^{i\varphi}\frac{\sinh\left(v-\frac{\theta}{2}-\frac{i\eta}{2}\right)}{\sinh\left(v-\frac{\theta}{2}+\frac{i\eta}{2}\right)}
\label{eq:ev}
\ee
The new parameter $v$ will be also called rapidity. It runs along the lines $\mathrm{Im}\,v=0$ (so called positive parity)
and $\mathrm{ Im}\,v=\frac{\pi}{2}$ (a.k.a negative parity).
Substituting \reqs{eq:theta-eta} and \rref{eq:ev} into \reqs{implicit}, we find
\be
\begin{split}
e^{iq}&=\frac{\sinh\left(v+\frac{\theta}{2}+\frac{i\eta}{2}\right)\sinh\left(v-\frac{\theta}{2}+\frac{i\eta}{2}\right)}
{\sinh\left(v+\frac{\theta}{2}-\frac{i\eta}{2}\right)\sinh\left(v-\frac{\theta}{2}-\frac{i\eta}{2}\right)};
\\
e^{i\epsilon}&=e^{i2\varphi}\frac{\sinh\left(v-\frac{\theta}{2}-\frac{i\eta}{2}\right)\sinh\left(v+\frac{\theta}{2}+\frac{i\eta}{2}\right)}
{\sinh\left(v-\frac{\theta}{2}+\frac{i\eta}{2}\right)\sinh\left(v+\frac{\theta}{2}-\frac{i\eta}{2}\right)}.
\end{split}
\label{eq:eqv}
\ee
It is important to emphasize that the quasi-momenta and quasi-energies of \req{eq:eqv} are still connected
with each other by \req{eq:cos-cos} if the angles satisfy \req{eq:theta-eta}.

Solutions of different parity corresponds to the different segments of spectrum  \rref{eq:cos-cos}:
\be
\begin{split}
  -2\alpha <\epsilon <2\varphi;\ & \mathrm{positive\ parity}
  \\
  2\varphi <\epsilon <2\alpha;\ & \mathrm{negative\ parity}
\end{split}
\label{eq:parity}
\ee
Different parity solutions match at $q=\pm q_*$, where
\be
q_*(\eta)=\ldb 2\eta\rdb.
\label{eq:qstar}
\ee
Significance of separation into parity branches will become clear later in study of the multi-particle bound states.

Negative parity solution vanishes for $\phi\geq \alpha$. The latter regime is described by the replacement $v\to iv;\ \theta\to i\theta;\ \eta\to i\eta$:
\be
\tan\varphi=\frac{\sin\alpha\, \cosh\eta}{\sqrt{\cos^2\alpha-\sin^2\alpha\sinh^2\eta}},
\ \tan\alpha=\frac{\sin\theta}{\sinh\eta}.
\label{eq:etatthetatrig}
\ee
and
\be
\begin{split}
e^{iq}&=\frac{\sin\left(v+\frac{\theta}{2}+\frac{i\eta}{2}\right)\sin\left(v-\frac{\theta}{2}+\frac{i\eta}{2}\right)}
{\sin\left(v+\frac{\theta}{2}-\frac{i\eta}{2}\right)\sin\left(v-\frac{\theta}{2}-\frac{i\eta}{2}\right)};
\\
e^{i\epsilon}&=e^{i2\varphi}\frac{\sin\left(v-\frac{\theta}{2}-\frac{i\eta}{2}\right)\sin\left(v+\frac{\theta}{2}+\frac{i\eta}{2}\right)}
{\sin\left(v-\frac{\theta}{2}+\frac{i\eta}{2}\right)\sin\left(v+\frac{\theta}{2}-\frac{i\eta}{2}\right)}.
\end{split}
\label{eq:eqvtrig}
\ee
In this case $\mathrm{Im}\,v=0,\ -\pi/2< v\leq \pi/2$.

\subsection{Two excitations, $\mathcal{N}=2$.}
\label{sec:xxz2}

We  look for the two-particle wave-function in a form
parameterized by two rapidities $\varepsilon_{1,2}$:
\be
\left|\psi_{2;\varepsilon_1,\varepsilon_2}\right\rangle
=\sum_{\substack{x_1=1\\x_1< x_2}}^{2L}
\sum_{\mathcal{P}}A_{\mathcal{P}}
\hat{\mathcal{P}}_\varepsilon
\prod_{j=1}^2\Upsilon_{x_j}(\epsilon_j)\hat{\sigma}^+_{x_j}
\left|0\right\rangle,
\label{eq:psi2}
\ee
where we use notation \rref{eq:psi1}. Symbol $\mathcal{P}$ stands for all possible permutations of set of
all indices. Here there are only two indices $\left\{1,2\right\}$ but, anticipating further use, we will consider the number of
those to be arbitrary.
In the word notation, the  permutations for the present case are $12$ and $21$.
Operators $\mathcal{P}_{\varepsilon}$ permute (according to the word)
the indices for the quasi-momenta $\varepsilon_{1,2,\dots}$ but
not the coordinates $x_{1,2}$.  Explicitly,
\be
\begin{split}
\widehat{123\dots n}_\varepsilon\ \varepsilon_1\varepsilon_2\dots \varepsilon_n= \varepsilon_1\varepsilon_2\dots\varepsilon_n;
\\
\widehat{213\dots n}_\varepsilon \  \varepsilon_1\varepsilon_2\dots \varepsilon_n=\varepsilon_2\varepsilon_1\dots\varepsilon_n;
\end{split}
\label{eq:permutation}
\ee
and so on.


Once again, we start from the translational property of the wavefunction \rref{eq:psi2}.
Similarly to \req{eq:psidagger-translation}, we find
\be
  \hat{t}^2\left|\psi_{2;\varepsilon_1,\varepsilon_2}\right\rangle
  =e^{iq(\varepsilon_1)+iq(\varepsilon_2)}
  \left[\left|\psi_{2;\varepsilon_1,\varepsilon_2}\right\rangle+\left|\left(\delta_1+\delta_2\right)\psi\right\rangle\right],
\nonumber
\ee
where the corrections violating the translational invariance are
\be
\begin{split}
 \left|\delta_1\psi\right\rangle &=
\sum_{\mathcal{P}}A_{\mathcal{P}}
\hat{\mathcal{P}}_\varepsilon
\left(e^{-i(q_1+q_2)L}-1\right)
\Upsilon_1\hat{\sigma}^+_{1}\hat{\sigma}^+_{2}\left|0\right\rangle;
    \\
     \left|\delta_2\psi_{\dots}\right\rangle &=
   \sum_{\mathcal{P}}A_{\mathcal{P}}\hat{\mathcal{P}}_\varepsilon
   \\
   &
   \times \Big\{
   \hat{\sigma}^+_{1} \sum_{x=2}^{2L-1}
   \left[\Upsilon_2\Upsilon_{x}(\varepsilon_1)
     e^{-iq_2L}
-\Upsilon_1\Upsilon_{x}(\varepsilon_2)
   \right]{\sigma}^+_{x}
   \\
   &
+\hat{\sigma}^+_{2} \sum_{x=3}^{2L}
   \left[\Upsilon_{x}(\varepsilon_1)
     e^{-iq_2L}
-\Upsilon_{x}(\varepsilon_2)
   \right]{\sigma}^+_{x}   \Big\}
\left|0\right\rangle.
\end{split}
\raisetag{3.1cm}\label{eq:2translation}
\ee
where hereinafter we will use the shorthand notation
\be
q_j\equiv q(\varepsilon_j);\quad \Upsilon_j\equiv \Upsilon(\varepsilon_j).
\label{eq:short}
\ee

One can easily see that $\left|\delta_2\psi\right\rangle$ vanishes provided that
\be
A_{\mathcal{P}}\hat{\mathcal{P}}_\varepsilon e^{-iq_2L}=A_{\mathcal{C}\cdot\mathcal{P}}.
\label{eq:quantization2}
\ee
Hereinafter, $\mathcal{C}$ is the cyclic permutation. Anticipating further use,  we define its action on an arbitrary $n$-letter word
\be
\mathcal{C}\cdot a_1a_2\dots a_{n-1}a_n=a_na_1a_2\dots a_{n-1}.
\label{eq:cyclic}
\ee  
Multiplying  equations \rref{eq:quantization2} for both permutations, we obtain the quantization condition
\be
e^{i(q_1+q_2)L}=1.\label{eq:q1q2}
\ee
Then, $\left|\delta_1\psi\right\rangle$ also vanishes and the wavefunction $\left|\psi_{2;\varepsilon_1,\varepsilon_2}\right\rangle$
satisfies the translational transformation \rref{eq:EQ}  with
\be
Q=\ldb  q_1+q_2\rdb,
\ee
and quantization  rule \req{eq:q-quantized}.

To diagonalize operator $\hat{\mathcal{V}}$ we apply $\left(\hat{\mathcal{V}}-e^{i\varepsilon_1+i\varepsilon_2}
\right)$ to wave-function \rref{eq:psi2}. We obtain using
\reqs{eq:vpsi1} -- \rref{eq:q-epsilon} that all the terms
for $x_1<x_2+1$ vanish so that only neighboring $|\ldcb x_2\rdcb -\ldcb x_1\rdcb|=1$ remain
\be
\begin{split}
&\hat{t}\left(\hat{\mathcal{V}}-e^{i\varepsilon_1+i\varepsilon_2}
\right)\left|\psi_{2;\varepsilon_1,\varepsilon_2}\right\rangle
= \sum_{x=1}^{L}
\sum_{\mathcal{P}}A_{\mathcal{P}}\hat{\mathcal{P}}_\varepsilon
 \hat{\Psi}_{2x}(1,2)
 \left|0\right\rangle;
 \\
 &\hat{\Psi}_{2x}^\dagger(\cdot )\equiv
 e^{-ix(q_1+q_2)} \left[
e^{2i\phi}\Upsilon_1
  -e^{iq_1+i\varepsilon_1+i\varepsilon_2}\Upsilon_2
 \right]
    {\sigma}^+_{ 2x-1 }
  {\sigma}^+_{2x}.
\end{split}
\label{eq:Vpsi2}
\ee
Right-hand-side of \req{eq:Vpsi2} vanishes if the amplitudes satisfy the relation
\be
\begin{split}
  &e^{2i\phi-i(\varepsilon_1+\varepsilon_2)}
  \left[\Upsilon_1A_{12}+\Upsilon_2A_{21}\right]
\\
&
= 
\left[A_{12}\Upsilon_2e^{iq_1}+A_{21}\Upsilon_1e^{iq_2}\right].
\end{split}
\label{eq:amp1}
\ee
Solution of \req{eq:amp1} gives the linear relation between the amplitudes,
\be
A_{21}=S_{12}A_{12}; \quad S_{12}S_{21}=1.
\label{eq:smatrix1}
\ee
The $S$-''matrix'' is found directly from \req{eq:amp1}. To make the connection with the $XXZ$-spin model
we use $\Upsilon$ from \req{eq:eqUpsilon} and parameterization \rref{eq:ev} -- \rref{eq:eqv}
for  $\varphi=\phi$,  and obtain
\be
S_{12}=\frac{\sinh\left(v_1-v_2+i\eta\right)}{\sinh\left(v_1-v_2-i\eta\right)}.
\label{eq:smatrixXXZ}
\ee
This is a standard form of  $S$-''matrix'' for the XXZ model in the gap-less regime. Here it corresponds to the condition $\phi\leq
\alpha$ in \req{eq:XXZ}. Opposite gapped regime, $\phi\geq \alpha$, is obtained by replacement of the hyperbolic functions to the trigonometric ones
\be
S_{12}=\frac{\sin\left(v_1-v_2+i\eta\right)}{\sin\left(v_1-v_2-i\eta\right)}
\label{eq:smatrixXXZgapped}
\ee
The momentum and the energy are given by
\rref{eq:eqvtrig} with $\varphi=\phi$.

Substitution of \req{eq:smatrix1} into \req{eq:quantization2} gives the non-linear equations for the spectrum for two particle
\be
\begin{split}
&e^{iq(v_2)L}=S_{21};\ e^{iq(v_1)L}=S_{12};\\
& E=\ldb \sum_{j=1}^2\epsilon(v_j)\rdb; \ Q=\ldb\sum_{j=1}^2 q(v_j)\rdb,
\end{split}
\label{eq:2BA}
\ee
where $S_{ij}$  are given by \reqs{eq:smatrixXXZ} or \rref{eq:smatrixXXZgapped}.

\subsection{Bethe Ansatz solution for arbitrary excitation number, $\mathcal{N}\geq 2$.}

We look for $\mathcal{N}$-particle wave-function [compare with \req{eq:psi2}]
\be
 \left|
\psi_{\mathcal{N};\{ \varepsilon_j \}_{j=1}^\mathcal{N}}
\right\rangle =
\!\!\!\!\!\!\!\!
\sum_{\substack{x_1=1\\x_1< x_2<\dots< x_\mathcal{N}}}^{2L}
\sum_{\mathcal{P}}A_{\mathcal{P}}
\hat{\mathcal{P}}_\varepsilon
\prod_{j=1}^\mathcal{N}
\Upsilon_{x_j}(\epsilon_j)\hat{\sigma}^+_{x_j}
\left|0\right\rangle.
\label{eq:psiN}
\ee

The translational property is established similarly to $\mathcal{N}=2$ case. Following derivation of \reqs{eq:2translation}, we obtain
\be
  \hat{t}^2\left|\psi_{\mathcal{N};\dots}\right\rangle
  =e^{i\sum_{j=1}^\mathcal{N}q_j}
  \left[\left|\psi_{\mathcal{N};\dots}\right\rangle+\left|\left(\delta_1+\delta_2\right)
      \psi\right\rangle\right]
  ,
\ee
where the corrections violating the translational invariance are

\be
\begin{split}
&\left|\delta_1\psi_{\dots}\right\rangle=
\sum_{\mathcal{P}}A_{\mathcal{P}}\hat{\mathcal{P}}_\varepsilon
\!\!\!\!\!
\sum_{\substack{x_1=3\\ x_1< x_2<\dots x_{\mathcal{N}-2}}}^{2L}
\\
&\ \times
\Bigg[
 e^{-i(q_\mathcal{N}+q_{\mathcal{N}-1})L}
 \Upsilon_{\mathcal{N}-1}
 \hat{\sigma}^+_{1}\hat{\sigma}^+_{2}
\left(\prod_{j=1}^{\mathcal{N}-2}
  \Upsilon_{x_j}(\epsilon_j)\hat{\sigma}^+_{x_j}\right)
\\
&\qquad
-
 \Upsilon_{1}
 \hat{\sigma}^+_{1}\hat{\sigma}^+_{2}
\left(\prod_{j=1}^{\mathcal{N}-2}
  \Upsilon_{x_j}(\epsilon_{j+2})\hat{\sigma}^+_{x_j}\right)
 \Bigg]
 \left|0\right\rangle;
 \\
 & \left|\delta_2\psi_{\dots}\right\rangle=\sum_{\mathcal{P}}A_{\mathcal{P}}\hat{\mathcal{P}}_\varepsilon
\!\!\!\!\!
\sum_{\substack{x_1=3\\ x_1< x_2<\dots x_{\mathcal{N}-1}}}^{2L}
\\
&\times
\Bigg\{\left[
e^{-i(q_\mathcal{N})}
  \Upsilon_{\mathcal{N}-1}
  \hat{\sigma}^+_{1}
 \left(\prod_{j=1}^{\mathcal{N}-2}
   \Upsilon_{x_j}(\epsilon_j)\hat{\sigma}^+_{x_j}\right)
 -
  \Upsilon_{1}
  \hat{\sigma}^+_{1}
 \left(\prod_{j=1}^{\mathcal{N}-2}
   \Upsilon_{x_j}(\epsilon_{j+1})\hat{\sigma}^+_{x_j}\right)
 \right]
 \\
 & +
 \left[
e^{-i(q_\mathcal{N})}
  \hat{\sigma}^+_{2}
 \left(\prod_{j=1}^{\mathcal{N}-2}
   \Upsilon_{x_j}(\epsilon_j)\hat{\sigma}^+_{x_j}\right)
 -
  \hat{\sigma}^+_{2}
 \left(\prod_{j=1}^{\mathcal{N}-2}
   \Upsilon_{x_j}(\epsilon_{j+1})\hat{\sigma}^+_{x_j}\right)
 \right]
\Bigg\}
 \left|0\right\rangle
 ;
\end{split}
\raisetag{0.5cm}
\label{eq:Ntranslation}
\ee
Correction $\left|\delta_2\psi_{\mathcal{N};\dots}\right\rangle$ vanishes provided that
[compare with \req{eq:quantization2}]
\begin{subequations}
  \label{eq:quantizationN}
\be
A_{\mathcal{P}}\hat{\mathcal{P}}_qe^{-iq_\mathcal{N}L}=A_{\mathcal{C}\cdot\mathcal{P}},
\label{eq:quantizationNa}
\ee
and the cyclic permutation $\mathcal{C}$ is defined in \req{eq:cyclic}. By the same token  $\left|\delta_1\psi_{\mathcal{N};\dots}\right\rangle$
also vanishes if
\be
A_{\mathcal{P}}\hat{\mathcal{P}}_qe^{-i\left(q_\mathcal{N}+q_\mathcal{N-1}\right)L}=A_{\mathcal{C}\cdot\mathcal{C}\cdot\mathcal{P}}.
\label{eq:quantizationNb}
\ee
One can immediately see that condition \rref{eq:quantizationNb} is obtained by using \req{eq:quantizationNa} twice and, therefore, it is not
an independent condition.
\end{subequations}

Under the conditions \rref{eq:quantizationN} wave-function \rref{eq:psiN} is an eigenfunction of translational operator \rref{eq:EQ}
with
\be
Q=\ldb\sum_{j=1}^\mathcal{N}q_j\rdb=\ldb\sum_{j=1}^\mathcal{N}q(v_j)\rdb.
\label{eq:Qq}
\ee
The quantization condition  \rref{eq:q-quantized} obviously follows from applying \req{eq:quantizationNa} $\mathcal{N}$ times.

To find the eigenfunction of operator $\hat{\mathcal{V}}$, we apply it to wave-function \rref{eq:psiN}. We obtain,
similarly to \req{eq:Vpsi2},
\be
\begin{split}
&\hat{t}
\left(
  \hat{\mathcal{V}}-e^{i\sum_j^\mathcal{N}\varepsilon_j}
\right)
  \left|\psi_{\mathcal{N},\dots}\right\rangle=
 %
  \sum_{
    \substack{
      x_1=1\\x_1< x_2<\dots < x_{\mathcal{N}}
      }
  }
  ^{2L}
  \\
  &
  \times
\sum_{\mathcal{P}}A_{\mathcal{P}}\hat{\mathcal{P}}_\varepsilon
  \Bigg[ \sum_{y=1}^L
  \sum_{k=1}^{\mathcal{N}-1}
  \left(
    \delta_{x_k,2y-1}\delta_{x_{k+1},2y}\right)
  \hat{\Psi}_{2y}^\dagger(k,k+1)
  \left(\prod_{\substack{j=1\\
      j\neq k,k+1}}^\mathcal{N}
\Upsilon_{x_j}(\epsilon_j)\hat{\sigma}^+_{x_j}\right)
\\
&
+
\sum_{
  \substack{y_1=1\\  y_1<y_2}}^L
  \sum_{\substack{k_1=1\\k_1<k_2}}^{\mathcal{N}-1}
   \left(
    \delta_{x_{k_1},2y_1-1}\delta_{x_{k_1+1},2y_1}\right)
    \left(
      \delta_{x_{k_2},2y_2-1}\delta_{x_{k_2+1},2y_2}\right)
    \\
    &\times
   \hat{\Psi}_{2y_1}^\dagger(k_1,k_1+1)\hat{\Psi}_{2y_2}^\dagger(k_2,k_2+1)
   \left(\prod_{\substack{j=1\\
         j\neq k_1,k_1+1\\
         j\neq k_2,k_2+1}}^\mathcal{N}
     \Upsilon_{x_j}(\epsilon_j)\hat{\sigma}^+_{x_j}\right)
  + \dots\Bigg]\left|0\right\rangle,
\end{split}
\label{eq:Vpsin}
\ee
where operator $\hat{\Psi}_{2x}^\dagger(\dots )$ is defined in \req{eq:Vpsi2}.
Each term in \req{eq:Vpsin} vanishes if the amplitudes are connected by generalization of \reqs{eq:amp1} -- \rref{eq:smatrix1}
\be
A_{(k,k+1)\cdot\mathcal{P}}=A_{\mathcal{P}}\hat{\mathcal{P}}_{v}S_{k,k+1};\  S_{k,k+1}\equiv S(v_k,v_{k+1})
\label{eq:smatrixN}
\ee
for each $k$. Hereinafter, notation $(k,k+1)$ stays for the permutations of the nearest neighboring letters $k,k+1$ in the nearest words.

Under the conditions \rref{eq:smatrixN}, equation \rref{eq:Vpsi2} shows that the wave-function \rref{eq:psiN}
indeed diagonalizes operator $\mathcal{V}$ and the eigenvalues for the total evolution operator $\mathcal{U}$
are given by
\be
E=\ldb\sum_{j=1}^\mathcal{N}\epsilon_j\rdb=\ldb\sum_{j=1}^\mathcal{N}\epsilon(v_j)\rdb.
\label{eq:eE}
\ee

Equations \rref{eq:quantizationNa} and \rref{eq:smatrixN} enable one to obtain the quantization condition on
the rapidities $v_j$. Representing cycling permutations as a combination of the pairwise permutation
\[
  \mathcal{C}=(1,2)\cdot(2,3)\cdot\dots\cdot)(\mathcal{N},\mathcal{N-1}),
\]
and using \req{eq:smatrixN}, we obtain from \req{eq:quantizationNa}
\be
\begin{split}
  &e^{-iq(v_j)L}=S_{1,j}S_{2,j}\dots S_{j-1,j}S_{j+1,j}S_{\mathcal{N},j},\\
  &j=1,2,\dots \mathcal{N},
\end{split}
\label{eq:BA1}
\ee
that is the standard form of the Bethe Ansatz equations. The $S$-matrices \rref{eq:smatrixXXZ}, \rref{eq:smatrixXXZgapped} are the familiar ones
for spin chains. The difference here are the expressions for the wave-vectors and energies in terms of rapidities see \reqs{eq:eqv}
and \rref{eq:eqvtrig} for $\varphi=\phi$.

We close this section by writing out the Bethe Ansatz equations in terms of rapidities $v_j$ only. The non-linear equations for rapidities are
\begin{subequations}
  \label{eq:BAv1}
  \be
  \begin{split}
   &\left( \prod_{\pm}\frac{\sinh\left(v_j\pm \frac{\theta}{2}+\frac{i\eta}{2}\right)}
{\sinh\left(v_j\pm\frac{\theta}{2}-\frac{i\eta}{2}\right)}\right)^L=
    \prod_{\substack{k=1\\k\neq j}}^\mathcal{N}
    \frac{\sinh\left(v_j-v_k+i\eta\right)}{\sinh\left(v_j-v_k-i\eta\right)},
    \\
    &j=1,2,\dots,\mathcal{N}.
\end{split}\raisetag{4mm}
\ee
In terms of the rapidities $v_j$ the quasi-energy and quasi-momentum is found as
\be
\begin{split}
 & e^{iE}=e^{i2\mathcal{N}\phi}\prod_{j=1}^{\mathcal{N}}
  \frac{\sinh\left(v_j-\frac{\theta}{2}-\frac{i\eta}{2}\right)\sinh\left(v_j+\frac{\theta}{2}+\frac{i\eta}{2}\right)}
  {\sinh\left(v_j-\frac{\theta}{2}+\frac{i\eta}{2}\right)\sinh\left(v_j+\frac{\theta}{2}-\frac{i\eta}{2}\right)};
  \\
  & e^{iQ}=\prod_{j=1}^{\mathcal{N}}
  \frac{\sinh\left(v_j-\frac{\theta}{2}+\frac{i\eta}{2}\right)\sinh\left(v_j+\frac{\theta}{2}+\frac{i\eta}{2}\right)}
  {\sinh\left(v_j-\frac{\theta}{2}-\frac{i\eta}{2}\right)\sinh\left(v_j+\frac{\theta}{2}-\frac{i\eta}{2}\right)}.
\end{split}
\ee
and parameters $\theta,\eta$ are defined  in terms of parameters of the model \rref{eq:XXZ} as
\be
\tan\phi=\frac{\sin\alpha\, \cos\eta}{\sqrt{\cos^2\alpha+\sin^2\alpha\sin^2\eta}},
\ \tan\alpha=\frac{\sinh\theta}{\sin\eta}.
\ee
\end{subequations}

Equations \rref{eq:BAv1} are applicable for the case $\phi\leq \alpha$,  (gapless case for the related Hamiltonian $XXZ$ chain).
Opposite (gapped) case is described by trigonometric functions \cite{Lamacraft}
\begin{subequations}
   \label{eq:BAv2}
  \be
  \begin{split}
   &\left[ \prod_{\pm}\frac{\sin\left(v_j\pm \frac{\theta}{2}+\frac{i\eta}{2}\right)}
{\sin\left(v_j\pm\frac{\theta}{2}-\frac{i\eta}{2}\right)}\right]^L=
    \prod_{\substack{k=1\\k\neq j}}^\mathcal{N}
    \frac{\sin\left(v_j-v_k+i\eta\right)}{\sin\left(v_j-v_k-i\eta\right)}
    \\
    &j=1,2,\dots,\mathcal{N}.
\end{split}
\raisetag{4mm}
\ee

\be
\begin{split}
 & e^{iE}=e^{i2\mathcal{N}\phi}\prod_{j=1}^{\mathcal{N}}
  \frac{\sin\left(v_j-\frac{\theta}{2}-\frac{i\eta}{2}\right)\sin\left(v_j+\frac{\theta}{2}+\frac{i\eta}{2}\right)}
  {\sin\left(v_j-\frac{\theta}{2}+\frac{i\eta}{2}\right)\sin\left(v_j+\frac{\theta}{2}-\frac{i\eta}{2}\right)};
  \\
  & e^{iQ}=\prod_{j=1}^{\mathcal{N}}
  \frac{\sin\left(v_j-\frac{\theta}{2}+\frac{i\eta}{2}\right)\sin\left(v_j+\frac{\theta}{2}+\frac{i\eta}{2}\right)}
  {\sin\left(v_j-\frac{\theta}{2}-\frac{i\eta}{2}\right)\sin\left(v_j+\frac{\theta}{2}-\frac{i\eta}{2}\right)},
\end{split}
\ee
where $-\pi/2\leq \mathrm{Re}v_j<\pi/2$, and
\be
\tan\phi=\frac{\sin\alpha\, \cosh\eta}{\sqrt{\cos^2\alpha-\sin^2\alpha\sinh^2\eta}},
\ \tan\alpha=\frac{\sin\theta}{\sinh\eta}.
\ee

\end{subequations}

\subsection{Two particle bound states solution.}
\label{sec:xxz-b2}

There are two kind of solutions of \reqs{eq:2BA}: (i) both rapidities $v_1$ and $v_2$ are real and so are quasi-momenta $q_1$ and $q_2$;
(ii) $v_{12}$ are  complex and $v_1=v_2^*$. In the first case \reqs{eq:2BA}   lead to the size dependent quantization of both
rapidities $v_{1,2}$; such type of states are called two-particle continuum and have meaning of two almost independent particles going
through each other and acquiring only a phase due to their collision. In the second case, $\mathrm{Im}v_1=\mathrm{Im}v_2$ is independent
of $L$ at the limit $L\gg 1$ and only $\mathrm{Re}\ v_{1}=\mathrm{Re}\ v_{2}$ is quantized depending on the system length $L$. Then, the emerging structure can be viewed as a new single-particle excitation appearing on the background of two particle excitations and we concentrate on the spectrum of such excitations.

Let us look for a solution with
\[
  \left|e^{iq_1}\right|= \left|e^{-iq_2}\right|<1.
\]
Then the periodicity condition \rref{eq:quantization2} for $\mathrm{Im}\, q_1\,L \gg 1$ yields
\be
A_{12}=1;\quad A_{21}=0,
\label{eq:A12=1}
\ee
and one obtains from \req{eq:smatrix1} that the bound state is determined by zeros of the s-matrix,
\be
S_{12}=0.
\label{eq:bs2}
\ee
Then, using \reqs{eq:smatrixXXZ} -- \rref{eq:smatrixXXZgapped}, we obtain
the equation for rapidities (so-called $2$-string solution),
\be
v_j=v^{(2)}+\left(j-\frac{3}{2}\right)i\eta;\quad j=1,2.
\ee
The spectrum of this excitation $ \epsilon^{(2)}(q^{(2)})$ is implicitly given by
$q^{(2)}=q(v_1)+q(v_2)$, $\epsilon^{(2)}=\epsilon(v_1)+\epsilon(v_2)$
with the quantization condition, see \req{eq:q1q2}, $e^{iq^{(2)}L}=1$.

For the gapless case, $\phi< \alpha$, we find from \req{eq:eqv}
\be
\begin{split}
&e^{iq^{(2)}}\!\!=\frac{\sinh\left(v^{(2)}+\frac{\theta}{2}+\frac{i2*\eta}{2}\right)\sinh\left(v^{(2)}-\frac{\theta}{2}+\frac{i2*\eta}{2}\right)}
{\sinh\left(v^{(2)}+\frac{\theta}{2}-\frac{i2*\eta}{2}\right)\sinh\left(v^{(2)}-\frac{\theta}{2}-\frac{i2*\eta}{2}\right)};
\\
&e^{i\epsilon^{(2)}}\!\!=e^{i2*2\phi}\frac{\sinh\left(v^{(2)}-\frac{\theta}{2}-\frac{i2*\eta}{2}\right)
  \sinh\left(v^{(2)}+\frac{\theta}{2}+\frac{i2*\eta}{2}\right)}
{\sinh\left(v^{(2)}-\frac{\theta}{2}+\frac{i2*\eta}{2}\right)\sinh\left(v^{(2)}+\frac{\theta}{2}-\frac{i2*\eta}{2}\right)}.
\end{split}
\label{eq:eqv2}
\ee
Remarkably, this apparently complicated expression can be reduced to the expression similar to the one particle spectrum \rref{eq:cos-cos}.
Let us write the energy as
\be
\epsilon^{(2)}=
\ldb \tilde{\epsilon}^{(2)}+{\epsilon}^{(2)}(\pi)\rdb;\quad {\epsilon}^{(2)}(\pi)\equiv 4\phi-2\varphi_2,
\label{eq:shift2}
\ee
where we introduced the notation similar to \req{eq:phialpha}:
\be
\tan\varphi_2=\frac{\tanh\theta}{\tan 2\eta},
\quad \tan\alpha_2=\frac{\sinh\theta}{\sin 2\eta}.
\label{eq:phialpha2}
\ee
Then, the second of \reqs{eq:eqv2} takes the form
\be
e^{i\tilde{\epsilon}^{(2)}}\!\!=e^{i2\varphi_2}\frac{\sinh\left(v^{(2)}-\frac{\theta}{2}-\frac{i2*\eta}{2}\right)
  \sinh\left(v^{(2)}+\frac{\theta}{2}+\frac{i2*\eta}{2}\right)}
{\sinh\left(v^{(2)}-\frac{\theta}{2}+\frac{i2*\eta}{2}\right)\sinh\left(v^{(2)}+\frac{\theta}{2}-\frac{i2*\eta}{2}\right)}.
\label{eq:tildeepsilon2}
\ee
Equation \rref{eq:tildeepsilon2} and the first of \reqs{eq:eqv2} are nothing but \reqs{eq:eqv} with the obvious substitutions of
$\eta\to 2\eta,\ \varphi\to \varphi_2$. Therefore, the relation \rref{eq:cos-cos} holds
\[
\cos\left[ \tilde\epsilon^{(2)}\left(q^{(2)}\right)\right] =\frac{1-\tan^2\alpha_2\cos q^{(2)}}{1+\tan^2\alpha_2},
\]
Finally, excluding $\alpha_2,\varphi_2$ with the help of \req{eq:phialpha2}, we obtain the expression for the spectrum of the bound state
(2-string solution)
\be
\begin{split}
&{\epsilon}^{(2)} = \ldb \tilde{\epsilon}^{(2)}+{\epsilon}^{(2)}(\pi)\rdb ;\\
&\cos\left[ \tilde{\epsilon}^{(2)}(q^{(2)})\right] =\frac{\cos^2\alpha\sin^22\eta-\sin^2\alpha\sin^2\eta\cos q^{(2)}}
{\cos^2\alpha\sin^22\eta+\sin^2\alpha\sin^2\eta},
\\
&{\epsilon}^{(2)}(\pi)\equiv 4\phi-2\arctan\left[\left(\tan\phi\right)\left(\tan\eta\right)\left(\cot 2 \eta\right)\right].
\end{split}
\label{eq:cos-cos2}
\ee

The gapped case $\phi>\alpha$ is considered analogously and it gives
\be
\begin{split}
&\cos\left[ \tilde{\epsilon}^{(2)}(q^{(2)})\right] =\frac{\cos^2\alpha\sinh^22\eta-\sin^2\alpha\sinh^2\eta\cos q^{(2)}}
{\cos^2\alpha\sinh^22\eta+\sinh^2\alpha\sin^2\eta},\\
\label{eq:cos-cos2-trig}
&{\epsilon}^{(2)}(\pi)\equiv 4\phi-2\arctan\left[\left(\tan\phi\right)\left(\tanh\eta\right)\left(\coth 2 \eta\right)\right].
\end{split}
\ee

Equations \rref{eq:cos-cos2}--\rref{eq:cos-cos2-trig} are the main results of this subsection. They show that the
spectrum of 2-string has the functional form similar to that of the one
particle excitation. The control phase $\phi$ enters quite non-trivially into the quasi-energy shift as well as into the rescaling of the spectrum.
It is worth emphasizing that after the single particle and the 2-string spectra are measured, all the parameters of the model are fixed,
i.e. the spectra of the larger string solutions considered later do not have any fitting parameters at all.

To complete the study of 2-string, we need to specify at which regions of the momenta the bound states are physical.
In terms of the periodic Bloch amplitudes \rref{eq:psi1},
the bound state solution \rref{eq:psi2}, \rref{eq:A12=1} has the form 
\be
 \left|\psi_{2;\cdot}\right\rangle=\sum_{\substack{x_1=1\\x_1\leq x_2}}^{2L}
e^{-\frac{iq^{(2)}x_2}{2}}e^{\frac{iq(v_1)(x_2-x_1)}{2}}
\prod_{j=1}^2\tilde{\Upsilon}_{x_j}(v_j)\hat{\sigma}^+_{x_j}
\left|0\right\rangle,
\label{eq:psi2-bound}
\ee
the requirement for the pair to be bound reads
\be
\left|e^{iq(v_1)}\right|^2 <1,
\ee
(this is, of course, equivalent to what we assumed in the very beginning).
With the help of \req{eq:eqv} it translates, for $\phi < \alpha$,
into the restriction
\be
\cos\left(2\mathrm{Im}\ v^{(2)}\right) >  \cos\left(2 \mathrm{Im}\ v^{(2)}-2\eta\right).
\ee
so that only $\mathrm{Im}\ v^{(2)}=0$   is allowed.
According to to \req{eq:parity}, it corresponds to the
positive parity solution
with
\[
 -2\alpha_2 <\tilde{\epsilon}^{(2)} <2\varphi_2.
\]
and the spectrum is terminated at points $\pm q_*^{(2)}(\eta);  q_*^{(2)}(\eta) =\ldb 4 \eta\rdb$.

The same calculation for the gapped case, $\phi\geq\alpha$, gives the condition
\be
\cosh 2\eta >1, 
\ee
that is always valid so that the whole spectrum \rref{eq:cos-cos2-trig} is physical.

\subsection{$n$ - particle bound state solution ($n$-string).}

As in the two-particle case, rapidities found from \reqs{eq:BAv1}--\rref{eq:BAv2} may be complex. In this subsection
we will concentrate on the solution where quantization of only one parameter in the limit $L\gg 1$ depends on the system length $L$.
In other words, the emerging solution can be viewed as a new elementary excitation comprised of $n$ single spin excitations bound to each other.
The results for a more general case will be given in the next subsection.

Similarly to two-particle case \rref{eq:A12=1}, we look for the solution where only one amplitude is present:
\[
  A_{12\dots n}=1;\ A_{P}=0 \ \mathrm{otherwise}.
\]
Then, \req{eq:smatrixN} requires
\be
S_{j,j+1}=0;\ j=1,2\dots n-1.
\label{eq:bsn}
\ee

Then, using \reqs{eq:smatrixXXZ} -- \rref{eq:smatrixXXZgapped}, we obtain
the equation for rapidities (so-called $n$-string solution),
\be
v_j=v^{(n)}+\left(j-\frac{n+1}{2}\right)i\eta;\ j=1,2,\dots,n.
\label{eq:string}
\ee
Similarly  to the two particle case of \reqs{eq:eqv2} we obtain for
the quasi-momentum
$q^{(n)}=\sum_{j=1}^nq(v_j)$ and quasi-energy $\epsilon^{(n)}=\sum_{j=1}^n\epsilon(v_j)$
(for the gapless case $\phi< \alpha$)
\be
\begin{split}
 e^{iq^{(n)}}&=\frac{\sinh\left(v^{(n)}+\frac{\theta}{2}+\frac{in\eta}{2}\right)\sinh\left(v^{(n)}-\frac{\theta}{2}+\frac{in\eta}{2}\right)}
 {\sinh\left(v^{(n)}+\frac{\theta}{2}-\frac{in\eta}{2}\right)\sinh\left(v^{(n)}-\frac{\theta}{2}-\frac{in\eta}{2}\right)};
 \\
 e^{i\epsilon^{(n)}}&=e^{i2n\phi}\frac{\sinh\left(v^{(n)}-\frac{\theta}{2}-\frac{in\eta}{2}\right)
 \sinh\left(v^{(n)}+\frac{\theta}{2}+\frac{in\eta}{2}\right)}
{\sinh\left(v^{(n)}-\frac{\theta}{2}+\frac{in\eta}{2}\right)\sinh\left(v^{(n)}+\frac{\theta}{2}-\frac{in\eta}{2}\right)}.
\end{split}
\label{eq:eqvn}
\ee

Further manipulations are similar to the $2$-string solution and aim to reduce the expression for the spectrum to the shift and renormalization of
a single-excitation spectrum.
Similarly to \req{eq:shift2}, we write
\be
\epsilon^{(n)}=
\ldb \tilde{\epsilon}^{(n)}+{\epsilon}^{(n)}(\pi)\rdb;\quad {\epsilon}^{(n)}(\pi)\equiv 2n\phi-2\varphi_n,
\label{eq:shiftn}
\ee
where
\be
\tan\varphi_n=\frac{\tanh\theta}{\tan n\eta},
\quad \tan\alpha_n=\frac{\sinh\theta}{\sin n\eta}.
\label{eq:phialphan}
\ee
Then, the second of \reqs{eq:eqvn} takes the form
\be
e^{i\tilde{\epsilon}^{(n)}}=e^{i2\varphi_n}\frac{\sinh\left(v^{(n)}-\frac{\theta}{2}-\frac{in\eta}{2}\right)
  \sinh\left(v^{(n)}+\frac{\theta}{2}+\frac{in\eta}{2}\right)}
{\sinh\left(v^{(n)}-\frac{\theta}{2}+\frac{in\eta}{2}\right)\sinh\left(v^{(n)}+\frac{\theta}{2}-\frac{in\eta}{2}\right)}.
\label{eq:tildeepsilon-n}
\ee
Equation \rref{eq:tildeepsilon-n} and the first of \reqs{eq:eqv2} are nothing by \reqs{eq:eqv} with the obvious substitutions of
$\eta\to n\eta,\ \varphi\to \varphi_n$. Therefore, the relation \rref{eq:cos-cos} holds
\[
  \cos\left[ \tilde{\epsilon}^{(n)}(q^{(n)})\right] =
  \frac{\cos^2\alpha\sin^2n\eta-\sin^2\alpha\sin^2\eta\cos q^{(n)}}{\cos^2\alpha\sin^2n\eta+\sin^2\alpha\sin^2\eta}.
\]
Finally, excluding $\alpha_n,\varphi_n$ with the help of \req{eq:phialphan}, we obtain the expression for the spectrum of the $n$-particle bound state
($n$-string solution)
\be
\begin{split}
  &{\epsilon}^{(n)} = \ldb \tilde{\epsilon}^{(n)}+{\epsilon}^{(n)}(\pi)\rdb ;\\
  &\cos\left[ \tilde{\epsilon}^{(n)}(q^{(n)})\right] =\frac{\cos^2\alpha\sin^2n\eta-\sin^2\alpha\sin^2\eta\cos q^{(n)}}
{\cos^2\alpha\sin^2n\eta+\sin^2\alpha\sin^2\eta},
  \\
&{\epsilon}^{(n)}(\pi)\equiv 2n\phi-2\arctan\left[\left(\tan\phi\right)\left(\tan\eta\right)\left(\cot n \eta\right)\right].
\end{split}
\label{eq:cos-cosn}
\ee
At $n=2$, one recovers \req{eq:cos-cos2} for the two-particle bound state.

The gapped case $\phi>\alpha$ is considered analogously and it gives
\be
\begin{split}
 e^{iq^{(n)}}&=\frac{\sin\left(v^{(n)}+\frac{\theta}{2}+\frac{in\eta}{2}\right)\sin\left(v^{(n)}-\frac{\theta}{2}+\frac{in\eta}{2}\right)}
 {\sin\left(v^{(n)}+\frac{\theta}{2}-\frac{in\eta}{2}\right)\sin\left(v^{(n)}-\frac{\theta}{2}-\frac{in\eta}{2}\right)};
 \\
 e^{i\epsilon^{(n)}}&=e^{i2n\phi}\frac{\sin\left(v^{(n)}-\frac{\theta}{2}-\frac{in\eta}{2}\right)
 \sin\left(v^{(n)}+\frac{\theta}{2}+\frac{in\eta}{2}\right)}
{\sin\left(v^{(n)}-\frac{\theta}{2}+\frac{in\eta}{2}\right)\sin\left(v^{(n)}+\frac{\theta}{2}-\frac{in\eta}{2}\right)}.
\end{split}
\label{eq:eqvn-trig}
\ee

We find
\be
\begin{split}
&\cos\left[ \tilde{\epsilon}^{(n)}(q^{(n)})\right] =\frac{\cos^2\alpha\sinh^2n\eta-\sin^2\alpha\sinh^2\eta\cos q^{(n)}}
{\cos^2\alpha\sinh^2n\eta+\sin^2\alpha\sinh^2\eta}.
\label{eq:cos-cosn-trig}
\\
&
{\epsilon}^{(n)}(\pi)\equiv 2n\phi-2\arctan\left[\left(\tan\phi\right)\left(\tanh\eta\right)\left(\coth n \eta\right)\right].
\end{split}
\ee

It is instructive to consider the limiting case of the long strings $n\eta\gg 1$ in the gapped case. We
obtain
\be
\begin{split}
&\cos\left[ \tilde{\epsilon}^{(n)}(q^{(n)})\right] =\frac{\cos^2\alpha-e^{-2n\eta}\sin^2\alpha\cos q^{(n)}}
{\cos^2\alpha+e^{-2n\eta}\sin^2\alpha }.
\label{eq:cos-cosn-trig-limit}
\\
&
{\epsilon}^{(2)}(\pi)\equiv 2n\phi-2\arctan\left[\left(\tan\phi\right)\left(\tanh\eta\right)\right].
\end{split}
\ee
This equation gives the intuition about the structure of the bound state.
The  energy shift is contributed by the term proportional to $n$ and the constant term.
The former has the interpretation of the excitation packed closely, the constant term has the meaning of the two domain walls separating the vacuum and the excited
state. The exponential with $n$ scaling of the bandwidth indicates the order of the perturbation theory needed to shift the $n$-particle complex by one lattice
constant.

The very important case is $\phi=\alpha$, i.e. $\eta=0$, see \req{eq:phialpha}. It corresponds to
the isotropic $XXX$ Heisenberg spin model. Taking this limit in \req{eq:cos-cosn-trig}, we find
\be
\begin{split}
&\cos\left[ \tilde{\epsilon}^{(n)}(q^{(n)})\right] =\frac{\cos^2\alpha\ n^2-\sin^2\alpha\cos q^{(n)}}
{\cos^2\alpha\ n^2 +\sin^2\alpha\sinh^2\eta}.
\label{eq:cos-cosn-XXX}
\\
&
{\epsilon}^{(2)}(\pi)\equiv 2n\phi-2\arctan\left[\left(\tan\phi\right)/n\right].
\end{split}
\ee
Equations \rref{eq:cos-cosn}--\rref{eq:cos-cosn-XXX} are the main results of this subsection.
They give the complete description for the spectrum of string of arbitrary length.

What remains is to specify the domains at which the bound state solutions are physical.
To accomplish this task, we write the expression for the wave function
similarly to \req{eq:psi2-bound},
\be
\begin{split}
 \left|\psi_{n;\cdot}\right\rangle&=\sum_{\substack{x_1=1\\x_1\leq x_2\leq\dots\leq x_n}}^{2L}
e^{-iq^{(n)}x_n/2}\prod_{j=1}^{n-1}e^{iq^{(n)}_j(x_{j+1}-x_j)/2}
\\
&\times\prod_{j=1}^n
\tilde{\Upsilon}_{x_j}(v_j)\hat{\sigma}^+_{x_j}
\left|0\right\rangle,
\\
&
q^{(n)}_j\equiv \sum_{k=1}^nq(v_k);\quad q^{(n)}=q^{(n)}_n.
\end{split}
\label{eq:properboundstate}
\ee
The condition for the bound state to be physical is, therefore,
\be
\left|e^{iq^{(n)}_j}\right|^2 <1;\quad j=1,2,\dots, n-1.
\ee
With the help of \req{eq:eqv} it translates, for $\phi<\alpha$,
into the restrictions
\[
  \cos\left(2\mathrm{Im}\,v^{(n)}-(n-2j)\right) >  \cos\left(2 \mathrm{Im}\,v^{(n)}- n\eta\right);
\]
for $1\leq j\leq n$.
\begin{subequations}
  \label{eq:conditions}
For $\mathrm{Im}\ v^{(n)}=0$ it yields
\be
\sin j\eta\ \sin(n-j)\eta >0;\  j=1,2,\dots \left[\frac{n}{2}\right];
\ee
$\mathrm{Im}\ v^{(n)}=\frac{\pi}{2}$
\be
\sin j\eta\ \sin(n-j)\eta <0;\  j=1,2,\dots \left[\frac{n}{2}\right];
\ee
\end{subequations}
If neither of those conditions are satisfied the solution is non-physical.

\begin{figure}
  \includegraphics[width=\columnwidth]{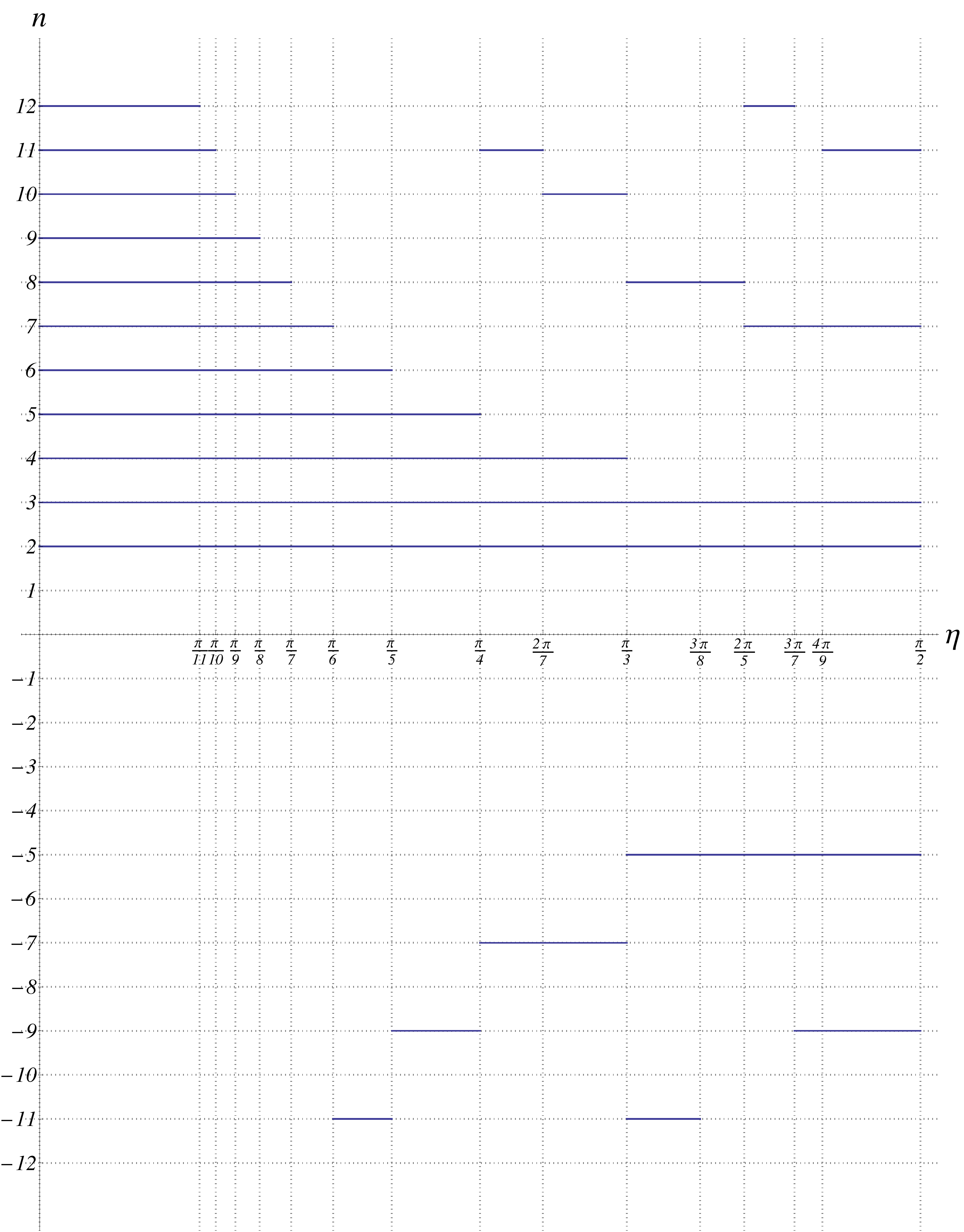}
  \caption{
Bound states existence regions for a few multi-particle  bound states, $2\leq n \leq 12$.
    Solid lines denote the region of existence of $|n|$-particle bound states for adjusting angle $\eta$.
    Positive (negative) $n$ correspond to the solutions of positive (negative) parity, see  \req{eq:Parityn}. For even $n$ only positive parity solutions
  exist.}
  \label{fig2}
\end{figure}

The regions of the existence of the positive, $\mathrm{Im}\ v^{(n)}=0$,  and negative parity $\mathrm{Im}\ v^{(n)}=\frac{\pi}{2}$,
\be
\begin{split}
  -2\alpha_n <\epsilon^{(n)} <2\varphi_n;\ & \mathrm{positive\ parity};
  \\
  2\varphi_n <\epsilon^{(n)} <2\alpha_n;\ & \mathrm{negative\ parity},
\end{split}
\label{eq:Parityn}
\ee
obtained from \reqs{eq:conditions} are shown on Fig.~\ref{fig2}.
Solutions are terminated $q=\pm q_*^{(n)}$, where
\be
q_*^{(n)}(\eta)=\ldb 2n\eta\rdb.
\label{eq:qstarn}
\ee
One can see that the spectrum is dramatically reconstructed at each rational $\eta/\pi$, ``devil staircase'' structure for the gapless phase.
This phenomenon was studied in detail in Ref.~\cite{TakahashiSuzuki1972}.

For the gapped phase, $\phi\geq\alpha$, the state is physical for
\[
 \sinh j\eta\ \sinh(n-j)\eta >0;\  j=1,2,\dots \left[\frac{n}{2}\right],
\]
which is always true, i.e. spectrum \rref{eq:cos-cosn-trig} is valid in the whole region.

\subsection{Bethe-Gaudin-Takahashi equations.}
\label{sec:BGD}

The purpose of this subsection is to derive the system of non-linear equations describing the multiple string solutions
and operating only with positions of the string $v^{(\mu)}$, where $\mu=1,2,\dots$ labels the number of single-particle excitations comprising the
string ($\mu=1$ corresponds to the single particle excitations).

The number of each of $\mu$-strings is connected to the total number of excitations $\mathcal{N}$ by
\be
\sum_{\mu=1}^{\mathcal{N}}\mu\, n_\mu=\mathcal{N}.
\label{eq:stringnumber}
\ee
The previous subsection is a particular case of \req{eq:stringnumber} with $n_{\mu}=\delta_{\mu \mathcal{N}}$.
The total quasi-energy and the quasi-momentum of the state are
given by addition of that for all the strings,
\be
E=\ldb\sum_{\mu=1}^{\mathcal{N}}
\sum_{j=1}^{n_\mu}
\epsilon^{(\mu) }\left(v_j^{(\mu )}\right)\rdb;
\ Q=\ldb\sum_{\mu=1}^{\mathcal{N}}
\sum_{j=1}^{n_\mu}
q^{(\mu) }\left(v_j^{(\mu )}\right)\rdb.
\label{eq:EQexpression}
\ee
where the quasi-energies and quasi-momenta of constituting excitations are given by \reqs{eq:eqvn}/\rref{eq:eqvn-trig}
for the gapless/gapped cases.

We will adopt the standard string hypothesis for the description of all possible solutions.
The further derivation is absolutely equivalent to that for $XXZ$ chains
and consists
of the substituting of the string solutions  \rref{eq:string} to the original Bethe Ansatz equations \rref{eq:BAv1} or \rref{eq:BAv2}
and multiplications of all the original Bethe Ansatz equations within the same string (with the same $v^{(n)}$) by each other.

For the gapless case one finds
\be
\label{eq:Tak}
\begin{split}
  &\left[ \prod_{\pm}\frac{\sinh\left(v_j^{(\mu)}\pm \frac{\theta}{2}+\frac{i\mu\eta}{2}\right)}
    {\sinh\left(v_j^{(\mu)}\pm\frac{\theta}{2}-\frac{i\mu \eta}{2}\right)}\right]^L=
\prod_{\nu=1}^{\mathcal{N}}\prod_{k=1}^{n_\nu}
G^{\mu\nu}
\left(v_j^{(\mu)}-v_k^{(\nu)}\right);
     \\
    &1\leq j<n_\mu; \quad \mu=1,\dots,\mathcal{N};
     \\
     &
     G^{\mu\nu}\left(v\right)\equiv
     \prod_{m=\left|\mu-\nu\right|/2}^{(\mu+\nu-2)/2}
     \frac{\sinh \left(v+im\eta\right)}{\sinh \left(v-im\eta\right)}
      \frac{\sinh \left(v+i(m+1)\eta\right)}{\sinh \left(v-i(m+1)\eta\right)}
      ;
      \\
      &
      G^{\mu=\nu}\left(v=0\right)=1;
\end{split}
\ee
The variable $m$ in the product takes integer (half-integer) values for integer (half-integer) limits.
Each rapidity may of even, $\mathrm{Im} v_j^{(\mu)}=0$, or odd  $\mathrm{Im} v_j^{(\mu)}=\frac{\pi}{2}$ or odd parity.
The possibility for the corresponding string solutions are discussed in the previous subsection.
Equations \rref{eq:Tak} (known as Bethe-Gaudin-Takahashi equations) describe that the
collision of any strings with each other leave the strings intact and may acquire only the phase factors
$G^{\mu\nu}$. In this respect each string behaves as a particle and \reqs{eq:Tak} gives the complete description of the multi-string
continuum.

For the gapped case, all the rapidities $v_j^{(\mu)}$ are real and the Bethe-Gaudin-Takahashi equations are
\be
\begin{split}
  &\left[ \prod_{\pm}\frac{\sin\left(v_j^{(\mu)}\pm \frac{\theta}{2}+\frac{i\mu\eta}{2}\right)}
    {\sin\left(v_j^{(\mu)}\pm\frac{\theta}{2}-\frac{i\mu \eta}{2}\right)}\right]^L=
\prod_{\nu=1}^{\mathcal{N}}\prod_{k=1}^{n_\nu}
G^{\mu\nu}
\left(v_j^{(\mu)}-v_j^{(\nu)}\right);
     \\
    &1\leq j<n_\mu; \quad \mu=1,\dots,\mathcal{N};
     \\
     &
     G^{\mu\nu}\left(v\right)\equiv
     \prod_{m=\left|\mu-\nu\right|/2}^{(\mu+\nu-2)/2}
     \frac{\sin \left(v+im\eta\right)}{\sin \left(v-im\eta\right)}
      \frac{\sin \left(v+i(m+1)\eta\right)}{\sin \left(v-i(m+1)\eta\right)}
      ;
      \\
      &
      G^{\mu=\nu}\left( v= 0\right)=1.
\end{split}
\label{eq:Tak2}
\ee

\subsection{What could be observed in modern quantum computer experiments?}
\label{sec:QCXXZ}

The conventional way to proceed with the solution of \reqs{eq:Tak}--\rref{eq:Tak2} is to look for the ground state (minimal energy) and spectrum of excitations
around the ground state or study the thermodynamics of the system. This route is meaningless for the Floquet system as the quasi-energy is not an
extensive quantity. Therefore, there is no mechanism of the relaxation leading to the minimal quasienergy even in the presence of the external bath.

What remains, however, handy for the quantum computer is to study the correlation functions of many particle operators acting on the vacuum state.
Let us define the Pauli string operator
\be
\hat{\Sigma}^{(k)}_x=
\prod_{m=0}^{k-1}\hat{\sigma}_{\ldcb x+m\rdcb}^{x}.
\label{eq:Sigma}
\ee
where $\hat{\sigma}^{x,y,z}_x$ are the standard Pauli matrices acting
in the Hilbert space of a single qubit $\mathcal{h}_x$.
The Pauli string operator is both Hermitian and unitary, $\left[\hat{\Sigma}^{(k)}_x\right]^2=\hat{\openone}$.

  
\begin{figure}

    \captionsetup[subfigure]{width=0.4\columnwidth}
  %
  \subfloat
  [Spectral function for $k=1$. Only single-particle excitations are present (blue).]{
    \begin{picture}(60,70)  
      \put(0,6){    \includegraphics[width=65\unitlength]{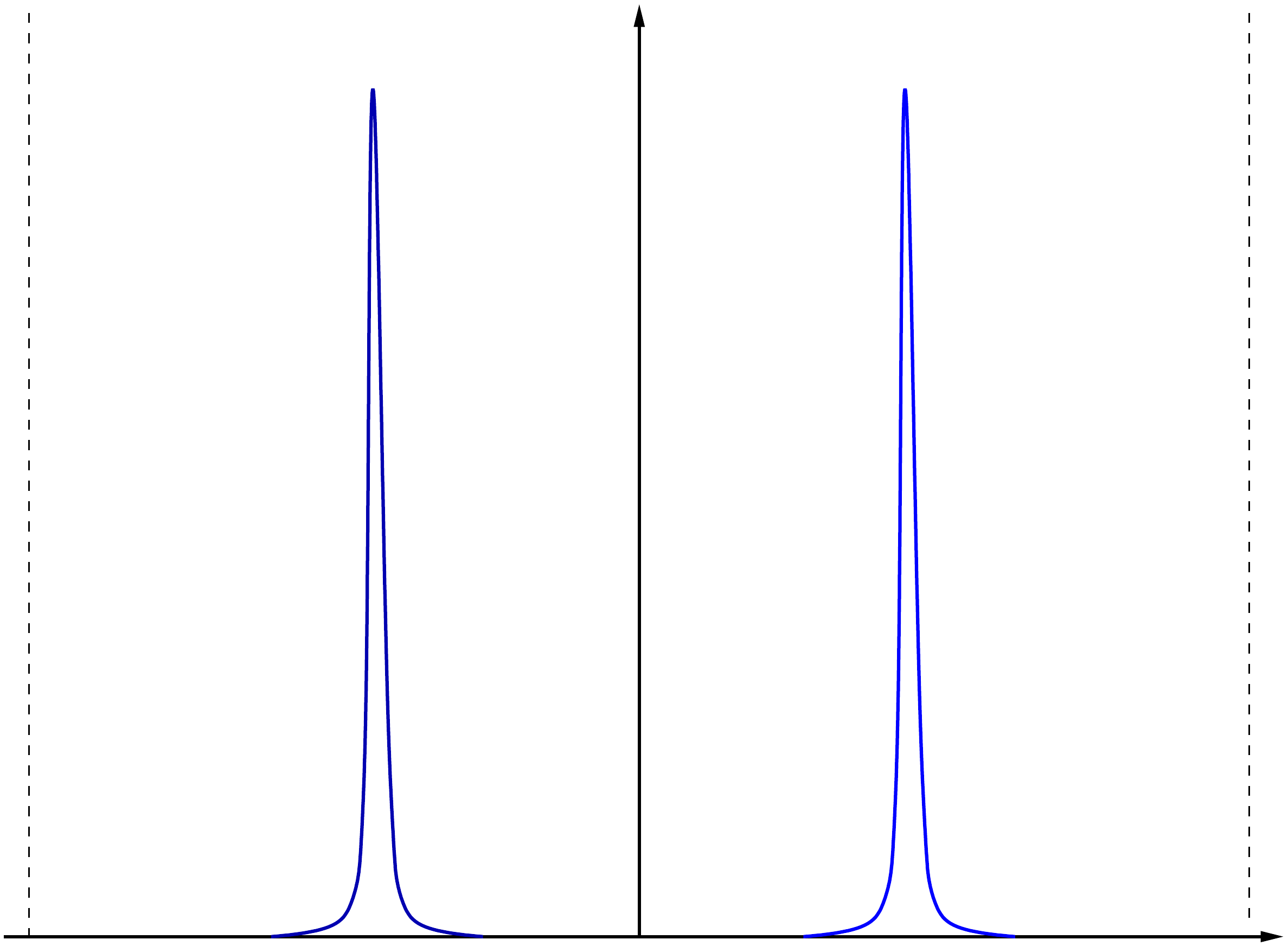}}
      \put(-1,3){$-\pi$}
      \put(65,3){$\pi$}
      \put(67,9){$\omega$}
      \put(11,52){$\epsilon^{(1)}_-(Q)$} \put(42,52){$\epsilon^{(1)}_+(Q)$}
      \put(28,66){$\mathcal{D}^{(1)}(\omega,Q)$}
    \end{picture}
\label{fig:D1}
}
\hfill
\subfloat[Spectral function for $k=2$. Single excitations (2-string) give peaks non-vanishing in the thermodynamic limit  (blue).
Heights of each peak in the
continuum of one strings $n_1=2$  (red) scale as $1/L$.]
{
    \begin{picture}(60,70)  
      \put(0,6){    \includegraphics[width=65\unitlength]{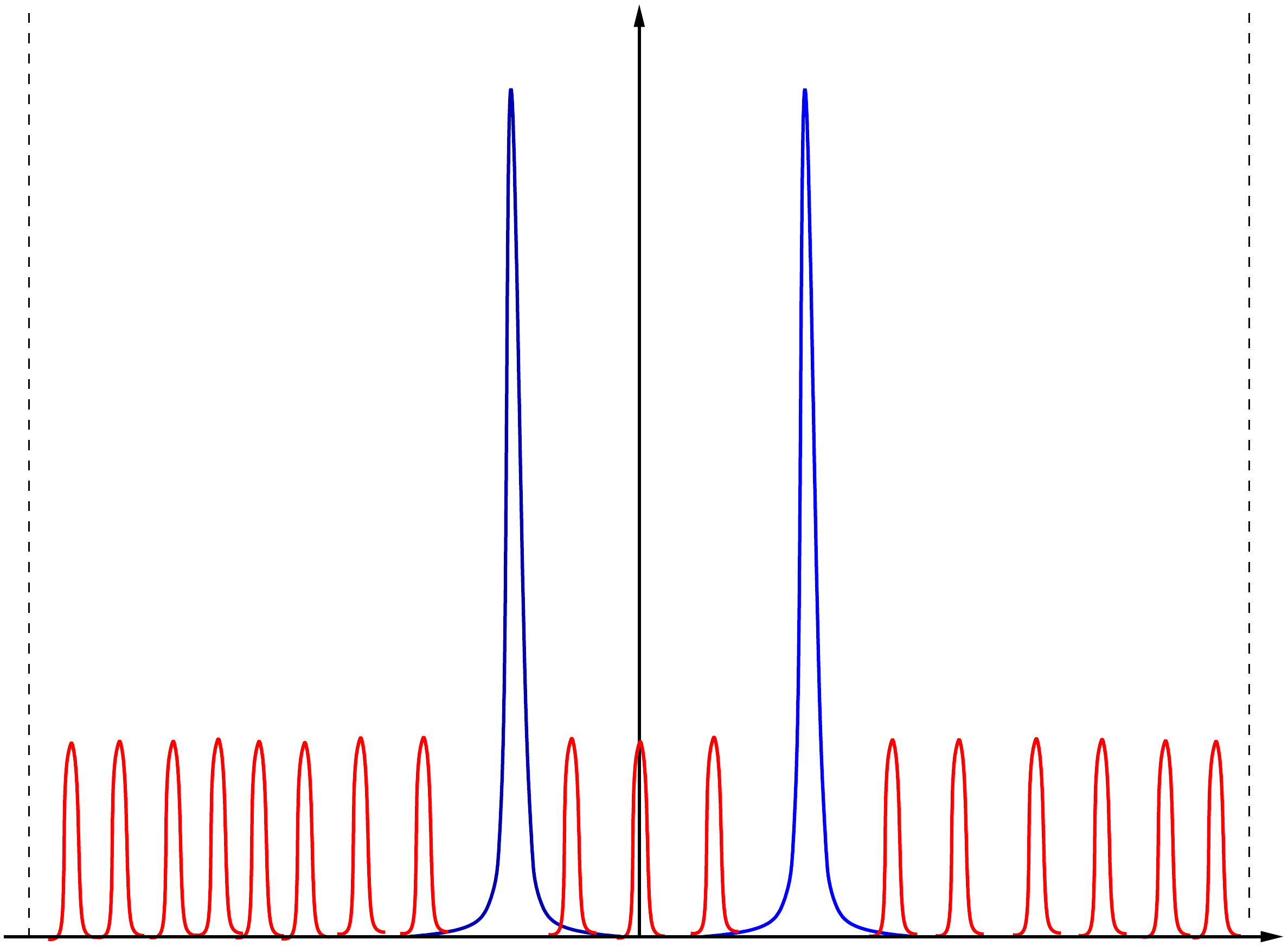}}
      \put(-1,3){$-\pi$}
      \put(65,3){$\pi$}
      \put(67,9){$\omega$}
      \put(17,52){$\epsilon^{(2)}_-(Q)$} \put(40,52){$\epsilon^{(2)}_+(Q)$}
      \put(28,66){$\mathcal{D}^{(2)}(\omega,Q)$}
    \end{picture}
  \label{fig:D2}
}
\hspace*{\fill}
\\
\subfloat[Spectral function for $k=3$. Single excitations (3-string) give peaks non-vanishing in the thermodynamic limit  (blue).
Heights of each peak in the continuum of one- and two- strings $n_1=1,n_2=1$  (red) scale as $1/L$.
Heights of each peak in the three particle continuum of one strings $n_1=3$  (gold) scale as $1/L^2$.]
{
    \begin{picture}(60,70)  
      \put(0,6){    \includegraphics[width=65\unitlength]{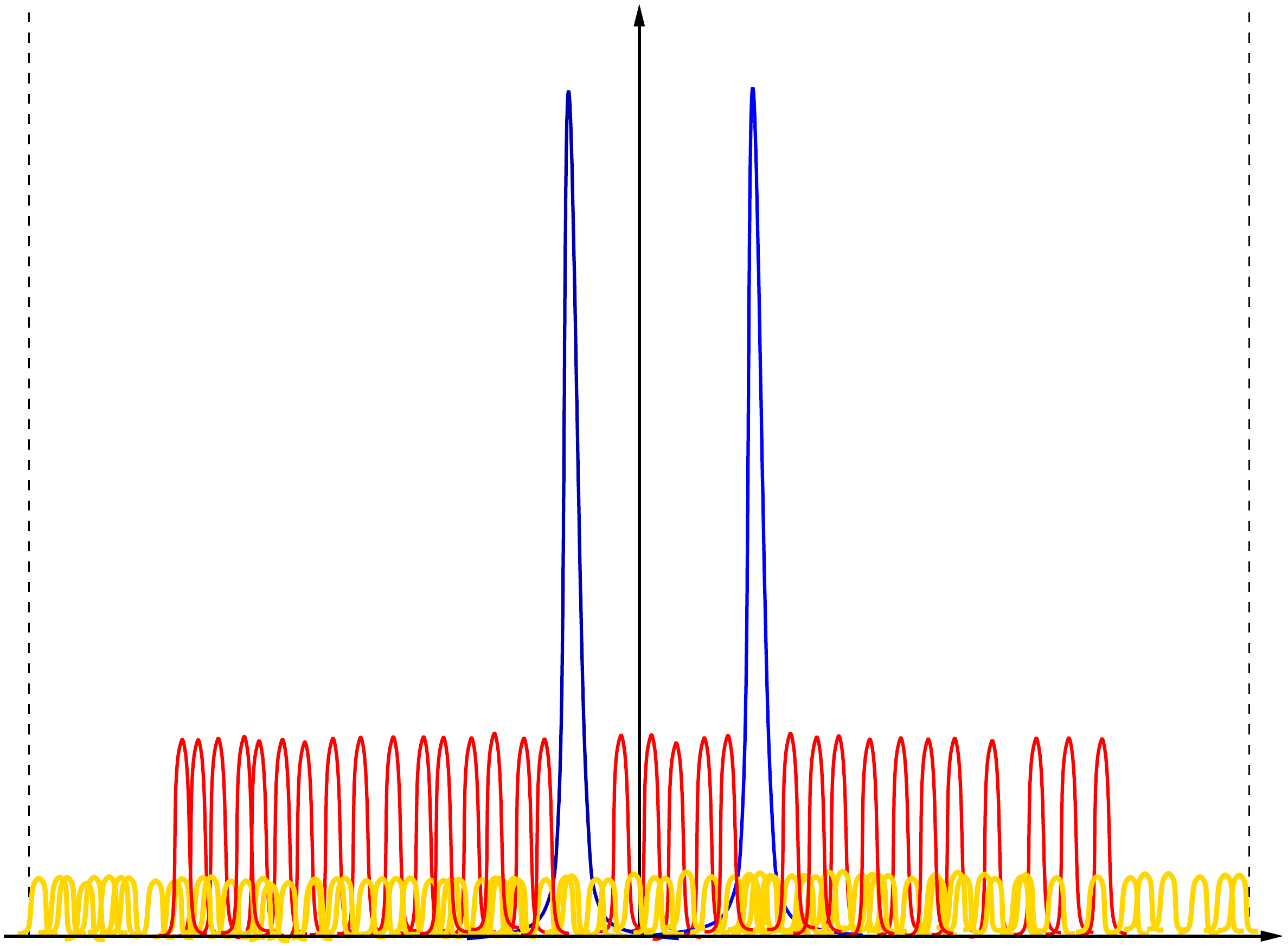}}
      \put(1,3){$-\pi$}
      \put(65,3){$\pi$}
      \put(67,9){$\omega$}
      \put(18,52){$\epsilon^{(3)}_-(Q)$} \put(41,52){$\epsilon^{(3)}_+(Q)$}
      \put(28,66){$\mathcal{D}^{(3)}(\omega,Q)$}
    \end{picture}
  }
  \hfill
  \subfloat[Spectral function for $k=4$. Single excitations (4-string) give peaks non-vanishing in the thermodynamic limit  (blue).
     Heights of each peak in the continuum of  $n_1=1,n_3=1$  (red) and $n_2=2$ scale as $1/L$.
     Heights of each peak in the three particle continuum $n_1=2,n_2=1$  (gold) scale as $1/L^2$.
   The contribution of the four particle continuum, $n_1=4$, is hardly resolvable (green).]{
   \begin{picture}(60,70)
     \put(0,6){    \includegraphics[width=65\unitlength]{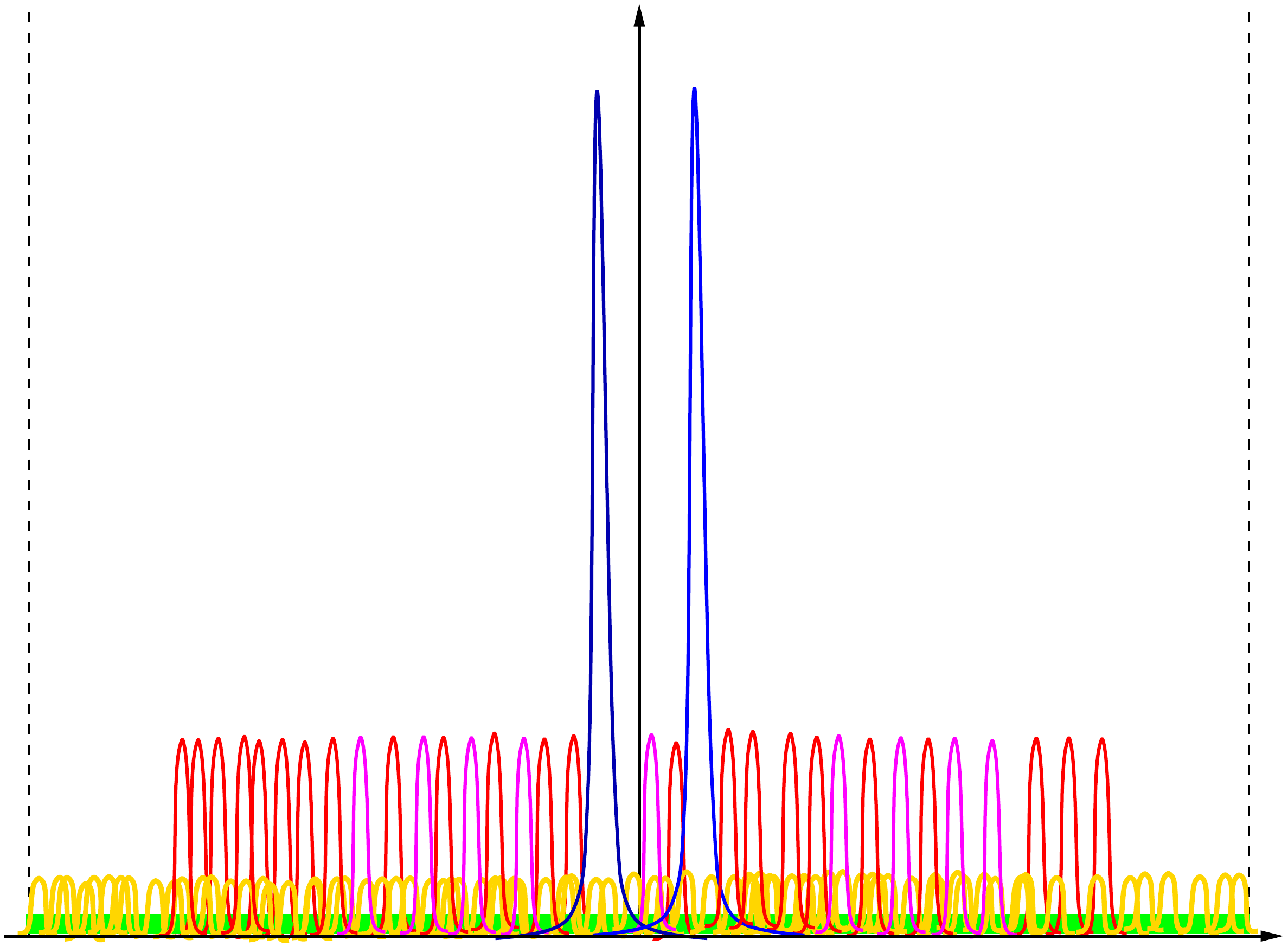}}
     \put(-1,3){$-\pi$}
       \put(65,3){$\pi$}
       \put(67,9){$\omega$}
      \put(18,52){$\epsilon^{(4)}_-(Q)$} \put(39,52){$\epsilon^{(4)}_+(Q)$}
       \put(28,66){$\mathcal{D}^{(4)}(\omega,Q)$}
     \end{picture}
     }
  \hspace*{\fill}
\caption{Schematic draft of the spectral density $\mathcal{D}^{(k)}(\omega,Q)$ for the fixed quasi-momentum $q$. Notice that the scale of
the quasi-energies is not linear in either size of the system or the number of excitations.}
\label{fig:D}
\end{figure}


Let us define the correlation function for the Pauli string operators
\be
\mathcal{D}^{(k)}\left(n,x\right)
=\left\langle 0\right|\hat{\Sigma}^{(k)}_x\hat{\mathcal{U}}^n
  \hat{\Sigma}^{(k)}_0\left|0\right\rangle.
\label{eq:function}
\ee
The procedure of measuring such a function is discussed in ~\ref{sec:ap1}.

Using the properties of the eigenfunctions of the evolution operator, see Section \ref{sec1}, we find
\be
\mathcal{D}^{(k)}\left(n,2 x\right)
=\frac{1}{L}
\sum_Qe^{-i x Q}
\sum_{\bv} w_{\bv}(Q,k)e^{in E(Q;k;\bv )},
\label{eq:intermsofeigenfunctions}
\ee
where the oscillator strengths are  given by
\be
w_{\bv}(Q,k)=
L
\left|\left\langle\psi_{Q,\mathcal{N}=k,\bv}\right|\hat{\Sigma}^{(k)}_0
\left|0\right\rangle\right|^2.
\ee
Notation ${\bv}$ means all the rapidities found from \reqs{eq:Tak} or \rref{eq:Tak2}.

Next, let us calculate the Fourier transform
\be
\mathcal{D}^{(k)}\left(\omega,Q\right)=
\sum_{n=0}^{\infty}e^{-n(i\omega+\gamma)}\sum_{x=1}^{L}e^{iQx}
\mathcal{D}^{(k)}\left(n,2x\right)
\label{eq:F1}
\ee
where the quasi-momenta are quantized as $e^{iQL}=1$.
Line-width, $\gamma\ll 1$, is introduced in order to provide proper convergence for the time Fourier transform. It can be also given
the physical meaning of the decay of due to the effect of {\em e.g.} external noise on the Quantum computer.

Substituting \req{eq:intermsofeigenfunctions} into \req{eq:F1} we find
\be
\mathcal{D}^{(k)}\left(\omega,Q\right)=
\sum_{\bv}\frac{ w_{\bv}(Q,k)}{2} \coth
\left[\frac{\gamma+i\left(\omega-E(Q;k;\bv )\right)}{2}\right],
\ee
notice the periodicity of this result.

The precise information about the oscillator strength is not available. However, the scaling of the
matrix elements with the size of the system can be easily related to the total number of strings in the solution
\be
w_{\bv}(Q,k)\propto L^{1-\sum_{\nu}n_\mu}.
\label{eq:scaling}
\ee
According to constrain \rref{eq:stringnumber},
only the $k$-particle bound state ($k$-string) contribution remains finite
at $L\to\infty$. All the other discrete states give vanishing contribution separately -- all together producing
a featureless background. We emphasize however that the oscillator strength of the $k$-string remains finite
and the width of the corresponding peak is determined by only extrinsic parameter $\gamma$. This is a feature of the
integrable model preventing the decay into the continuous spectrum.
The schematic picture of the spectral density is sketched on Fig.~\ref{fig:D}.

Therefore, the bound state manifests itself as a profound peak in the spectral density and can be readily measured.
The advantage of the quantum computer is that any needed correlation function can be quite easily manufactured unlike the
standard condensed matter physics system where the toolbox of the physical measurements is quite limited.
The most recent studies of the effects of the strings on the response functions can be found in Ref.~\cite{bsStructureFactor}.

\section{Chiral Hubbard model.}

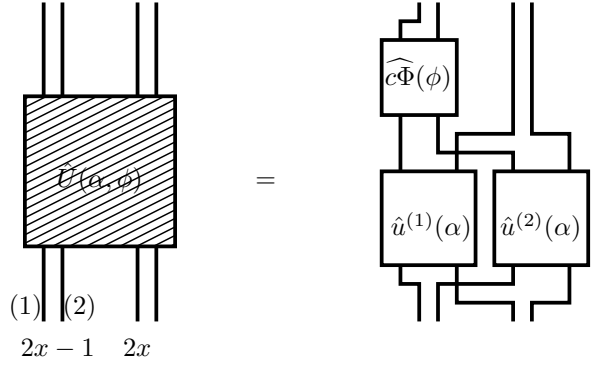
\begin{figure}
\qquad\qquad
  \begin{picture}(130,60)
    \put(0,5){\includegraphics[width=95\unitlength]{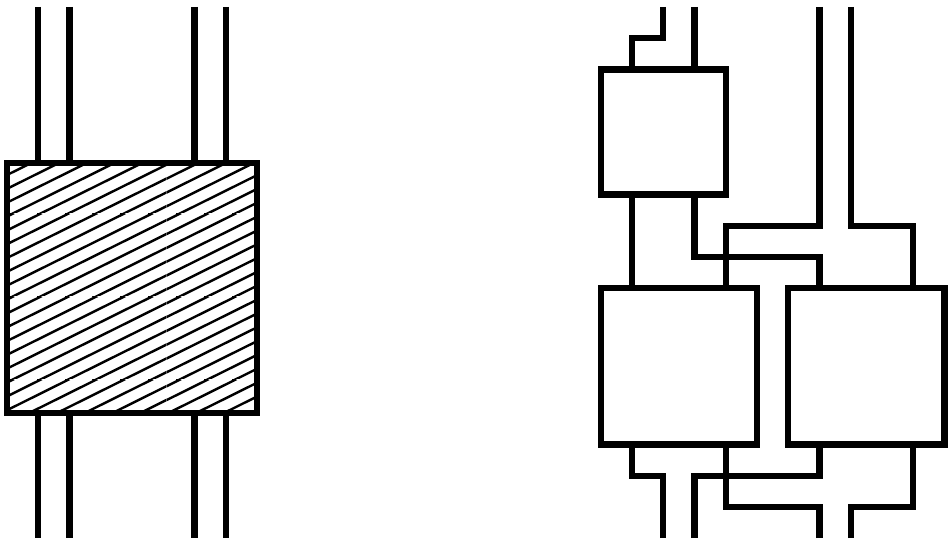}}
    \put(0,0){$2x-1$} \put(17,0){$2x$}
    \put(6,28){$\hat{U}(\alpha,\phi)$}  \put(39,28){$=$}
    \put(61.5,20){$\hat{u}^{(1)}(\alpha)$} \put(79.7,20){$\hat{u}^{(2)}(\alpha)$}\put(60.5,45){$\widehat{c\Phi}(\phi)$}
    \put(-2,7){$(1)$}\put(7,7){$(2)$}
\end{picture}
\hspace*{\fill}
\caption{The quantum circuit for the Chiral Hubbard gate \rref{eq:FHgate}. Chirality appears in the asymmetry of the odd and even
    site in the first slice. Second slice is shifted by one lattice constant as shown in Fig.~\ref{Fig1}.}
  \label{Fig:FHgate}
\end{figure}

To obtain the Hubbard model \cite{Hubbard-book}, we need to simulate the additional (``spin'') degree of freedom. To achieve it, we replicate
the qubit on each site $x$. Replica labels $r=1,2$ are analogous (with some subtlety to be displayed later) to the spin degrees of freedom
so we will introduce Pauli matrices $\tau^{x,y,z}_{r_1r_2}$ acting in the replica space of one site.
The Hilbert space of site $x$ is now four-dimensional $\mathcal{h}_x= \mathcal{h}_{x,1}\otimes \mathcal{h}_{x,2}$,
so that inter-site gate $\hat{U}_{2x-1/2}$ analogous to \rref{eq:XXZ}, acts in $16$-dimensional Hilbert space
$\mathcal{ h}_{\ldcb 2x\rdcb}\otimes\mathcal{ h}_{\ldcb 2x-1\rdcb}$. Similarly to charge and spin conservation in the Hubbard model,
the number of excitations in each of replicas is conserved. To guarantee the integrability
(by its relation to Shastry $R$-matrix \cite{Shastry} for the Hubbard model \cite{prosenH}), we choose this gate as,
 see also Fig.~\ref{Fig:FHgate},
\be
\hat{U}_{2x-1/2}(\alpha,\phi)=
\widehat{c\Phi}(\phi)\left[\hat{u}^{(1)}(\alpha)\otimes \hat{u}^{(2)}(\alpha)\right],
\label{eq:FHgate}
\ee
where each operator $\hat{u}^{(r=1,2)}(\alpha)$ acts in four-dimensional space of corresponding replicas
spanned by basis $\left(
  \left|0_r0_r\right\rangle,
\
\left|0_r1_r\right\rangle,
\
\left|1_r0_r\right\rangle,
\
\left|1_r1_r\right\rangle
\right)$,
\be
\hat{u}^{(r)}(\alpha)=
\begin{pmatrix}
1 & 0&0&0\\
0& \cos\alpha & i \sin\alpha & 0
\\
0&  i \sin\alpha & \cos\alpha  & 0\\
0&0&0& 1
\end{pmatrix},
\ee
i.e. corresponds to \req{eq:XXZ} with $\phi=0$ (no-interaction within one replica).

Operator $\widehat{c\Phi}(\phi)$ acts only in four-dimensional Hilbert space of odd-sites $\mathcal{h}_{2x-1}$ and glues replicas $r=1,2$.
In the basis $\left(\left|0_10_2\right\rangle,
\
\left|0_11_2\right\rangle,
\
\left|1_10_2\right\rangle,
\
\left|1_11_2\right\rangle
\right)$,
\be
\widehat{c\Phi}(\phi)=\begin{pmatrix}
1 & 0&0&0\\
0& 1 & 0 & 0
\\
0& 0 & 1 & 0\\
0&0&0& e^{i2\phi}
\end{pmatrix}
\ee
Clearly, the parameter $\phi$ is equivalent to the on-site interaction of the electrons with opposite spin in the Hamiltonian Hubbard model.
The chirality means that this interaction breaks the mirror symmetry of the model,
i.e. the spectrum in general is not invariant with respect to $Q\to -Q$.

Similarly to $XXZ$ model, it is sufficient to consider only the certain signs of the phases, $\phi,\alpha >0$.
Indeed, $\alpha\to -\alpha$ corresponds to the transformation
\[ \hat{\mathcal{U}}
  \to \left[\bigotimes_{r=1,2}\bigotimes^{L}_{x=1}\hat{\sigma}_{2x,r}^z\right]\hat{\mathcal{U}}
  \left[\bigotimes_{r=1,2}\bigotimes^{L}_{x=1}\hat{\sigma}_{2x,r}^z\right]
\]
leaving the spectra intact.

On the other hand, replacement $\alpha \to -\alpha, \phi \to -\phi$ together with the mirror reflection
leads to the parity transformation \rref{eq:parity} with $\hat{W}=\widehat{c\Phi}$.
We thus obtain the rule
\be
\phi \to -\phi; \quad \mathcal{E} \to -\mathcal{E}; \quad Q\to -Q;
\label{eq:minusphi}
\ee

The conservation of the number of excitations in each replicas enables to reorder the basis of $16$-dimensional space for two sites.
\be
\mathcal{ h}_{\ldcb 2x\rdcb}\otimes\mathcal{ h}_{\ldcb 2x-1\rdcb}
=\mathcal{ h}_{2x}^{(0)}\otimes \mathcal{ h}_{2x}^{(1)}
\otimes \mathcal{ h}_{2x}^{(2)}
\otimes \mathcal{ h}_{2x}^{(3)}\otimes \mathcal{ h}_{2x}^{(4)},
\label{eq:reorder}
\ee
where the superscript labels the total number of the excitations in the involved qubits.
The Hilbert subspaces $\mathcal{ h}_{2x}^{(0)},\mathcal{ h}_{2x}^{(4)}$ are one dimensional.
The corresponding sub-blocks of the unitary gates are
\begin{subequations}
\be
\hat{\mathcal{u}}_0=1;\qquad \hat{\mathcal{u}}_4=e^{i2\phi}.
\ee
Each of the subspaces $\mathcal{ h}_{2x}^{(1)}, \mathcal{ h}_{2x}^{(3)}$ are four-dimensional.
In the basis $\left|0,1_r\right\rangle$, $\left|1_r,0\right\rangle$
the sub-block is $\delta_{r_1r_2}\hat{\mathcal{u}}_1$, where
\be
\hat{\mathcal{u}}_1(\alpha)=\begin{pmatrix}
\cos\alpha & i \sin\alpha
\\
 i \sin\alpha & \cos\alpha 
\end{pmatrix}.
\ee
Analogously, the $\mathcal{ h}_{2x}^{(3)}$ is characterized by basis
$\left|1_11_2,1_r\right\rangle$, $\left|1_r,1_11_2\right\rangle$
the sub-block is $\delta_{r_1r_2}\hat{\mathcal{u}}_3$,
with
\be
\hat{\mathcal{u}}_3(\alpha,\phi)=\begin{pmatrix}
\cos\alpha & i \sin\alpha
\\
 i e^{2i\phi}\sin\alpha &  e^{2i\phi}\cos\alpha 
\end{pmatrix}.
\ee
Finally, the subspace with two excitations is six-dimensional $\mathcal{ h}_{2x}^{(2)}$ (we chose not
to separate it into singlet and triplet subspace).
In the basis $\left|1_{r_1},1_{r_2}\right\rangle,  \left|1_11_2,0\right\rangle,  \left|0,1_11_2\right\rangle$
the sub-block has the form
\be
\hat{\mathcal{u}}_2(\alpha,\phi)=\begin{pmatrix}
\begin{matrix}&\delta_{r_1r_1^\prime}\delta_{r_2r_2^\prime} \cos^2\alpha
 \\ & +(-1)^{r_1+r_2} \delta_{r_1r_2^\prime}\delta_{r_2r_1^\prime} \sin^2\alpha;
\end{matrix}
  & i\tau^x_{r_1r_2}\cos\alpha\sin\alpha;
& i\tau^x_{r_1r_2}\cos\alpha\sin\alpha
\\
\\
 i\tau^x_{r_1r_2}\cos\alpha\sin\alpha; & \cos^2\alpha; & - \sin^2\alpha\\ \\
  i\tau^x_{r_1r_2} e^{2i\phi}\sin\alpha\cos\alpha; & - e^{2i\phi}\sin^2\alpha; & e^{2i\phi} \cos^2\alpha
\end{pmatrix}
\ee
\label{eq:UHubbard}
\end{subequations}

\subsection{Single excitation, $\mathcal{N}=1$}

The discussion here is closely related to that of Sec.~\ref{sec:xxz1}.

We look for the single excitation wave function in replica $r=1,2$ in the form,
\be
\left|\psi_{1_r;\varepsilon}\right\rangle=
\sum_{x=1}^{2L}\Upsilon_x(\epsilon)\hat{\sigma}^+_{x,r}\left|0\right\rangle
.
\label{eq:psi1Hubbard}
\ee
where all the entries are defined in \req{eq:psi1}.

Translational property of wavefunction \rref{eq:psi1Hubbard} is investigated similarly to \req{eq:psidagger-translation} leading to
quantization condition \rref{eq:q-q}.
To diagonalize the first of \reqs{eq:EQ}, we use
\reqs{eq:UeUo}, \rref{eq:UHubbard}, and \rref{eq:psi1Hubbard}. We find
\be
\begin{split}
 & \left(\hat{\mathcal{V}}-e^{i\varepsilon}
\right)
\left|\psi_{1_r,\varepsilon}\right\rangle
\\
&
    =\sum_{x=1}^{L}e^{-iq(\varepsilon)x}
    \left[1,\Upsilon(\varepsilon)\right]\left\{\hat{\mathcal{u}}^{(1)}-
    e^{i\varepsilon}\hat{\mathbb{T}}^{(1)}(\varepsilon)\right\}
    \begin{bmatrix}\hat{\sigma}^+_{2x,r}\\ \hat{\sigma}^+_{2x-1,r}\end{bmatrix}
    \left|0\right\rangle,
\end{split}
\label{eq:vpsi1Hubbard}
\ee
with all the notation of \req{eq:vpsi1}.

The condition for the left-hand-side of this equation to vanish
leads to \req{eq:eqUpsilon} and the identical equation for spectrum
that we copy to make the section self-contained.
\begin{subequations}
\be
e^{iq}=e^{-i2\varepsilon}\frac{1+i\sin\alpha\, e^{i\varepsilon}}{1-i\sin\alpha\, e^{-i\varepsilon}}.
\label{eq:q-epsilon-H}
\ee
The quasi-energy of the original circuit \rref{eq:quasienergy} is than obtained using \req{eq:quasienergies-e-q} 
\be
e^{i\epsilon }=\frac{1+i\sin\alpha\, e^{i\varepsilon}}{1-i\sin\alpha\, e^{-i\varepsilon}}.
\label{eq:e-epsilon-H}
\ee
\label{implicit-H}
\end{subequations}
It is needless to say  that \req{eq:cos-cos}  remains valid.
We will see shortly that the $S$-matrices will be conveniently expressed in terms of rapidities $\varepsilon$ so no further re-parameterization
is required.

\subsection{Two excitations, $\mathcal{N}=2$.}
We  look for the two-particle wave-function in a form,
\be
\begin{split}
 &\left|\psi_{\varepsilon_1,\varepsilon_2}\right\rangle=\sum_{\substack{x_1=1\\x_1\leq x_2}}^{2L}
\sum_{\mathcal{P}}A_{\mathcal{P}}^{r_1r_2}
\hat{\mathcal{P}}_\varepsilon
\\
&\times
{B}_{x_1,x_2}(\varepsilon_1,\varepsilon_2)
\left(\prod_{j=1}^2
\Upsilon_{x_j}(\epsilon_j)\hat{\sigma}^+_{x_j,r_j}\right)
\left|0\right\rangle,
\end{split}
\label{eq:psi2H}
\ee
where the notation for permutations is defined in Sec.~\ref{sec:xxz2}.
The summation over the repeated replica indices $r_\cdot$ is always assumed unless stated otherwise.

An unusual object here is the
two particle vortex ${B}_{x_1,x_2}(\varepsilon_1,\varepsilon_2)={B}_{x_1,x_2}(\varepsilon_2,\varepsilon_1)$
defined as
\be
\begin{split}
&{B}_{x_1,x_2}(\varepsilon_1,\varepsilon_2)
=
\left\{
\begin{matrix}
  1, &x_1\neq x_2;\\
   {B}_x\left(\varepsilon_1,\varepsilon_2\right), &x_1=x_2;
  \end{matrix}
\right.
\\
&
{B}_x\left(\varepsilon_1,\varepsilon_2\right)
=\left\{
  \begin{matrix}
 {b}_e\left(\varepsilon_1,\varepsilon_2\right),  & x &\mathrm{even}\\
 {b}_o\left(\varepsilon_1,\varepsilon_2\right),
     & x &\mathrm{odd},
\end{matrix}
\right.
\end{split}
\label{eq:B}
\ee
and it expresses the renormalization of the amplitude of the wavefunction at the sites of double
occupation. Significance of such renormalization and explicit form of
$b_{o,e}\left(\varepsilon_1,\varepsilon_2\right)$ will become clear later.

The translational property is derived similarly to \req{eq:2translation} and we obtain similarly to \req{eq:quantization2}
\be
A_{\mathcal{P}}^{r_1r_2}\hat{\mathcal{P}}_qe^{-iq_2L}=A^{r_2r_1}_{\mathcal{C}\cdot\mathcal{P}},
\label{eq:quantization2H}
\ee
and
\be
Q=\ldb  q_1+q_2\rdb,
\ee
quantized according to \req{eq:q-quantized}.

Next step is to diagonalize the operator $\hat{\mathcal{V}}$. We apply  $\left(\hat{\mathcal{V}}-e^{i\mathcal{E}}\right)$
  to wave-function \rref{eq:psi2H}. Using
\reqs{eq:vpsi1Hubbard} we obtain that all the terms with $|x_1-x_2|>1$ vanish for ${\cal E}=\epsilon_1+\epsilon_2$
and arbitrary amplitudes $A_{\mathcal{P}}^{r_1r_2}$ so
that [compare with \req{eq:Vpsi2}]
\be
\begin{split}
 &\hat{t} \left(\hat{\mathcal{V}}-e^{i\varepsilon_1+i\varepsilon_2}\right)
\left|\psi_{\varepsilon_1,\varepsilon_2}\right\rangle =
\sum_{x=2}^{L}
\sum_{\mathcal{P}}A_{\mathcal{P}}^{r_1r_2}\hat{\mathcal{P}}_\varepsilon
 \hat{\Psi}_{2x,r_1r_2}^{(2)}(1,2)
 \left|0\right\rangle;
\\
&\hat{\Psi}_{2x,r_1r_2}^{(2)}(1,2)
=e^{-i(q_1+q_2)x}\times
\\
&
\left[{\Upsilon_1},b_e(1,2),b_o(1,2)\Upsilon_1\Upsilon_2\right]
 \left[
     \hat{\mathcal{u}}^{(2)}- e^{i\varepsilon_1+i\varepsilon_2}
    \hat{\mathbb{T}}^{(2)}(1,2)
          \right]
          \begin{bmatrix}\hat{\sigma}^+_{2x-1,r_1^\prime}
            \hat{\sigma}^+_{2x,r_2^\prime}\\[3mm]
\hat{\sigma}^+_{2x,r_1}\hat{\sigma}^+_{2x,r_2}\\[3mm] 
  \hat{\sigma}^+_{2x-1,r_1}\hat{\sigma}^+_{2x-1,r_2}
  \end{bmatrix};
\\
&\hat{\mathbb{T}}^{(2)}(1,2)
=\begin{pmatrix}
\delta_{r_1r_2^\prime}
\delta_{r_2r_1^\prime}
\frac{e^{iq_1}\Upsilon_2}{\Upsilon_1};
    & \hat{0}_{1\times 2}\\[3mm]
     \hat{0}_{2\times 1};
     &
     \begin{pmatrix}
0&e^{i(q_1+q_2)} \\[3mm]
1 & 0
\end{pmatrix}
\end{pmatrix},
   \end{split}
   \raisetag{0.3cm}
   \label{eq:vpsi2Hubbard}
\ee
where we used the short-hand notation \rref{eq:short}. The sub-block in the space of two particle excitations
is defined in \req{eq:UHubbard}. The $r$-dependent signs are introduced in order to accommodate all six states of this subspace.


In the case of $\tau^x_{r_1r_2}A_{\mathcal{P}}^{r_2r_1}=0$ (as we will see  it  translates to the triplet state of the conventional Hubbard model)
the right-hand-side of \req{eq:vpsi2Hubbard} vanishes if
\be
\sum_{\mathcal{P}}\hat{\mathcal{P}}_\varepsilon
\left[\Upsilon_1-e^{i(\varepsilon_1+\varepsilon_2+q_1)}\Upsilon_2\right]A_{\mathcal{P}}^{r_1r_2}=0;
\label{eq:HSmatrixtriplet1}
\ee

It follows from \req{eq:eqUpsilon} that
\be
\begin{split}
e^{i(\varepsilon_1+\varepsilon_2+q_1)}\Upsilon_2
&=\left(\cos^2\alpha- \sin^2\alpha\right) \Upsilon_1
\\
&+i\sin\alpha\cos\alpha \left[1+ \Upsilon_1 \Upsilon_2\right].
\end{split}
 \label{eq:eqUpsilon1}
\ee
Substituting \req{eq:eqUpsilon1} into \req{eq:HSmatrixtriplet1}, we find that the only possible
solution is
\be
 \sum_{\mathcal{P}}A_{\mathcal{P}}^{r_1r_2}
 =\frac{\tau^x_{r_1r_2}}{2} \sum_{\mathcal{P}}
 \left(\tau^x_{r_3r_4}A_{\mathcal{P}}^{r_4r_3}\right).
 \label{eq:HSmatrixtriplet2}
  \ee
where formula is written to  include the condition $\tau^x_{r_1r_2}A_{\mathcal{P}}^{r_2r_1}=0$ automatically.

Remaining case to consider is
\be
A_{\mathcal{P}}^{r_1r_2}=\tau^x_{r_1r_2}A_{\mathcal{P}}^{s};\quad A_{\mathcal{P}}^{s}=
\frac{1}{2}\left(\tau^x_{r_3r_4}A_{\mathcal{P}}^{r_4r_3}\right).
 \label{eq:singlet1}
 \ee
 Equation \rref{eq:HSmatrixtriplet2} is clearly satisfied,
 and vanishing of the left-hand-side in \req{eq:vpsi2Hubbard}
 requires three conditions
 \be
 \begin{split}
   &\sum_{\mathcal{P}}\hat{\mathcal{P}}_\varepsilon
   \left\{
     \Upsilon_1\cos 2\alpha+
i\sin 2 \alpha \left[ b_o(1,2)\Upsilon_1\Upsilon_2+b_e(1,2)\right]
-e^{i(\varepsilon_1+\varepsilon_2+q_1)}\Upsilon_2
\right\}A_{\mathcal{P}}^{s}=0;
   \\
   &
   \sum_{\mathcal{P}}\hat{\mathcal{P}}_\varepsilon
   \left\{
     b_e (1,2)\cos^2\alpha- b_o(1,2)\Upsilon_1\Upsilon_2\sin^2\alpha
   \right.
   \\
   & \qquad\qquad \left.
+ i\sin \alpha \cos \alpha\Upsilon_1
-e^{i(\varepsilon_1+\varepsilon_2)}\Upsilon_1\Upsilon_2
b_o(1,2)
\right\}
A_{\mathcal{P}}^{s}=0;
\\
&
\sum_{\mathcal{P}}\hat{\mathcal{P}}_\varepsilon
   \left\{
     b_o (1,2)\Upsilon_1\Upsilon_2\cos^2\alpha- b_e(1,2)\sin^2\alpha
     + i\sin \alpha \cos \alpha\Upsilon(q_1)
     \right.
       \\
       &\qquad\qquad\left.
-e^{i(\varepsilon_1+\varepsilon_2+q_1+q_2)-2 i\phi}
b_e(1,2)
\right\}
A_{\mathcal{P}}^{s}=0,
\\
\end{split}
 \label{eq:singlet2}
 \ee
 to be met.

Using \req{eq:eqUpsilon1} and similar expressions
\be
\begin{split}
 &e^{i(\varepsilon_1+\varepsilon_2)}\Upsilon_1\Upsilon_2
 =\cos^2\alpha - \sin^2\alpha \Upsilon_1\Upsilon_2
 \\
& \quad
 +i\sin\alpha\cos\alpha \left[\Upsilon_1+ \Upsilon_2\right];
 \\
  &e^{i(\varepsilon_1+\varepsilon_2+q_1+q_2)}
 =\cos^2\alpha\Upsilon_1\Upsilon_2  - \sin^2\alpha 
 \\
 & \quad
 +i\sin\alpha\cos\alpha \left[\Upsilon_1+ \Upsilon_2\right],
\end{split}
 \label{eq:eqUpsilon2}
\ee
we find the relations
 \be
 \begin{split}
   &b_o (1,2)=\frac{1}{2};
   \quad
   b_e (1,2)=\frac{1}{2}\frac{e^{i(\varepsilon_1+\varepsilon_2+q_1+q_2)}+1}
 {e^{i(\varepsilon_1+\varepsilon_2+q_1+q_2)-2i\phi}+1};
 \\
&\sum_{\mathcal{P}}
\hat{\mathcal{P}}_\varepsilon\Bigg\{\sin\alpha
\left[\Upsilon_1-\Upsilon_2\right]
\\
&\hspace*{2cm}
+i\cos\alpha \left[1-2b_e(1,2)\right]
\Bigg\}A_{\mathcal{P}}^{s}=0.
 \end{split}
\label{eq:singlet3}
 \ee

Identical transformations of \req{eq:singlet3} yield
\be
\begin{split}
  &A_{21}^s=S_{12}^s A_{12}^s;\\
  &S_{12}^s=\frac{\sin\tilde{\varepsilon}_1-\sin\tilde{\varepsilon}_2+2iu}{\sin\tilde{\varepsilon}_1-\sin\tilde{\varepsilon}_2-2iu},
\end{split}\label{eq:singlet4}
\ee
where we introduced the modified rapidities
\be
\tilde{\varepsilon}\equiv
\varepsilon+q(\varepsilon)-\phi,
\label{eq:tilde}
\ee
and the analogue of the interaction strength
\be
2u(\phi,\alpha)=\sin\phi\frac{\cos^2\alpha}{\sin\alpha}.
\label{eq:u}
\ee
In such a form, the scattering phase in the singlet channel $S_{12}^s$ is equivalent to that of the Hubbard model \cite{Lieb-Wu}.

Finally, we summarize results \rref{eq:HSmatrixtriplet2} and \rref{eq:singlet4} in one matrix
equation
\be
A_{21}^{r_1r_2}=S^{r_1r_2}_{r_{1'}r_{2'}}
(s_{12})
A_{12}^{r_{1'}r_{2'}},
\quad s_{ij}\equiv \sin\tilde{\varepsilon}_i-\sin\tilde{\varepsilon}_j.
\label{eq:SH1}
\ee
where $S$-matrix in the basis $\left[A^{11},A^{12},A^{21},A^{22}\right]$
reads
\be
\hat{S}(s)=-
\begin{pmatrix}
  1 &0&0&0\\[3pt]
  0& \frac{-2iu}{s-2iu}&
  \frac{-s}{s-2iu}&0\\[3pt]
  0&\frac{-s}{s-2iu}&
  \frac{-2iu}{s-2iu}&0\\[3pt]
  0&0&0& 1
\end{pmatrix}.
\label{eq:SH2}
\ee

$S$-matrix is analogous to that of the Hubbard model \cite{Lieb-Wu,Hubbard-book} with some important distinctions
reflecting the difference of fermions in the Hubbard model to the Quantum circuit under consideration: (i)
overall sign; (ii) the sign of the non-diagonal matrix element related to the different definition of the singlet.
Both distinctions do not preclude the system from integrability as they do not break the Yang-Baxter relation however
we will see that the feature (ii) breaks the $SU(2)$ symmetry so that the spectrum of the Quantum circuit has less degeneracies than the Hamiltonian
Hubbard model.

In the original treatment of the Hubbard model \cite{Lieb-Wu} (as well as $XXZ$ model of the previous section) the form of the two-particle $S$ matrix
was sufficient to write the Bethe ansatz wave function as the Hamiltonian operated only with two 
neighboring lattice sites. The gate \req{eq:UHubbard} operates with all four sites (including replicas),
and that is why the explicit check is required for three and four particle cases. This will be done in the following
two subsections.

\subsection{Three excitations, $\mathcal{N}=3$.}
We  look for the three-particle wave-function in a form,
 \be
 \begin{split}
  &\left|\psi_{\varepsilon_1,\varepsilon_2,\varepsilon_3}\right\rangle=\sum_{\substack{x_1=1\\x_1\leq x_2\leq x_3}}^{2L}
\sum_{\mathcal{P}}A_{\mathcal{P}}^{r_1r_2r_3}
\hat{\mathcal{P}}_\varepsilon
\\
&\times
{B}_{x_1,x_2}(\varepsilon_1,\varepsilon_2){B}_{x_2,x_3}(\varepsilon_2,\varepsilon_3)
\left(\prod_{j=1}^3
\Upsilon_{x_j}(\epsilon_j)\hat{\sigma}^+_{x_j,r_j}\right)
\left|0\right\rangle,
\end{split}
\raisetag{2.4cm}
 \label{eq:psi3H}
 \ee
 where the notation for permutations is defined in Sec.~\ref{sec:xxz2},
and other notation follows \req{eq:psi1}. Two particle vortex functions are given by \req{eq:B}.



The translational property is established in a complete analogy of \reqs{eq:quantization2H}
and \rref{eq:quantizationN}. It requires
\be
\begin{split}
A_{\mathcal{P}}^{r_1r_2r_3}\hat{\mathcal{P}}_\varepsilon e^{-iq_3L}=A^{r_3r_1r_2}_{\mathcal{C}\cdot\mathcal{P}},
\\
A_{\mathcal{P}}^{r_1r_2r_3}\hat{\mathcal{P}}_\varepsilon e^{-i(q_2+q_3)L}=A^{r_2r_3r_1}_{\mathcal{C}\cdot\mathcal{C}\cdot\mathcal{P}},
\end{split}
\label{eq:quantization3H}
\ee
where the cyclic permutation is defined in \req{eq:cyclic}

The total quasi-momentum is then given by
\be
Q=\ldb  q_1+q_2+q_3\rdb,
\label{eq:Q3H}
\ee
quantized according to \req{eq:q-quantized}.

Let us turn to the diagonalization the operator $\hat{\mathcal{V}}$. We apply $\left(\hat{\mathcal{V}}-e^{i
    \mathcal{E}}\right)$ to wave-function \rref{eq:psi3H}. Using
\reqs{eq:vpsi1Hubbard} we obtain that all the terms with $x_1+1< x_2<x_3-1$ vanish for $\mathcal{E}=\ldb \epsilon_1+\epsilon_2+\epsilon_3\rdb$
and arbitrary amplitudes $A_{\mathcal{P}}^{r_1r_2}$. The terms with $x_1-1<x_2=x_3$ and $x_1=x_2<x_3+1$ vanish if \req{eq:SH1} holds for all permutations
of neighboring indices 
 \be
 \begin{split}
&A^{\dots r_{k-1},
  r_k,r_{k+1},r_{k+2},\dots}_{(k,k+1)\cdot\mathcal{P}}
\\
&
\qquad
=A_{\mathcal{P}}^{\dots r_{k-1},
  l_1,l_2,r_{k+2},\dots}\hat{\mathcal{P}}_{\varepsilon}S\left(s_{k,k+1}\right)^{r_k,r_{k+1}}_{l_1,l_2},
 \end{split}
 \label{eq:smatrixNH}
\ee
in notation of \req{eq:smatrixN}. Here, $k=1,2$, however, \req{eq:smatrixNH} will be applied for an arbitrary number of excitations.

What remains is to check  that the relation  \req{eq:smatrixNH} leads to also vanishing of the terms
$x_1=x_2=x_3-1;\ x_1-1=x_2=x_3$. This statement is anything but trivial and we will
check it by a straightforward calculation. We find
similarly to \req{eq:vpsi2Hubbard}
\be
\begin{split}
&\hat{t} \left(\hat{\mathcal{V}}-e^{i\varepsilon_1+i\varepsilon_2+i\varepsilon_3}
\right)
\left|\psi_{\varepsilon_1,\varepsilon_2,\varepsilon_3}\right\rangle
\\
& \quad
=\sum_{x=2}^{L}e^{-i(q_1+q_2+q_3)x}
\sum_{\mathcal{P}}\hat{\mathcal{P}}_\varepsilon
A_{\mathcal{P}}^{r_1r_2r_3}\hat{\Psi}_{2x,r_1r_2r_3}^{(3),r}(1,2,3)\left|0\right\rangle;\quad r=1,2;
\\
&\hat{\Psi}_{2x,r_1r_2r_3}^{(3),r}(1,2,3)
=
\left[\Upsilon_1b_e(2,3);  b_o(1,2)\Upsilon_1\Upsilon_2\right]
\\
&\times
\left[
 \hat{\mathcal{u}}^{(3)}- e^{i\varepsilon_1+i\varepsilon_2+i\varepsilon_3}
 \hat{\mathbb{T}}^{(3)}(1,2,3)
 \right]
\begin{bmatrix}
\delta_{r_1r}\hat{\sigma}^+_{2x,r_3}
  \hat{\sigma}^+_{2x,r_2}\hat{\sigma}^+_{2x-1,r_1}
  \\
\delta_{r_3r} \hat{\sigma}^+_{2x,r_3}
 \hat{\sigma}^+_{2x-1,r_2}\hat{\sigma}^+_{2x-1,r_1}\\
\end{bmatrix};
\\ \\
 &\hat{\mathbb{T}}^{(3)}(1,2,3)
 =
   \begin{pmatrix}
   0&
  e^{iq_1+iq_2}\frac{\Upsilon_3 b_e(1,2)}{\Upsilon_1b_e(2,3)}
   \\
   e^{iq_1}\frac{\Upsilon_3 b_o(2,3)}{ \Upsilon_1 b_o(1,2)} & 0
\end{pmatrix},
\end{split}
\label{eq:vpsi3Hubbard}
\ee
where three excitation gate sub-block $\hat{\mathcal{u}}^{(3)}$ is defined in \req{eq:UHubbard}.
Vanishing of the right-hand-side of \req{eq:vpsi3Hubbard} requires two
conditions to be met:
 \be
 \begin{split}
&\sum_{\mathcal{P}}\hat{\mathcal{P}}_\varepsilon
\begin{pmatrix}
  \left[
    \cos\alpha \Upsilon_1b_e(2,3)
    - e^{i(\varepsilon_1+\varepsilon_2+\varepsilon_3+q_1)}\Upsilon_2\Upsilon_3b_o(2,3)
  \right];\
      i \sin\alpha \Upsilon_1 \Upsilon_2b_o(1,2) \\[5pt]
    i \sin\alpha \Upsilon_1b_e(2,3);  \
    \left[\cos\alpha \Upsilon_1\Upsilon_2b_0(1,2)-
e^{i(\varepsilon_1+\varepsilon_2+\varepsilon_3+q_1+q_2)-2i\phi}\Upsilon_3b_e(1,2)
    \right]
  \end{pmatrix}
  \\
  &
  \times
\begin{pmatrix}
A_{\mathcal{P}}^{rr_1r_2}\tau^x_{r_1r_2}\\[5pt]
A_{\mathcal{P}}^{r_1r_2r}\tau^x_{r_1r_2}
\end{pmatrix}=0.
\end{split}
\label{eq:3conditions}
 \ee

 Our purpose is to show that \reqs{eq:3conditions} are compatible with \reqs{eq:HSmatrixtriplet2} and \rref{eq:singlet3}.
 Indeed, as follows
 from \req{eq:eqUpsilon2}
 \be
   e^{i(\varepsilon_1+\varepsilon_2)}\Upsilon_1\Upsilon_2 =e^{i(\varepsilon_1+\varepsilon_2+q_1+q_2)}
    +\left[1-\Upsilon_1\Upsilon_2\right].
 \label{eq:Id1}
 \ee
 Using \req{eq:Id1} and \req{eq:singlet3}
 we reduce each of \req{eq:3conditions} to the form
 \be
\begin{split}
&\sum_{\mathcal{P}}\hat{\mathcal{P}}_\varepsilon
\Bigg\{\left[ G(1,2,3)+
2G_{12}(1,2,3)-G_{23}(1,2,3)
\right]A_{\mathcal{P}}^{rr_1r_2}
\\
&
+
\left[2
G_{23}(1,2,3)-G_{12}(1,2,3)
\right]
A_{\mathcal{P}}^{r_1r_2r}
\Bigg\}\tau^x_{r_1r_2}=0,
\end{split}
\label{eq:10}
\ee
 where the functions $G_\cdot$ (whose explicit form is known but not important) satisfy the symmetry relations 
 \be
 \begin{split}
 G_{12}(1,2,3)=G_{12}(2,1,3);
\ 
 G_{23}(1,2,3)=G_{23}(1,2,3),
 \label{eq:11}
\end{split}
 \ee
 and $G(1,2,3)$ is invariant with respect to all six permutations of its arguments.
 This symmetry and the direct consequences of 
\req{eq:HSmatrixtriplet2}
\be
 \begin{split}
&\sum_{\mathcal{P}}A_{\mathcal{P}}^{r_1r_2r_3}
=0;
\\
 &\left[A_{\mathcal{P}}^{r_1r_2r}+A_{(12)\cdot\mathcal{P}}^{r_1r_2r}\right]
 \tau^x_{r_1r_2}
 =2\left[A_{\mathcal{P}}^{r r_1r_2}+A_{(12)\cdot\mathcal{P}}^{r r_1r_2}\right]
 \tau^x_{r_1r_2};\\
 &\left[A_{\mathcal{P}}^{rr_1r_2}+A_{(23)\cdot\mathcal{P}}^{rr_1r_2}\right]
 \tau^x_{r_1r_2}
=2\left[A_{\mathcal{P}}^{r_1r_2r}+A_{(23)\cdot\mathcal{P}}^{r_1r_2r}\right]
 \tau^x_{r_1r_2};
\end{split}
\label{eq:HSmatrixtriplet20}
 \ee
 guarantees that  the relation \rref{eq:10} holds
 and the wavefunction \rref{eq:psi3H} is an eigenfunction of the operator $\hat{\mathcal{V}}$.

\subsection{Four excitations, $\mathcal{N}=4$.}

Let us  look for the four-particle wave-function in a form,
 \be
 \begin{split}
  &\left|\psi_{\varepsilon_1,\varepsilon_2,\varepsilon_3,\varepsilon_4}\right\rangle=\sum_{\substack{x_1=1\\x_1\leq x_2\dots \leq x_4}}^{2L}
\sum_{\mathcal{P}}A_{\mathcal{P}}^{r_1r_2r_3r_4}
\hat{\mathcal{P}}_\varepsilon
\\
&\times
\left(\prod_{k=1}^3
{B}_{x_k,x_{k+1}}(\varepsilon_k,\varepsilon_{k+1})\right)
\left(\prod_{j=1}^4
\Upsilon_{x_j}(\epsilon_j)\hat{\sigma}^+_{x_j,r_j}\right)
\left|0\right\rangle,
\end{split}
\raisetag{2.4cm}
 \label{eq:psi4H}
 \ee
 where the notation for permutations is defined in Sec.~\ref{sec:xxz2},
 and other notation follows \req{eq:psi1}. Two particle vortex functions are given by \req{eq:B}.

 As usual we start with the translational properties and find the requirement 
[compare with \req{eq:quantization3H}]
\be
\begin{split}
  A_{\mathcal{P}}^{r_1r_2 \dots r_{\mathcal{N}-1}
r_{\mathcal{N}}
  }\hat{\mathcal{P}}_\varepsilon e^{-iq_{\mathcal{N}}L}=A^{r_\mathcal{N}r_1r_2\dots r_\mathcal{N-1}}_{\mathcal{C}\cdot\mathcal{P}},
\\
A_{\mathcal{P}}^{r_1r_2 \dots r_{\mathcal{N}-1}
r_{\mathcal{N}}}\hat{\mathcal{P}}_\varepsilon e^{-i(q_{\mathcal{N}-1}+q_{\mathcal{N}})L}=A^{ r_\mathcal{N-1}r_\mathcal{N}r_1r_2\dots}_{\mathcal{C}\cdot\mathcal{C}\cdot\mathcal{P}},
\end{split}
\label{eq:quantization4N}
\ee
for $\mathcal{N}=4$.

The total quasi-momentum is then given by
\be
Q=\ldb  \sum_{j=1}^{\mathcal{N}}q(\varepsilon_j)\rdb,
\label{eq:QNH}
\ee
quantized according to \req{eq:q-quantized}, here $\mathcal{N}=4$.

We now diagonalize operator $\hat{\mathcal{V}}$. We apply $\left(\hat{\mathcal{V}}-e^{i\mathcal{E}}\right)$ to wave-function \rref{eq:psi4H}. Using
\reqs{eq:vpsi1Hubbard} we obtain that all the terms with $|x_k-x_{k+1}|>1$  (we will call those sites unpaired) vanish for
\be
\mathcal{E}
=\ldb  \sum_{j=1}^{\mathcal{N}}q(\varepsilon_j)\rdb,
\label{eq:varENH}
\ee
and arbitrary amplitudes $A_{\mathcal{P}}^{r_1r_2r_3r_4}$.
The terms with only one pair $|x_k-x_{k+1}|\leq 1$ (we call those pairs doublets ) and other unpaired vanish if \req{eq:SH1} holds for all permutations
involving neighboring indices $k,k+1$. By the same token
the contribution of independent doublets $|x_k-x_{k+1}|\leq 1,|x_l-x_{l+1}|\leq 1, x_{l}-x_{k+1}=1$ also vanishes.
Next configuration $ |x_k-x_{k+2}|\leq 1$, is called triplet configuration. It vanishes due to \req{eq:vpsi3Hubbard}
The only configuration remains is the quadruplet $ |x_k-x_{k+3}|\leq 1$ that has to be considered separately.
We find
  \be
 \begin{split}
&\hat{t} \left(\hat{\mathcal{V}}-e^{i\varepsilon_1+\dots+i\varepsilon_4}
\right)\left|\psi_{\varepsilon_1,\dots,\varepsilon_4}\right\rangle
\\
&\qquad
=\sum_{x=2}^{L}e^{-i(q_1+\dots+q_4)x}
\sum_{\mathcal{P}}\hat{\mathcal{P}}_\varepsilon
A_{\mathcal{P}}^{r_1r_2r_3r_4}\hat{\Psi}_{2x,r_1r_2r_3r_4}^{(4)}(1,\dots,4)\left|0\right\rangle;
\\
&\hat{\Psi}_{2x,r_1\dots r_4}^{(4)}(1,\dots,4)
\\
&\quad=
\left(b_e(3,4)b_o(1,2)\Upsilon_1 \Upsilon_2 e^{i2\phi}
  -e^{i(\varepsilon_1+\dots+\varepsilon_4+q_1+q_2)}
       b_e(1,2)b_o(3,4)\Upsilon_3 \Upsilon_4 
     \right)
     \\
     &\quad\times
\hat{\sigma}^+_{2x-1,r_1}\hat{\sigma}^+_{2x-1,r_2}\hat{\sigma}^+_{2x,r_3}\hat{\sigma}^+_{2x,r_4}
\end{split}
\label{eq:vpsi4Hubbard}
\ee
The right-hand-side of \req{eq:vpsi4Hubbard} vanishes provided that
\be
\begin{split}
  &
 \sum_{\mathcal{P}}\hat{\mathcal{P}}_\varepsilon
  \left[ b_e(3,4)b_o(1,2)\Upsilon_1 \Upsilon_2 
    -
      e^{i(\varepsilon_1+\varepsilon_2+\varepsilon_3+\varepsilon_4+q_1+q_2)-2i\phi}
       b_e(1,2)b_o(3,4)\Upsilon_3 \Upsilon_4 
     \right]
     \\
     &\times
A_{\mathcal{P}}^{r_1r_2r_3r_4}\tau^x_{r_1r_2}\tau^x_{r_3r_4}
=0.
\end{split}
\label{eq:4conditions}
\ee
Using  \reqs{eq:singlet3}, \rref{eq:Id1}, \rref{eq:eqUpsilon}, and \rref{eq:eqUpsilon2},
we reduce \req{eq:4conditions} to the form similar to \req{eq:10}:
\be
\sum_{\mathcal{P}}\hat{\mathcal{P}}_\varepsilon
\left[G_{4}(1\dots4)+G_{1}(1\dots 4)
\right]A_{\mathcal{P}}^{r_1\dots r_4}\tau^x_{r_1r_2}\tau^x_{r_3r_4}
=0,
\label{eq:4conditions1}
\ee
where the functions $G$ have the permutational symmetry
\be
\begin{split}
\hat{\mathcal{P}}_\varepsilon^{(234)}
G_1(1,2,3,4)=G_1(1,2,3,4);
\\ \hat{\mathcal{P}}_\varepsilon^{(123)}
G_4(1,2,3,4)=G_4(1,2,3,4),
\end{split}
\label{eq:G4}
\ee
where the subscript of type $(ijk)$ refers to any permutation among  $(ijk)$ and leaving everything else intact.

Generalization of first of three-particle \reqs{eq:HSmatrixtriplet20} to the arbitrary
number of particles yields
\be
\sum_{\mathcal{P}^{(k,k+1,k+2)}}A_{\mathcal{P}^{(k,k+1,k+2)}\cdot\mathcal{P}}^{r_1,\dots, r_{\mathcal{N}}}
=0;\ k=1,\dots \mathcal{N}-2,
\label{eq:HSmatrixtriplet200}
\ee
which together with symmetry relation \rref{eq:G4} guarantees the validity of condition \rref{eq:4conditions1}.
Thus, wavefunction \rref{eq:psi4H} is an eigenfunction of the operator $\hat{\mathcal{V}}$.

As doublet, triplet, and quadruplet exhaust all possible configurations we are now prepared to write down the wave-function
for an arbitrary number of excitations.

\subsection{Bethe Ansatz solution for an arbitrary number of excitations, $\mathcal{N}\geq 2$}.

The wave function is a straightforward generalization of \reqs{eq:psi3H} and \rref{eq:psi4H}.
\be
 \begin{split}
   &\left|\psi_{\varepsilon_1,\dots,\varepsilon_\mathcal{N}}\right\rangle
   =\sum_{\substack{x_1=1\\x_1\leq x_2\dots \leq x_\mathcal{N}}}^{2L}
\sum_{\mathcal{P}}A_{\mathcal{P}}^{r_1\dots r_\mathcal{N}}
\hat{\mathcal{P}}_\varepsilon
\\
&\times
\left(\prod_{k=1}^{\mathcal{N}-1}
{B}_{x_k,x_{k+1}}(\varepsilon_k,\varepsilon_{k+1})\right)
\left(\prod_{j=1}^\mathcal{N}
\Upsilon_{x_j}(\epsilon_j)\hat{\sigma}^+_{x_j,r_j}\right)
\left|0\right\rangle,
\end{split}
\raisetag{2.4cm}
 \label{eq:psiNH}
 \ee
The translational property is established in a complete analogy of \reqs{eq:quantization2H}
and \rref{eq:quantizationN}. It gives \reqs{eq:quantization4N}
\rref{eq:QNH}
for an arbitrary $\mathcal{N}$.

Next we diagonalize operator $\hat{\mathcal{V}}$. We apply $\left(\hat{\mathcal{V}}-e^{i\mathcal{E}}\right)$ to wave-function \rref{eq:psiNH}. Using
\reqs{eq:vpsi1Hubbard} we obtain that all the terms with unpaired sites vanish for
${\mathcal{E}}$ satisfying \req{eq:varENH} for an arbitrary $\mathcal{N}$, and arbitrary  amplitudes $A_{\mathcal{P}}^{\cdot}$.
All the doublets vanish if the amplitudes are related to each other by \req{eq:smatrixNH}.
Moreover all the triplets and quadruplets vanish as was shown explicitly in the two previous subsections.
It completes the proof that the wavefunction of form  \rref{eq:psiNH} is indeed an eigenfunction
of the operator $\hat{\mathcal{V}}$.

Bethe Ansatz equations are obtained from the compatibility of the boundary conditions \rref{eq:quantization4N}
with the relation \req{eq:smatrixNH}. In a complete
analogy with the derivation for XXZ model \req{eq:BA1}
we find
\be
\begin{split}
&A_{\mathcal{P}}^{r_1r_2 \dots r_{\mathcal{N}-1} r_{\mathcal{N}}} \hat{\mathcal{P}}_\varepsilon
   e^{-iq_{\cal N}L}
   \\
   &\qquad =
A_{\mathcal{P}}^{r_1^\prime r_2^\prime \dots r_{\mathcal{N}-1}^\prime r_{\mathcal{N}}^\prime }\hat{\mathcal{P}}_\varepsilon
\mathcal{S}(\dots)_{r_1^\prime r_2^\prime \dots r_{\mathcal{N}-1}^\prime r_{\mathcal{N}}^\prime }
^{ r_{\mathcal{N}} r_1r_2 \dots r_{\mathcal{N}-1}}
\end{split}
\label{eq:BA1H}
\ee
where the many-body $S$-matrix $\mathcal{S}(\dots)$ is given by
\be
\begin{split}
&\mathcal{S}_{r_1^\prime r_2^\prime \dots r_{\mathcal{N}-1}^\prime r_{\mathcal{N}}^\prime }
^{  r_1r_2 \dots r_{\mathcal{N}-1}r_{\mathcal{N}}}
\left(
  \left\{\varepsilon_{j}\right\}_{j=1}^{\mathcal{N}}
\right)
\\
&\quad
={S}^{r_1r_{2}}_{r_{1}^\prime \rho_{2}}
\left(s_{1,\mathcal{N}}\right)
{S}^{\rho_2 r_{3}}_{r_{2}^\prime \rho_{3}}
\left(s_{2,\mathcal{N}}\right)
\dots
{S}^{ \rho_{\mathcal{N}-2}r_{\mathcal{N}-1}}_{r_{\mathcal{N}-2}^\prime \rho_{\mathcal{N}-1}}
 \left(s_{\mathcal{N}-2,\mathcal{N}}\right)
{S}^{ \rho_{\mathcal{N}-1}r_{\mathcal{N}}}_{r_{\mathcal{N}-1}^\prime r_{\mathcal{N}}^\prime}
  \left(s_{\mathcal{N}-1,\mathcal{N}}\right)
\end{split}
  \label{eq:BA1HS}
\ee
Unlike for $XXZ$ model, equation \rref{eq:BA1H} is a matrix equation and still requires more work.
Fortunately, it is almost the same as for the standard Hubbard model and can be solved by the algebraic Bethe ansatz.

We represent the needed $\mathcal{S}$-matrix \rref{eq:BA1HS} as a trace of the monodromy matrix
\be
\begin{split}
&\mathcal{S}_{r_1^\prime r_2^\prime \dots r_{\mathcal{N}-1}^\prime r_{\mathcal{N}}^\prime }
^{ r_{\mathcal{N}} r_1r_2 \dots r_{\mathcal{N}-1}}
\left(\dots\right)=\mathcal{T}
_{r_1^\prime r_2^\prime \dots r_{\mathcal{N}-1}^\prime r_{\mathcal{N}}^\prime}
^{  r_1r_2 \dots r_{\mathcal{N}-1}r_{\mathcal{N}}}
\left(\sin\tilde{\varepsilon}_{\mathcal{N}}\right)
    \\
&
 \mathcal{T}_{r_1^\prime r_2^\prime \dots r_{\mathcal{N}-1}^\prime r_{\mathcal{N}}^\prime }
 ^{  r_1r_2 \dots r_{\mathcal{N}-1}r_{\mathcal{N}}}(\lambda)
 \\
 &
 =-
 {S}^{\rho_1r_{1}}_{r_{1}^\prime \rho_{2}}
 \left(\lambda_1\right)
 {S}^{\rho_2 r_{2}}_{r_{2}^\prime \rho_{3}}
 \left(\lambda_2\right)
 \dots
 {S}^{ \rho_{\mathcal{N}-2}r_{\mathcal{N}-2}}_{r_{\mathcal{N}-2}^\prime \rho_{\mathcal{N}-1}}
   \left(\lambda_{\mathcal{N}-2}\right)
  {S}^{ \rho_{\mathcal{N}-1}r_{\mathcal{N-1}}}_{r_{\mathcal{N}-1}^\prime \rho_{\mathcal{N}}}
  \left(\lambda_{\mathcal{N}-1}\right)
  {S}^{ \rho_{\mathcal{N}}r_{\mathcal
     {N}}}_{r_{\mathcal{N}-1} \rho_1}
   \left(\lambda_{\mathcal{N}}\right)
\\
&
\lambda_k\equiv \sin\tilde{\varepsilon}_k-\lambda;
\end{split}
\label{eq:BA2H}
\ee
where we used the property
$
{S}^{r_1r_{2}}_{r_{1}^\prime r_{2}^\prime}(0)
=-\delta_{r_1r_{1}^\prime}\delta_{r_2r_{2}^\prime}$.

Transfer matrices $\hat{\mathcal{T}}(\lambda)$ are known \cite{Yang1967,Lieb-Wu} to commute with each
other $\left[\hat{\mathcal{T}}(\lambda_1);\hat{\mathcal{T}}(\lambda_2)\right]=0$
and its eigenvalues $\Lambda(\lambda)$ are found from the algebraic Bethe ansatz equations \cite{KorepinBogolyubonIsergin}.
The eigenvalues $\Lambda$ are expressed in terms of spin rapidities $\lambda_j$
\begin{subequations}
\be
\Lambda\left(\sin\tilde{\varepsilon}_j\right)
  =(-1)^{\mathcal{N}_1+1}\prod_{k=1}^{\mathcal{N}_2}
  \frac{\sin\tilde{\varepsilon}_j-\lambda_k+iu}{\sin\tilde{\varepsilon}_j-\lambda_k-iu};\ j=1,\dots \mathcal{N}.
  \ee
  Spin rapidities are the solutions of the equations  
  \be
  \begin{split}
  &
  (-1)^{\mathcal{N}}
  \prod_{j=1}^{\mathcal{N}}
   \frac{\sin\tilde{\varepsilon}_j-\lambda_k+iu}{\sin\tilde{\varepsilon}_j-\lambda_k-iu}
  =
  \prod_{\substack{l=1,\\l\neq k}}^{\mathcal{N}_2}
  \frac{\lambda_l-\lambda_k+2iu}{\lambda_l-\lambda_k-2iu},
  \\
&\qquad k=1,\dots \mathcal{N}_2.
\end{split}
\raisetag{0.3cm}
\ee
where
$\mathcal{N}_{1,2}$ are the numbers of excitations in each replicas $\mathcal{N}=\mathcal{N}_{1}+\mathcal{N}_{2}$ 
(let us assume $\mathcal{N}_2\leq \cal{N}_1$).
\label{eq:BA3H}
\end{subequations}

Using \reqs{eq:BA2H} -- \rref{eq:BA3H} in \req{eq:BA1H}
and taking into account all the possible permutations,
we obtain the Bethe Ansatz equations for the spectrum of the
Chiral Hubbard circuit
\be
e^{-iq(\varepsilon_j)L}
=\Lambda\left(\sin\tilde{\varepsilon}_j\right);
\qquad j=1,\dots,\mathcal{N}.
\label{eq:BA4H}
\ee

The remaining step is to re-express quasi-momenta and quasi-energies
in terms of the modified rapidities $\tilde{\varepsilon}$ from \req{eq:tilde}.
Using \req{implicit-H} we find
\begin{subequations}
  \be
  e^{iq}=\frac{e^{i(\tilde{\varepsilon}+\phi)}-ie^{-\theta}}{e^{-i(\tilde{\varepsilon}+\phi)}+ie^{-\theta}}
  =ie^{i(\tilde{\varepsilon}+\phi)}F_{\theta,\phi}\left(\tilde{\varepsilon}\right);
  \quad \theta\equiv-\ln \sin\alpha,
\label{eq:q-epsilon-H2}
\ee
and we introduced the function
\be
F_{\theta,\phi}\left(\tilde{\varepsilon}\right)=\frac{\sinh\left(\frac{\theta}{2}+i\frac{\tilde{\varepsilon}+\phi}{2}
    -i\frac{\pi}{4}\right)}{\sinh\left(\frac{\theta}{2}-i\frac{\tilde{\varepsilon}+\phi}{2}
    +i\frac{\pi}{4}\right)}.
\label{eq:FH}
\ee
The quasienergy is given by
\be
e^{i\epsilon}=
-ie^{i(\tilde{\varepsilon}+\phi)}\frac{1}{F_{\theta,\phi}\left(\tilde{\varepsilon}\right)};
\label{eq:e-epsilon-H2}
\ee
and the interaction parameter \req{eq:u} is re-written as
\be
u(\phi,\alpha)=\sin\phi\sinh\theta.
\label{eq:Htheta}
\ee
\label{implicit-H2}
\end{subequations}

\be
\begin{split}
  &
  \left(\frac{e^{i(\tilde{\varepsilon}+\phi)}-ie^{-\theta}}{e^{-i(\tilde{\varepsilon}+\phi)}+ie^{-\theta}}\right)^L
  =(-1)^{\mathcal{N}_1+1}\prod_{k=1}^{\mathcal{N}_2}
  \frac{\sin\tilde{\varepsilon}_j-\lambda_k-iu}{\sin\tilde{\varepsilon}_j-\lambda_k+iu};
  \\
  &(-1)^{\mathcal{N}}
  \prod_{j=1}^{\mathcal{N}}
   \frac{\sin\tilde{\varepsilon}_j-\lambda_k+iu}{\sin\tilde{\varepsilon}_j-\lambda_k-iu}
  =
  \prod_{\substack{l=1,\\l\neq k}}^{\mathcal{N}_2}
  \frac{\lambda_l-\lambda_k+2iu}{\lambda_l-\lambda_k-2iu},
  \\
  &\qquad j=1,\dots \mathcal{N};
  \qquad k=1,\dots \mathcal{N}_2.
\end{split}
\raisetag{1cm}
\label{eq:BA5H}
\ee

\subsection{$2$ - particle bound state solution ($2$-string).}

The idea of search of two-particle bound state is analogous to that of 
Sec.~\ref{sec:xxz-b2}. We look for the (``replica singlet'') solution in the form
\be
A_{12}^{r_1r_2}=\tau^x_{r_1r_2}; \quad A_{21}^{r_1r_2}=0,
\label{eq:A12H}
\ee
According to \req{eq:singlet4}, it means that
\be
S_{12}^s=0.
\label{eq:bs2H1}
\ee
It is clear that the singlet state is the only possibility as the $S$-matrix connecting all the other components does not have any
rapidity dependence.

In terms of the rapidities \req{eq:bs2H1} acquires the form
\be
\sin\tilde{\varepsilon}_1=\lambda_1-iu;\quad
\sin\tilde{\varepsilon}_2=\lambda_1+iu
\label{eq:bs2H2}
\ee
where $\lambda_1$ is an arbitrary real number. To make it consistent with
the Lieb-Wu type equations \rref{eq:BA5H} one identifies $\lambda_1$ with the corresponding spin rapidities.

With the help of parameterization \rref{eq:Htheta}, we obtain from \req{eq:bs2H2} a useful relation
\be
\begin{split}
  \sin\tilde{\varepsilon}_1
  -\cosh\left(\theta-i\phi\right)
  =\sin\tilde{\varepsilon}_2
  -\cosh\left(\theta+i\phi\right).
\label{eq:bs2H20}
\end{split}
\ee

The spectrum of the two particle bound state $\epsilon^{(2)}\left(q^{(2)}\right)$ is given implicitly by $\epsilon^{(2)}
=\epsilon\left(\tilde{\varepsilon}_1\right)+\epsilon\left(\tilde{\varepsilon}_2\right)$ and  $q^{(2)}
=q\left(\tilde{\varepsilon}_1\right)+q\left(\tilde{\varepsilon}_2\right)$.
With the help of \req{eq:bs2H20}, we find for function \rref{eq:FH}
\be
\begin{split}
&  
F_{\theta,\phi}^{(2)}\left(\tilde{\varepsilon}_1,\tilde{\varepsilon}_2\right)=
F_{\theta,\phi}\left(\tilde{\varepsilon}_1\right)F_{\theta,\phi}\left(\tilde{\varepsilon}_2\right)
\\
&=
\frac{\cos \phi +i\sinh\left(\theta+i\tilde{\varepsilon}_1\right)}
{\cos \phi -i\sinh\left(\theta-i\tilde{\varepsilon}_2\right)}.
\end{split}
\label{eq:FH2}
\ee
Then, equations \rref{implicit-H2} give the parametric expression for the spectrum
\be
\begin{split}
  &\epsilon^{(2)}=
  \ldb\pi+2\phi+2\mathrm{Re}\,\tilde{\varepsilon}
  -2\mathrm{Arg}\,\left[\cos \phi +i\sinh\left(\theta+i\tilde{\varepsilon}\right)\right]
  \rdb;
  \\
  &q^{(2)}=
  \ldb\pi+2\phi+2\mathrm{Re}\,\tilde{\varepsilon}
  +2\mathrm{Arg}\,\left[\cos \phi +i\sinh\left(\theta+i\tilde{\varepsilon}\right)\right]
  \rdb;
  \\
  &\tilde{\varepsilon}\equiv \pi-\arcsin\left(\lambda-iu\right),
  \raisetag{4mm}
  \end{split}
\label{eq:spectrum2H}
\ee
where $\lambda$ runs along the whole real axis.
The branch of $\arcsin(x)$ is chosen here to have the proper bound state $|e^{iq_1}|^2<1$
for  $u>0$. The opposite case $u<0$ can be investigated immediately using the
parity transformation \rref{eq:minusphi}.

\begin{figure}
     \label{fig:2bsH}

  %
  \subfloat
  []{
     \includegraphics[width=40\unitlength]{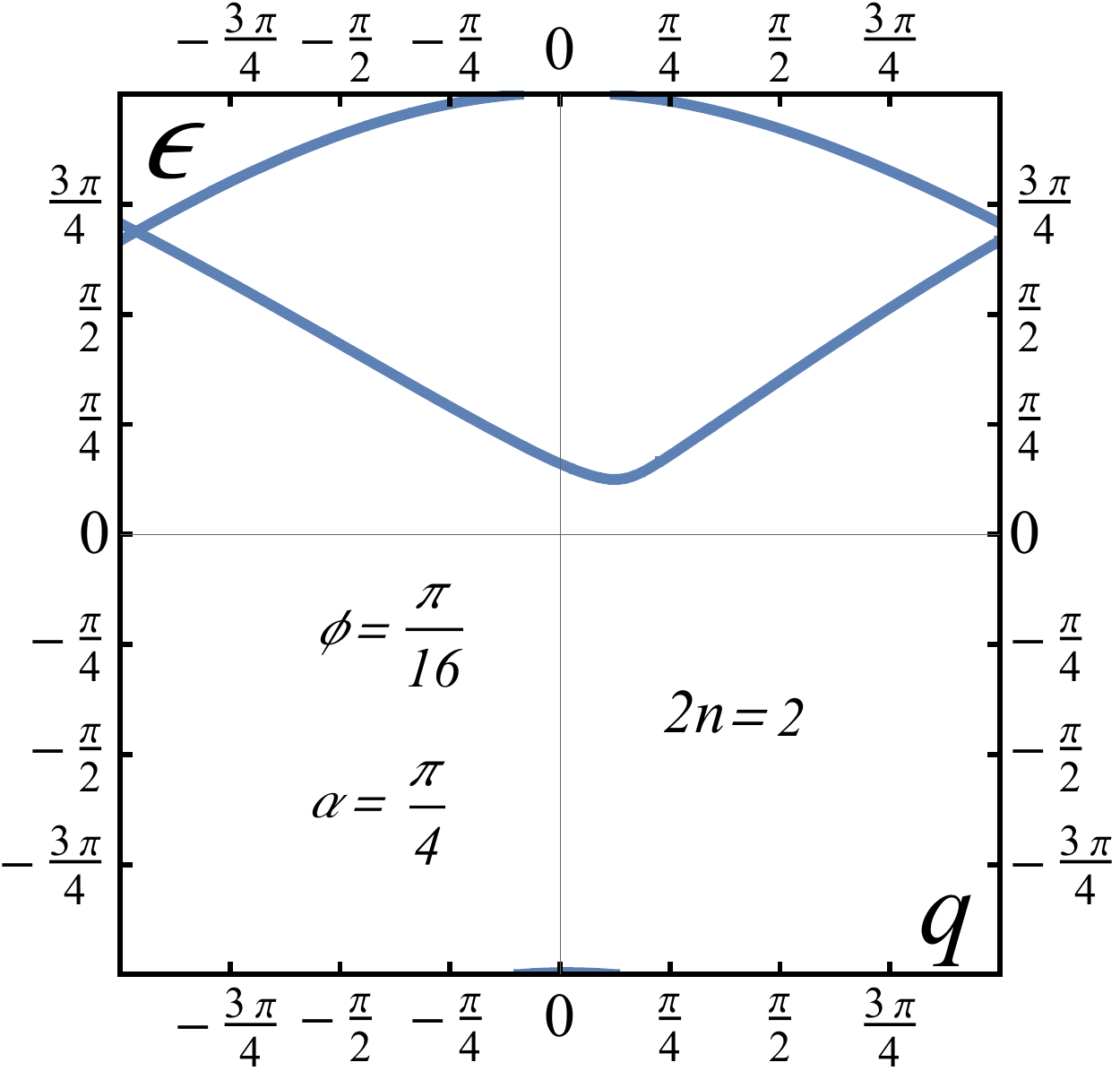}
\label{fig:2bsH-16}
}
  %
  \subfloat
  []{
     \includegraphics[width=40\unitlength]{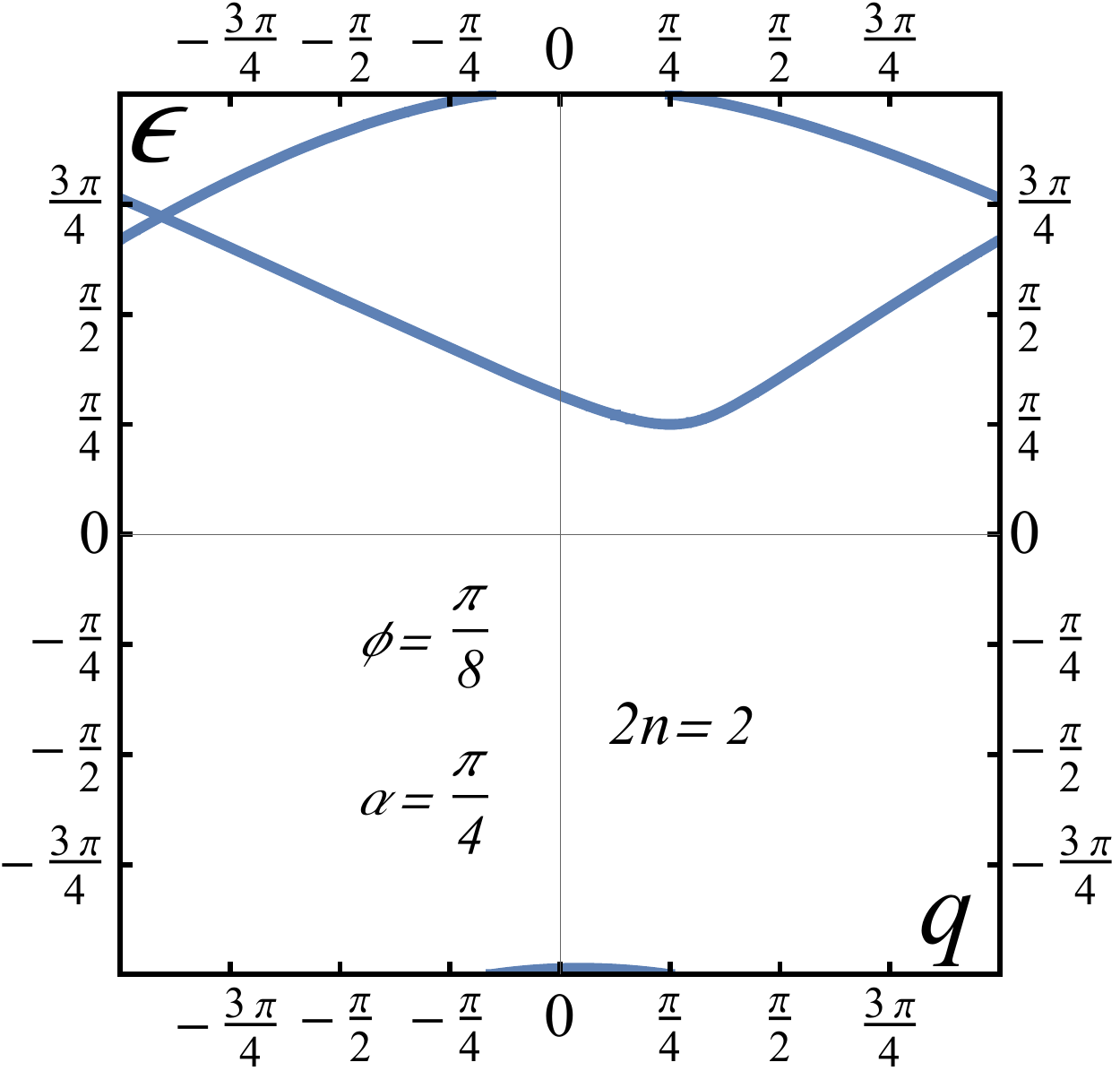}
\label{fig:2bsH-8}
}
  %
  \subfloat
  []{
     \includegraphics[width=40\unitlength]{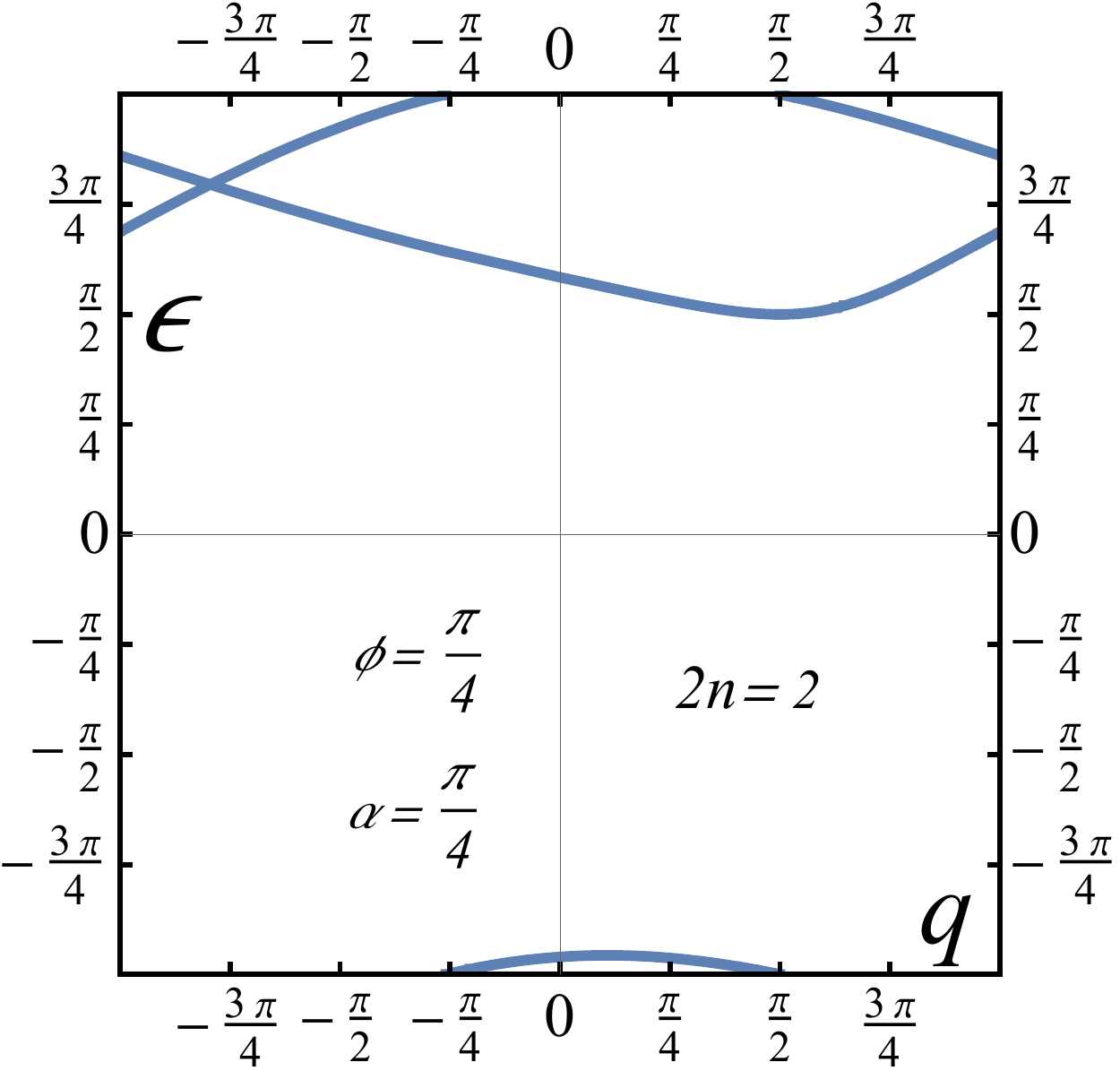}
\label{fig:2bsH-4}
}
  %
  \subfloat
  []{
     \includegraphics[width=40\unitlength]{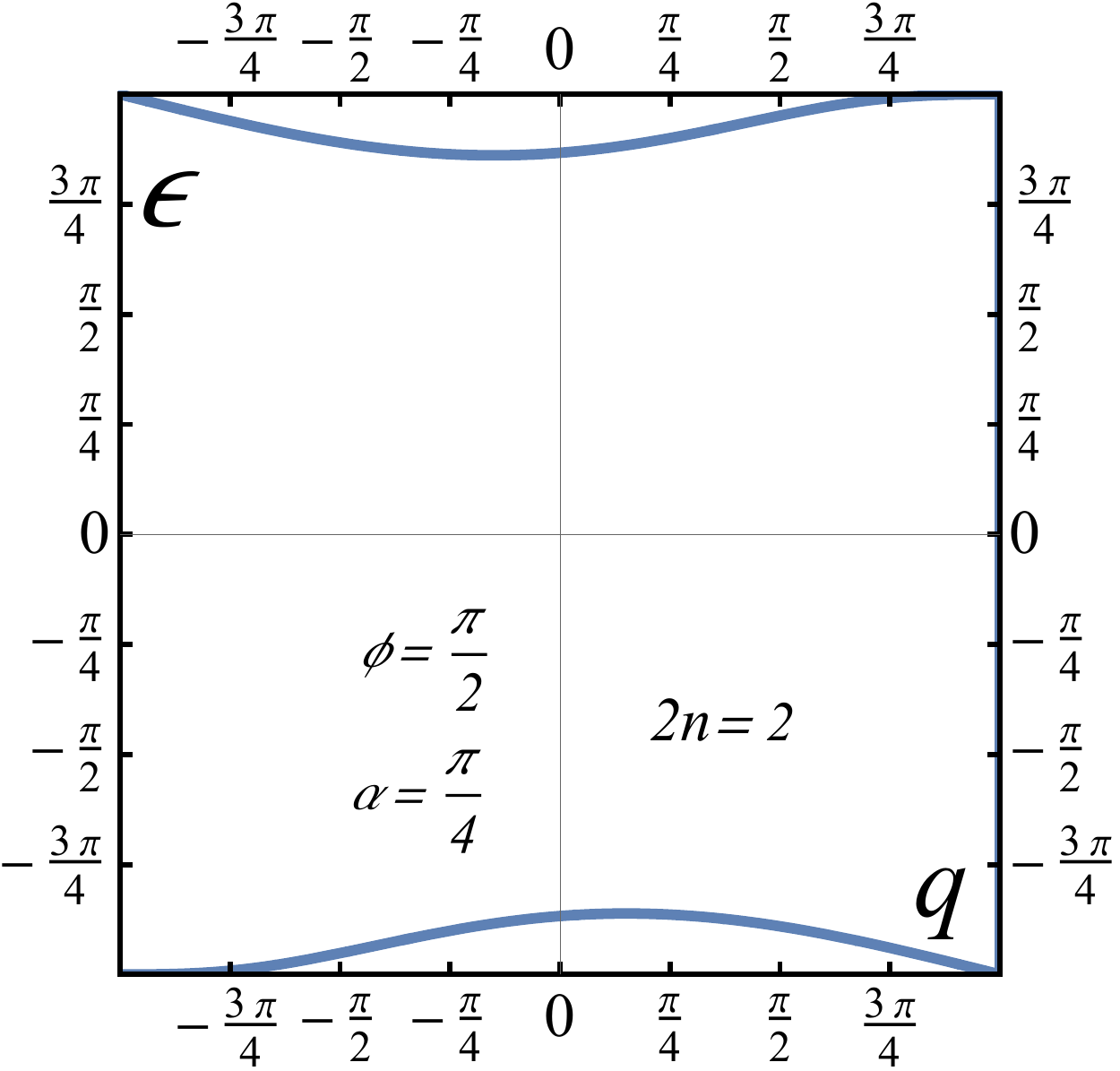}
\label{fig:2bsH-2}
}
  \caption{Spectra of two particle bound state for the Chiral Hubbard model found from \req{eq:spectrum2H},
    The swap angle $\alpha$ is fixed, $\alpha=\pi/4$. The phase $\phi$ characterizes the interaction strength, $\phi=\pi/2$ is the
  strongest possible interaction.}
\end{figure}

Unfortunately, the further simplifications of \req{eq:spectrum2H} are not possible for the general case and
we limit ourselves with the plotting of resulting curves for different values of phase $\phi$, see Fig.~\ref{fig:2bsH}


Notice, the curves are not symmetric with respect to $q\to -q$, this reflects the chirality of the interaction
in the model (spectra of the single-particle state remain symmetric).

\subsection{$2n$ - particle bound state solution ($2n$-string).}

We will look for the $2n$-particle bound state as the combination of the $2$-particle replica singlets
\rref{eq:A12H} bound to each other:
\be
\begin{split}
& A_{12\dots 2n-1;2n}^{r_1r_2\dots r_{2n-1}r_{2n}}=\tau^x_{r_1r_2}
\dots \tau^x_{r_{2n-1}r_{2n}}
;\\
& A_{(2k,2k+1)\cdot\mathcal{P}}^{r_1r_2\dots r_1r_2}=-A_{\mathcal{P}}^{r_1r_2\dots r_{2n-1}r_{2n}};
\quad k=1,\dots, n-1.
\\
&A_{\mathcal{P}}=0;\qquad \mathrm{all\ other\ permutations}.
\end{split}
\label{eq:A12=H}
\raisetag{1.6cm}
\ee
It significantly restricts the possible number of the non-vanishing amplitudes.
For instance, out of possible twenty four permutations for $4$-string
only $1234$ and $1324$ are allowed.

Compatibility of the ansatz \rref{eq:A12=H} with relations between the amplitudes \req{eq:smatrixNH}
imposes the conditions on $S$-matrices:
\be
S_{2k-1,2k}^s=0;\quad S^{r_1r_2}_{r_1^\prime r_2^\prime}\left(s_{2k,2k-1}\right)=-\delta_{r_1r_1^\prime}\delta_{r_2r_2^\prime}.
\label{eq:bsnH}
\ee

Equations \rref{eq:bsnH} and the explicit form of the $S$ matrix \rref{eq:SH2} fixes
the charge rapidities as following: 
\begin{subequations}
\be
\begin{split}
 & \sin\tilde{\varepsilon}_1=\lambda^{(n)}-inu;
 \\
 &\sin\tilde{\varepsilon}_2=\sin\tilde{\varepsilon}_3=\lambda^{(n)}-i(n-2)u;
 \quad \tilde{\varepsilon}_2=\pi-\tilde{\varepsilon}_3;
 \\
 & \qquad\vdots
\\
 &\sin\tilde{\varepsilon}_{2n-1}=\sin\tilde{\varepsilon}_{2n-2}=\lambda^{(n)}-i(2-n)u;
 \quad \tilde{\varepsilon}_2=\pi-\tilde{\varepsilon}_3;
 \\
 &
\sin\tilde{\varepsilon}_{2n}=\lambda^{(n)}+inu
\end{split}
\label{eq:kLambdacharge}
\raisetag{1.7cm}
\ee

Compatibility with the Bethe ansatz equations \rref{eq:BA5H} requires the spin rapidities to be on the straight line similarly to $XXZ$ model,
\be
\lambda_j^{(n)}=\lambda^{(n)}-iu\left(n-2j+1\right), \quad j=1,\dots n; 
\ee
\label{eq:kLambda}
\end{subequations}
The configuration \rref{eq:kLambda} is known in the Hubbard model literature as the
$k-\Lambda$ string.

The spectrum of the $2n$-particle bound state $\epsilon^{(2n)}\left(q^{(2n)}\right)$ is given implicitly by
$\epsilon^{(2n)}
=\sum_{j=1}^{2n}\epsilon\left(\tilde{\varepsilon}_j\right)$ and  $q^{(2n)}
=\sum_{j=1}^{2n}q\left(\tilde{\varepsilon}_1\right)$.

With the help of \req{eq:kLambdacharge}, we find for function \rref{eq:FH}
\[
F_{\theta,\phi}\left(\tilde{\varepsilon}_{2j}\right)F_{\theta,\phi}\left(\pi-\tilde{\varepsilon}_{2j}\right)
=\frac{\cosh\left(\theta+i\phi\right)-\sin \tilde{\varepsilon}_{2j} }{\cosh\left(\theta+i\phi\right)-\sin \tilde{\varepsilon}_{2j+2} }
\]
and
\[
  \prod_{j=1}^{2n}F_{\theta,\phi}\left(\tilde{\varepsilon}_{j}\right)=\frac{\cos \phi +i\sinh\left(\theta+i\tilde{\varepsilon}_1\right)}
{\cos \phi -i\sinh\left(\theta-i\tilde{\varepsilon}_{2n}\right)}.
\]

It yields the parametric form for the spectrum of $2n$-strings:
\be
\begin{split}
  &\epsilon^{(2n)}\!\! =
  \ldb n\pi+2\left\{\phi+\mathrm{Re}\,\tilde{\varepsilon}
  -\mathrm{Arg}\left[\cos \phi +i\sinh\left(\theta+i\tilde{\varepsilon}\right)\right]\right\}
  \rdb;
  \\
  &q^{(2n)}\!\! =
  \ldb n\pi+2\left\{\phi +\mathrm{Re}\,\tilde{\varepsilon}
  +\mathrm{Arg}\left[\cos \phi +i\sinh\left(\theta+i\tilde{\varepsilon}\right)\right]\right\}
  \rdb;
  \\
  &\tilde{\varepsilon} \equiv \pi-\arcsin\left(\lambda-in u\right),
  \raisetag{4mm}
  \end{split}
\label{eq:spectrumnH}
\ee
where $\lambda$ runs along the whole real axis. [For $n=1$, \req{eq:spectrumnH} reproduces \req{eq:spectrum2H}].
The branch of $\arcsin(x)$ is chosen here to have the proper bound state, see discussion after \req{eq:properboundstate},
for  $u>0$. The opposite case $u<0$ can be investigated immediately using the
parity transformation \rref{eq:minusphi}.

\begin{figure}
     \label{fig:nbsH}

  %
  \subfloat
  []{
     \includegraphics[width=40\unitlength]{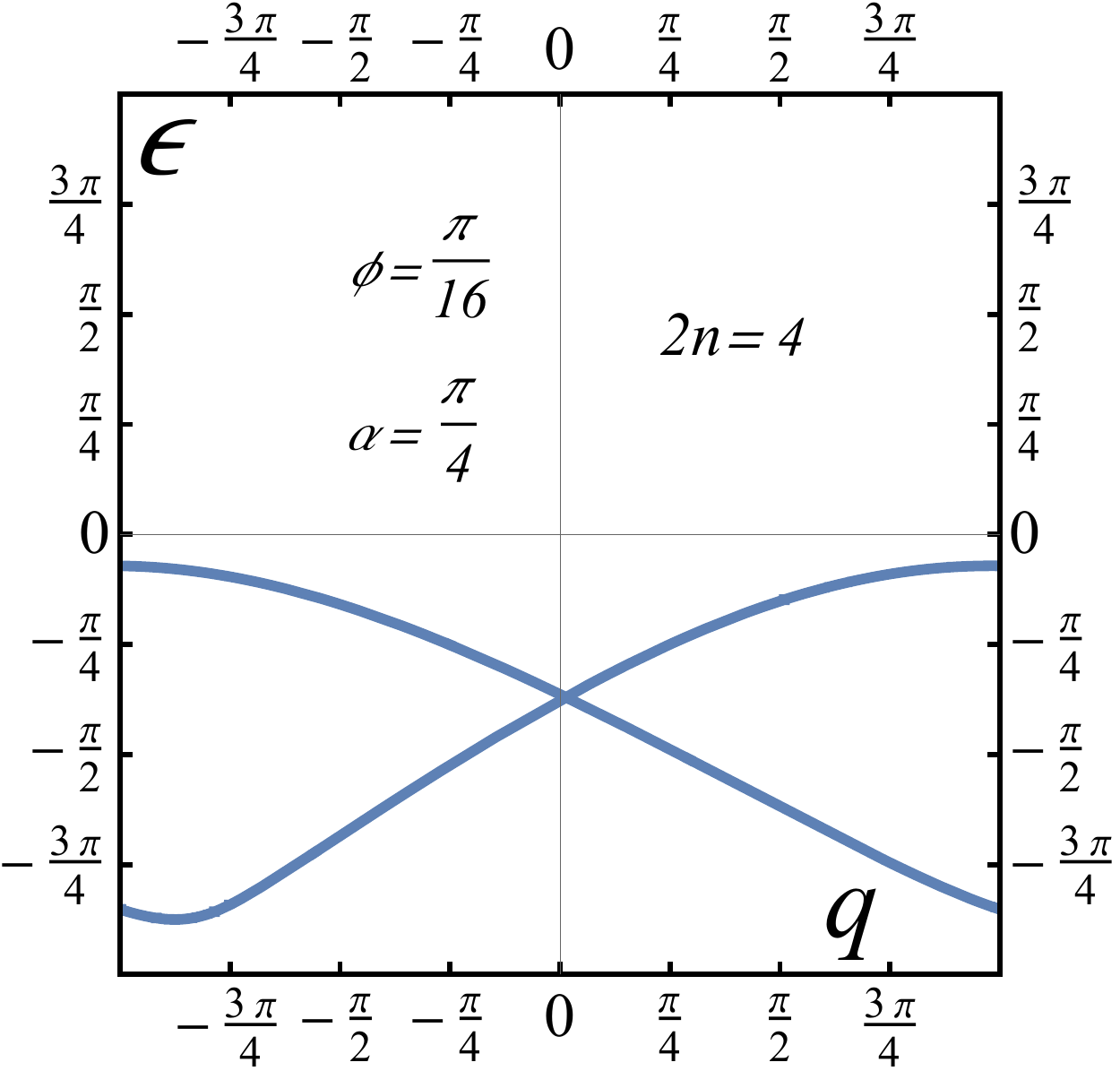}
\label{fig:4bsH-16}
}
  %
  \subfloat
  []{
     \includegraphics[width=40\unitlength]{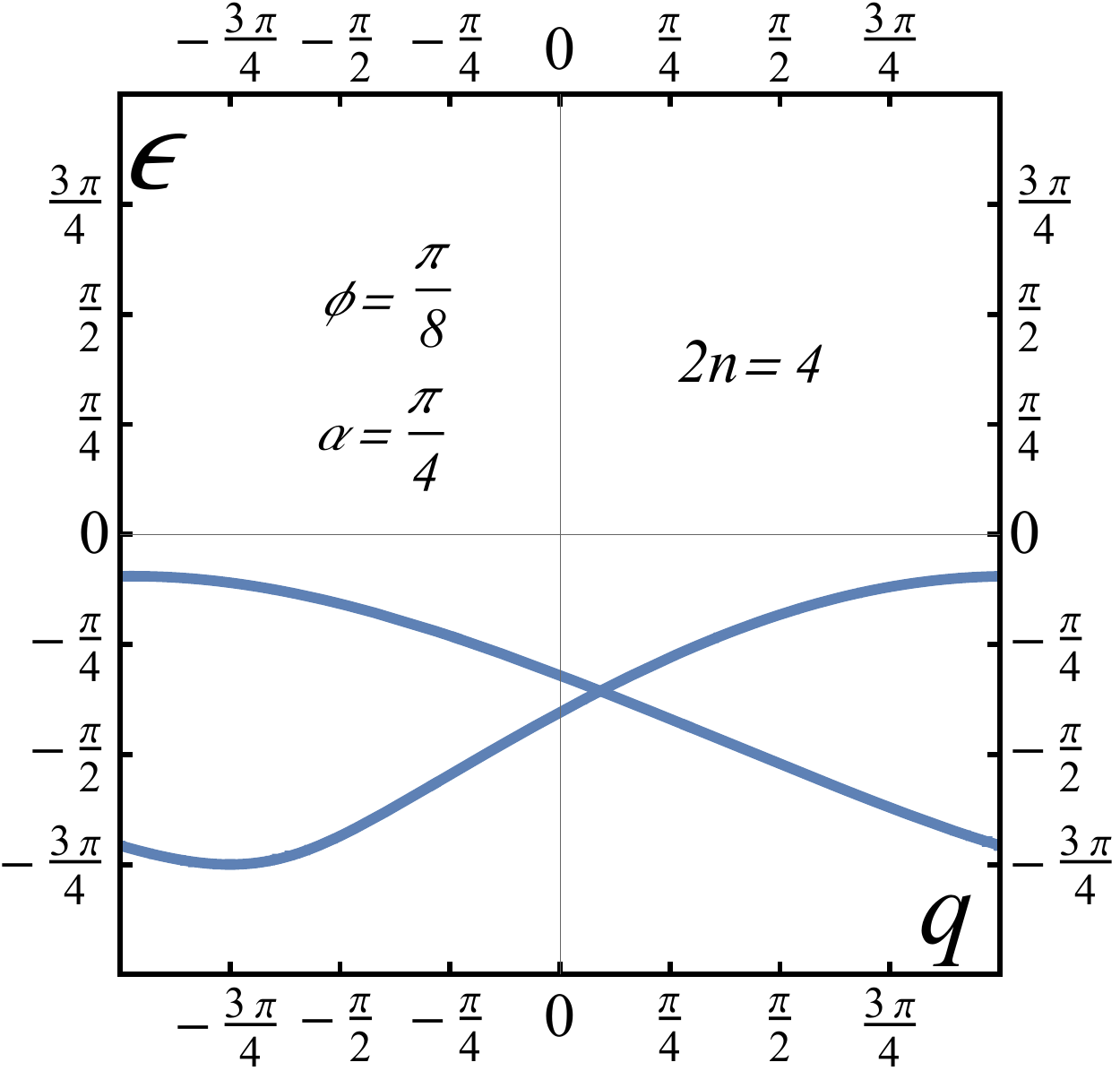}
\label{fig:4bsH-8}
}
  %
  \subfloat
  []{
     \includegraphics[width=40\unitlength]{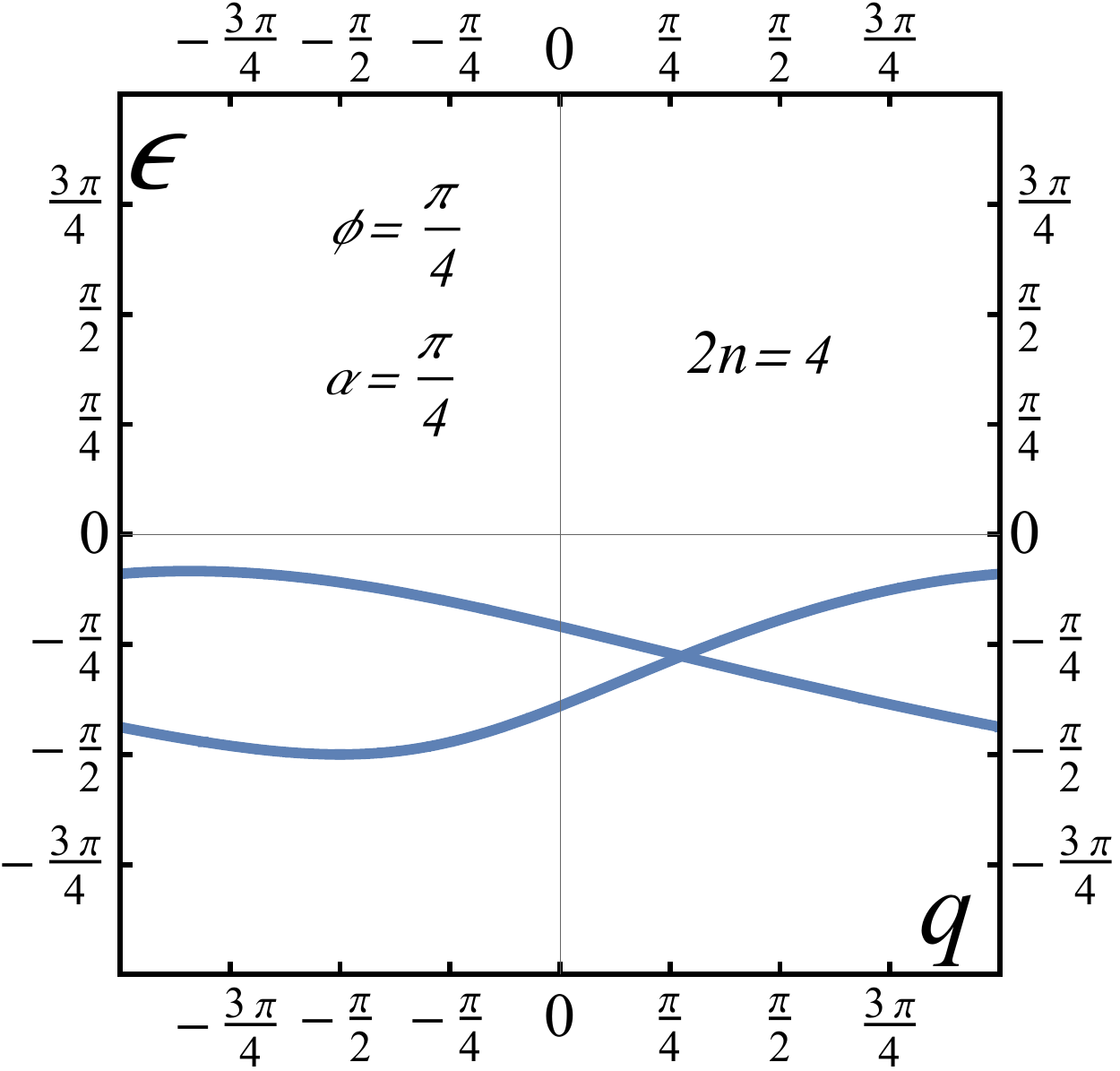}
\label{fig:4bsH-4}
}
  %
  \subfloat
  []{
     \includegraphics[width=40\unitlength]{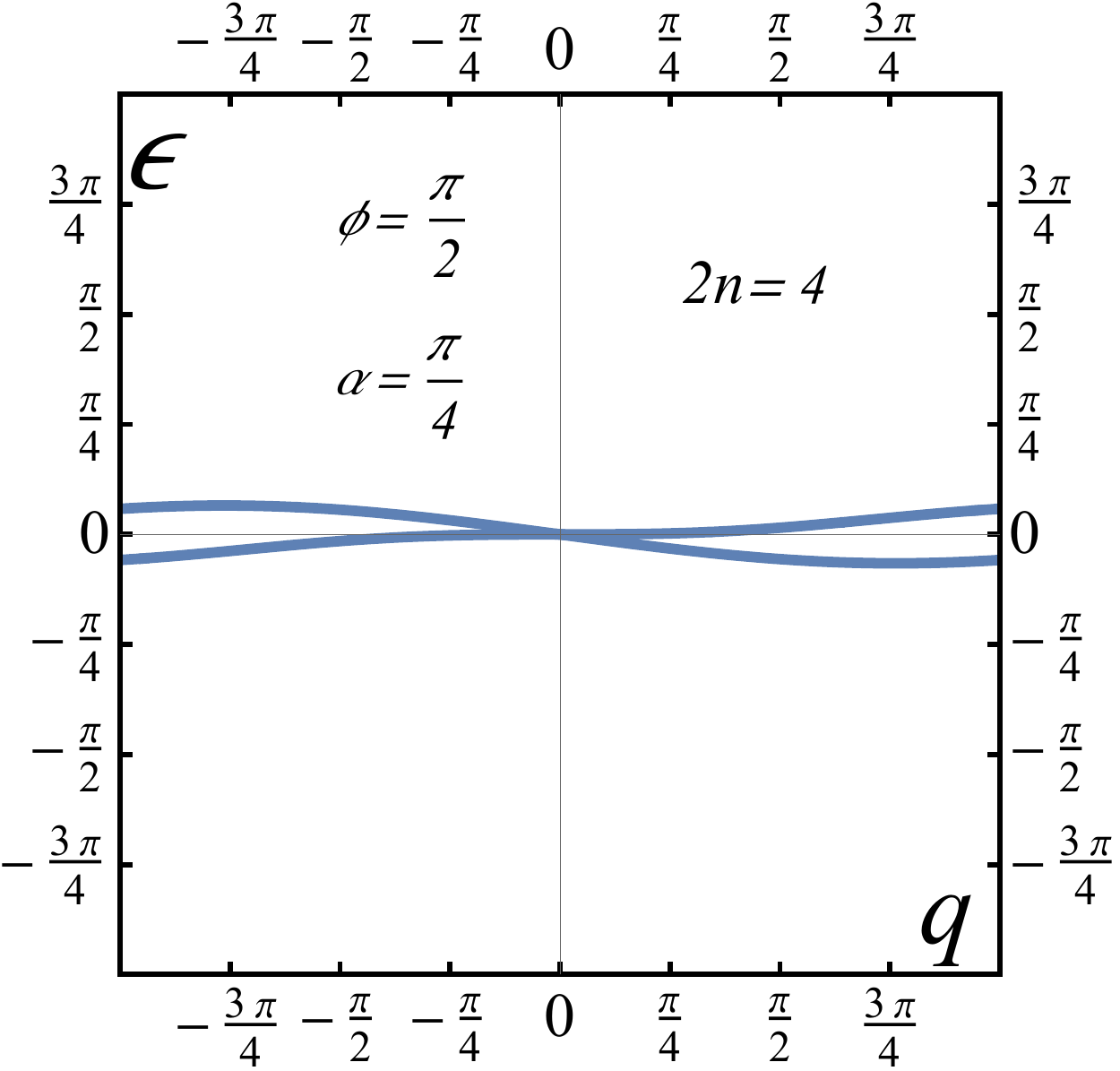}
\label{fig:4bsH-2}
}
\\
  \subfloat
  []{
     \includegraphics[width=40\unitlength]{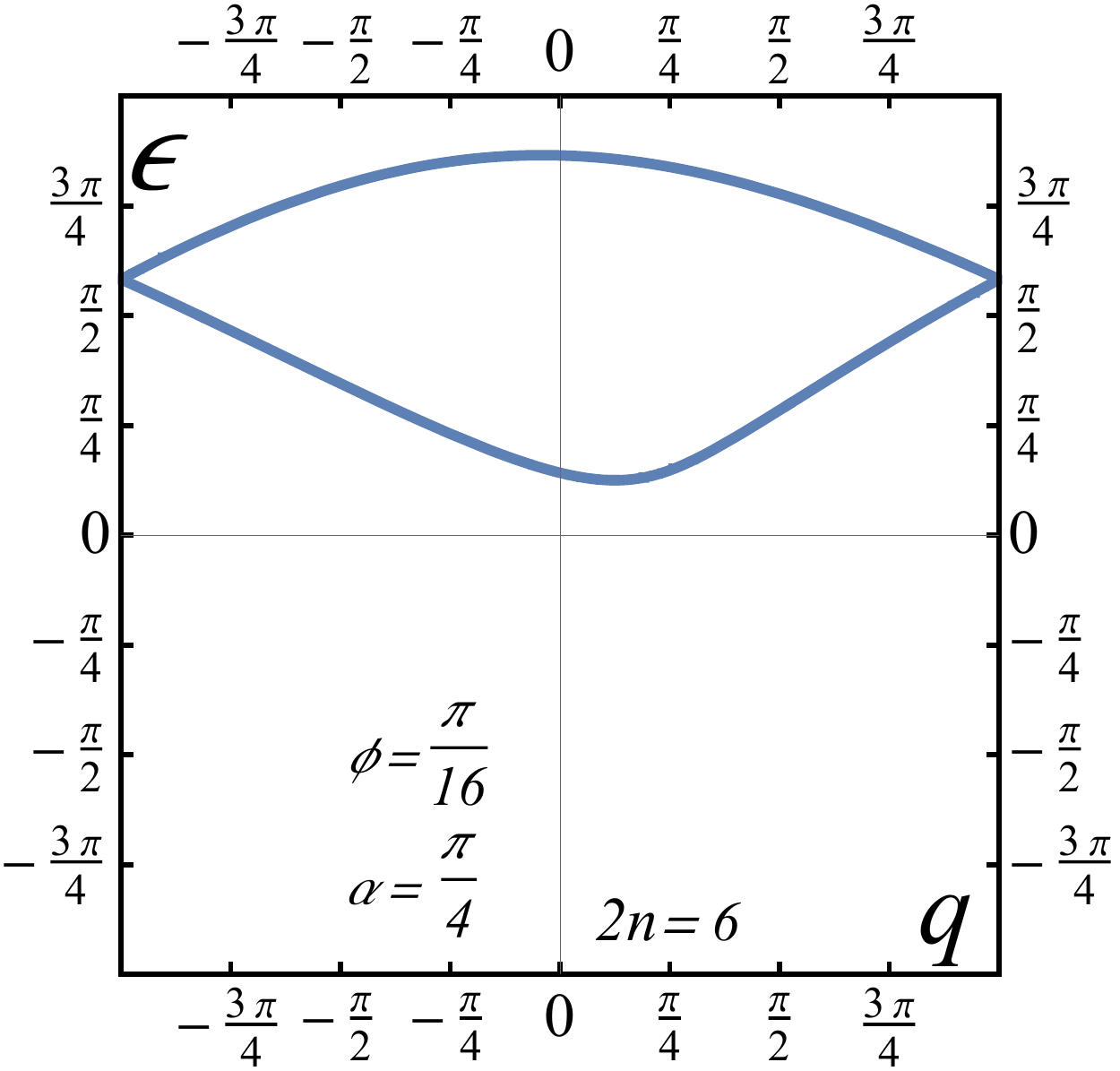}
\label{fig:6bsH-16}
}
  %
  \subfloat
  []{
     \includegraphics[width=40\unitlength]{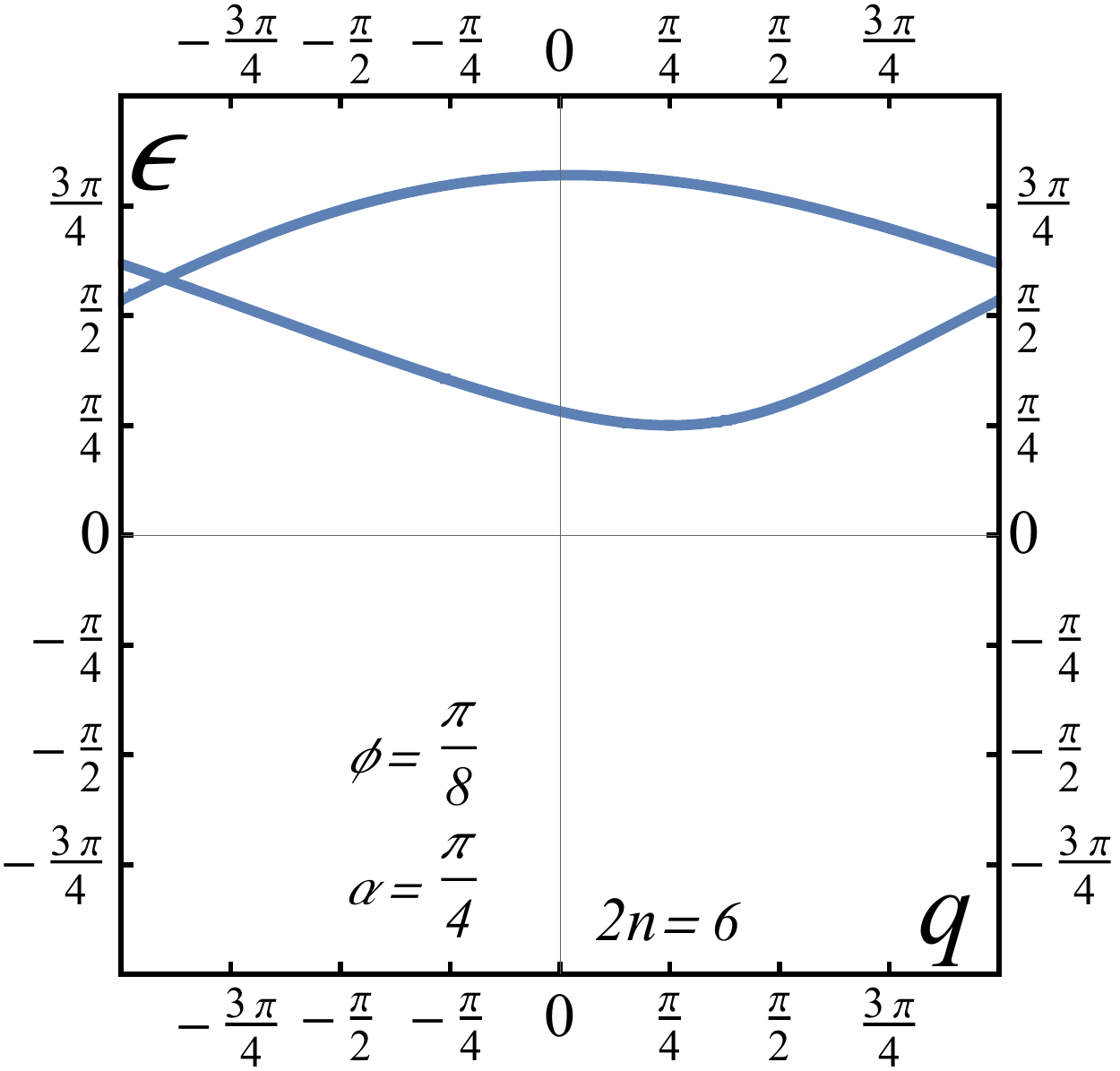}
\label{fig:6bsH-8}
}
  %
  \subfloat
  []{
     \includegraphics[width=40\unitlength]{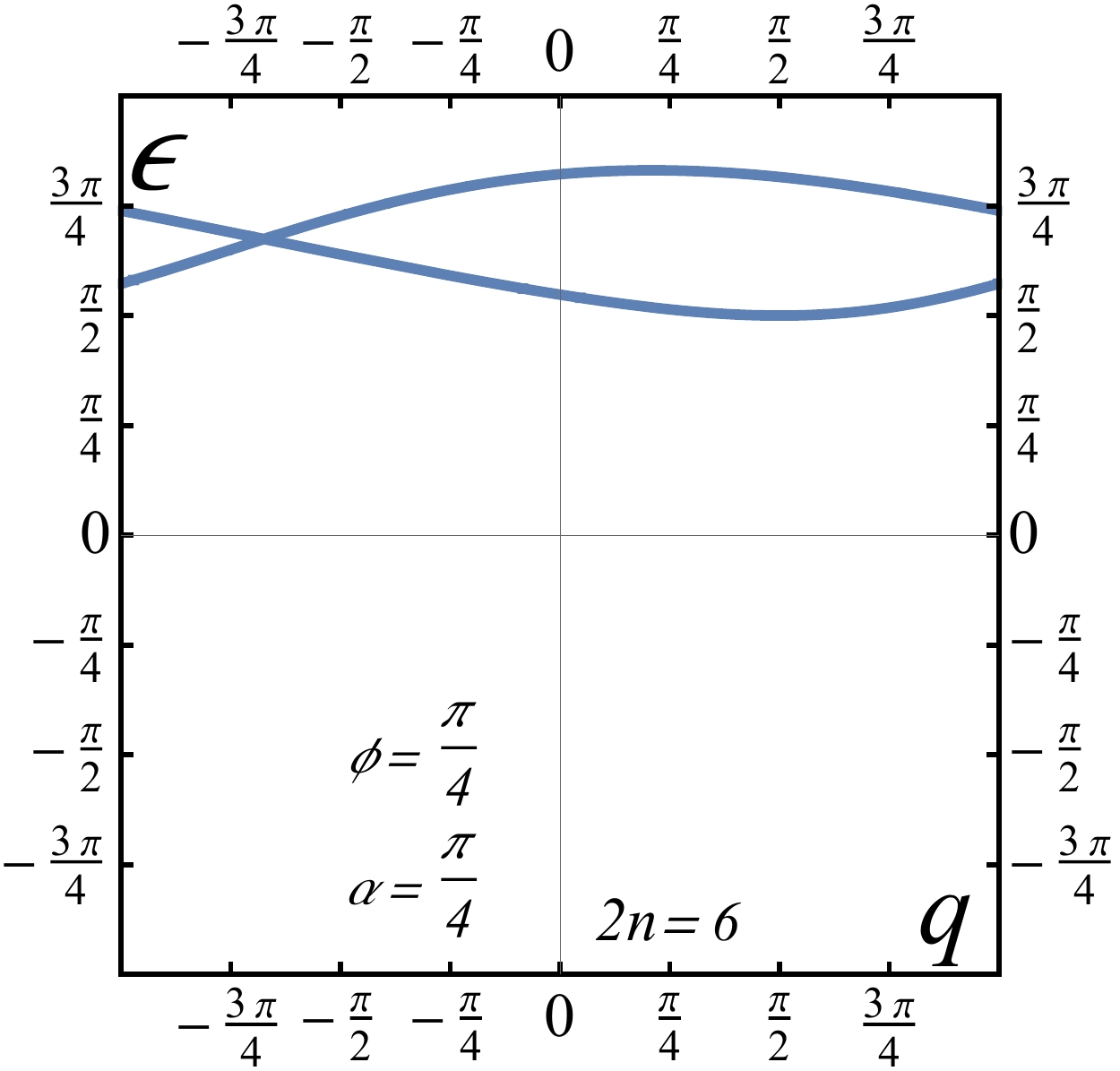}
\label{fig:6bsH-4}
}
  %
  \subfloat
  []{
     \includegraphics[width=40\unitlength]{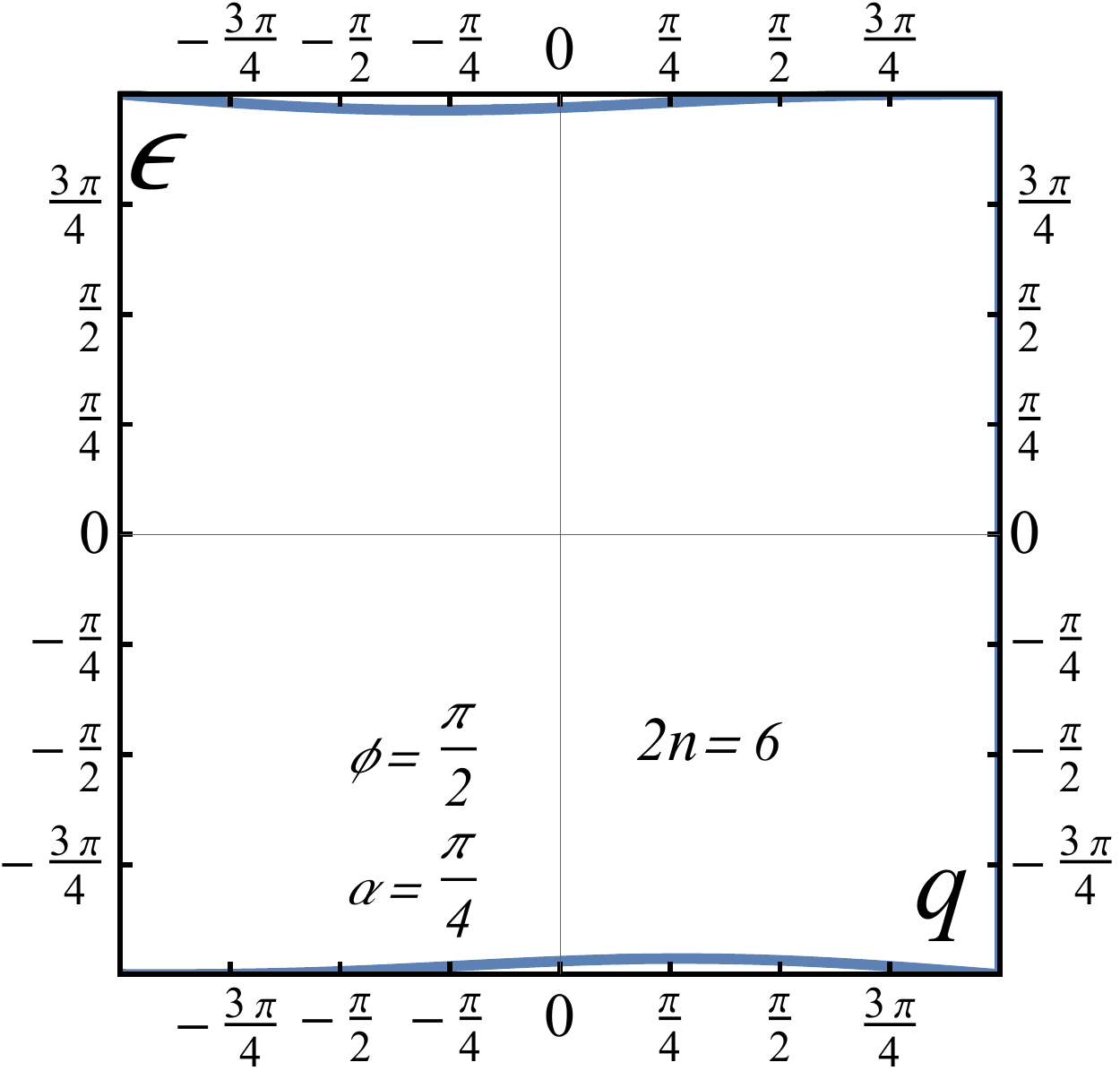}
\label{fig:6bsH-2}
}
\\
  \subfloat
  []{
     \includegraphics[width=40\unitlength]{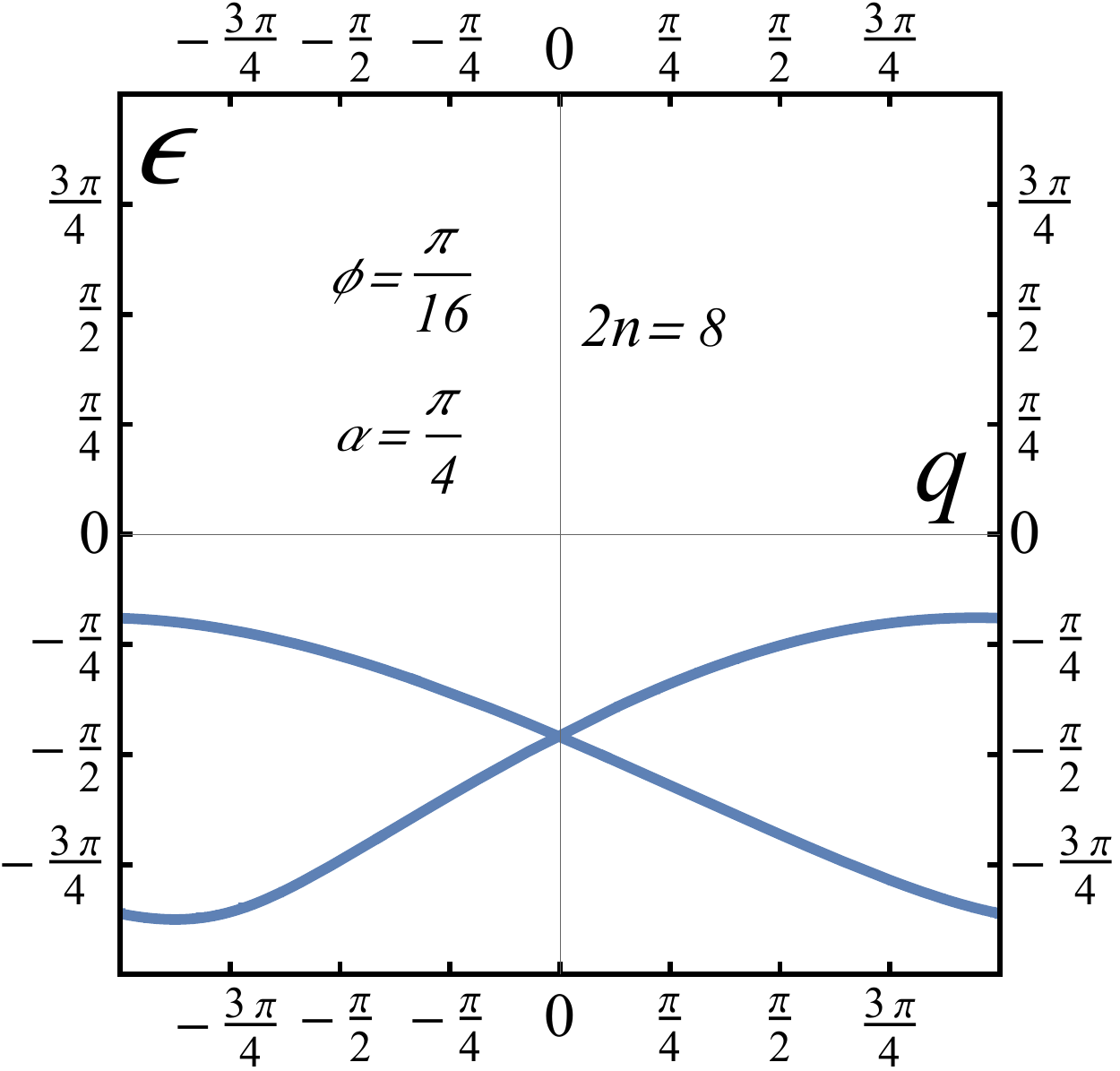}
\label{fig:8bsH-16}
}
  %
  \subfloat
  []{
     \includegraphics[width=40\unitlength]{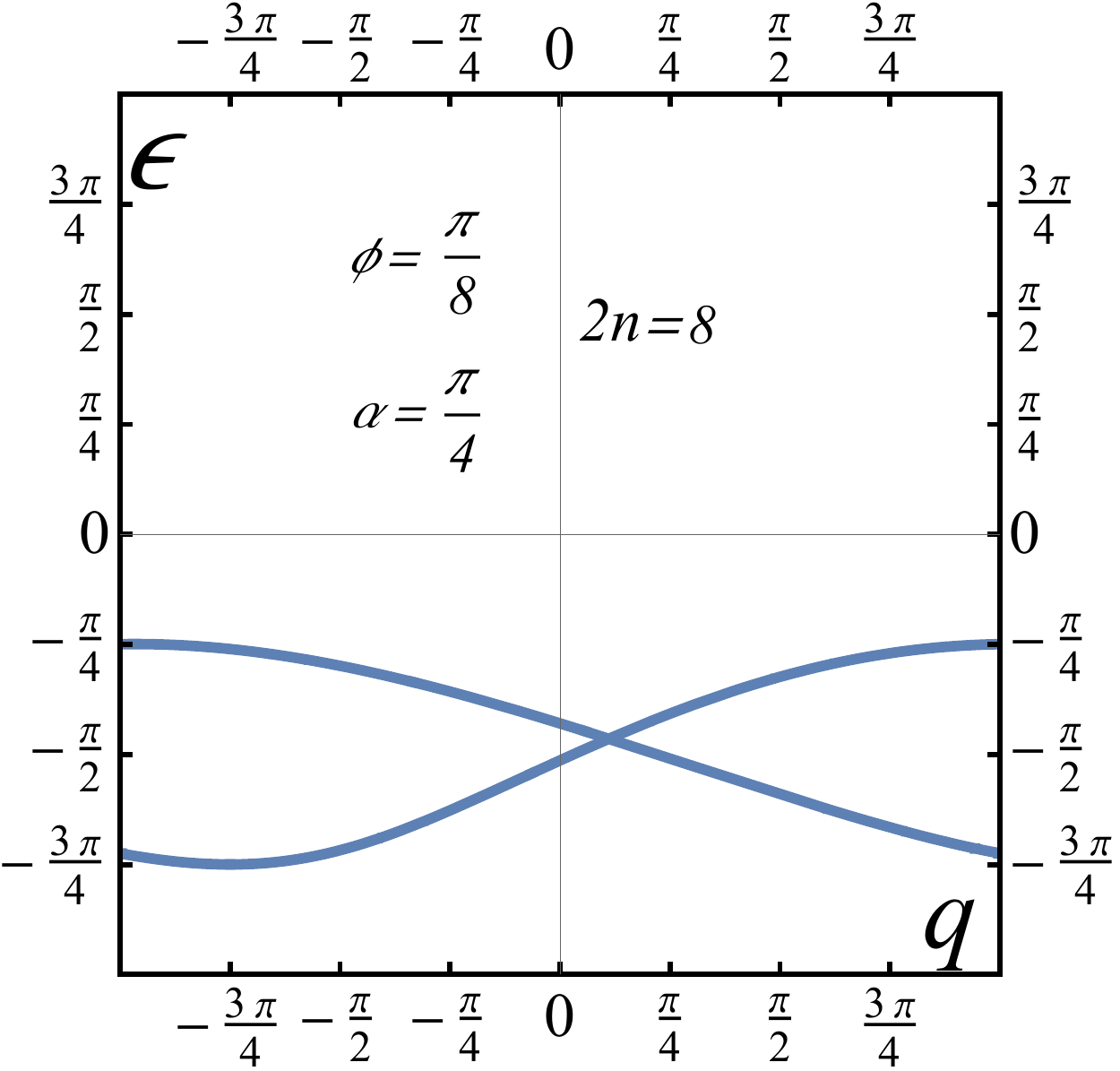}
\label{fig:8bsH-8}
}
  %
  \subfloat
  []{
     \includegraphics[width=40\unitlength]{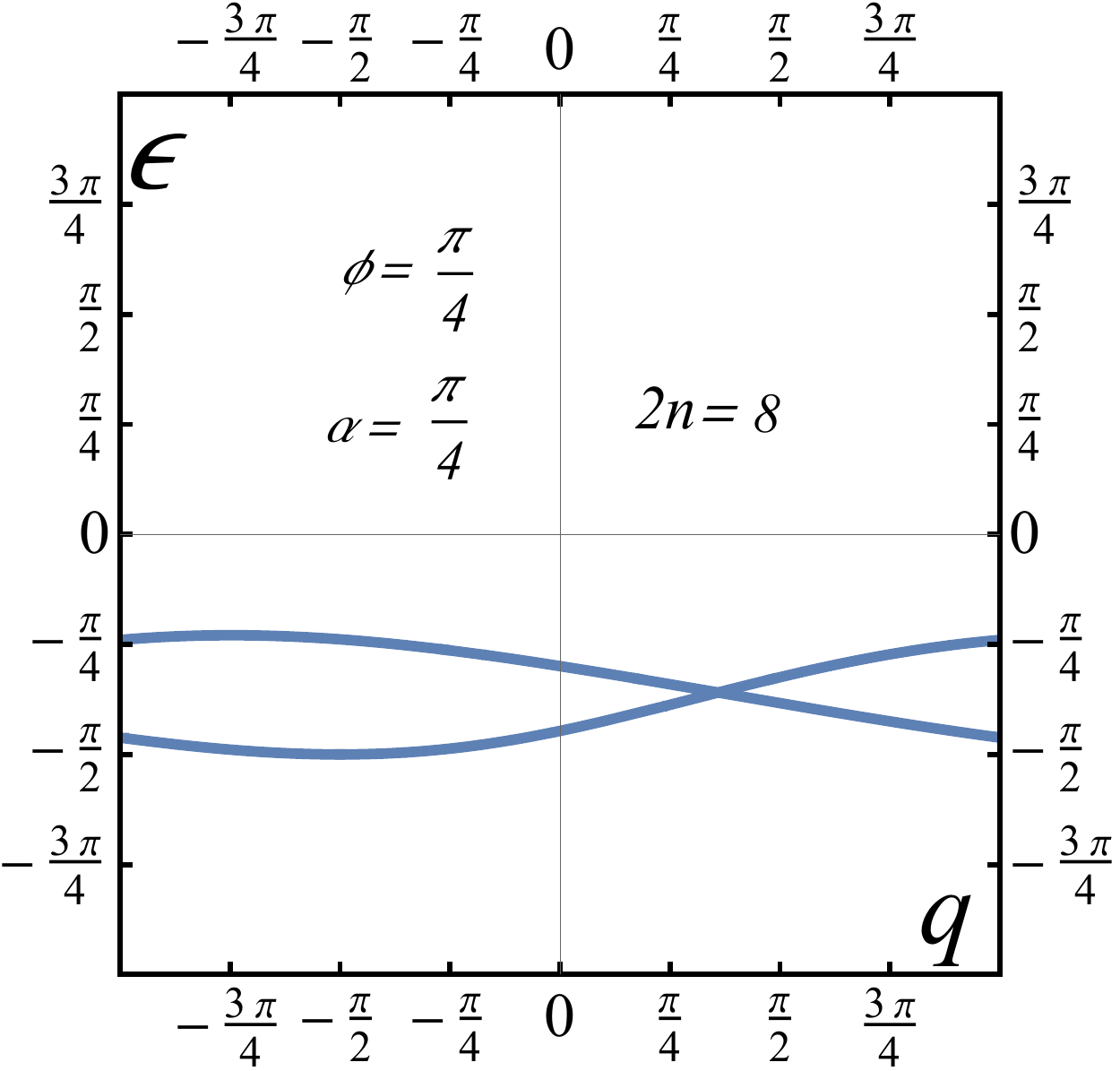}
\label{fig:8bsH-4}
}
  %
  \subfloat
  []{
     \includegraphics[width=40\unitlength]{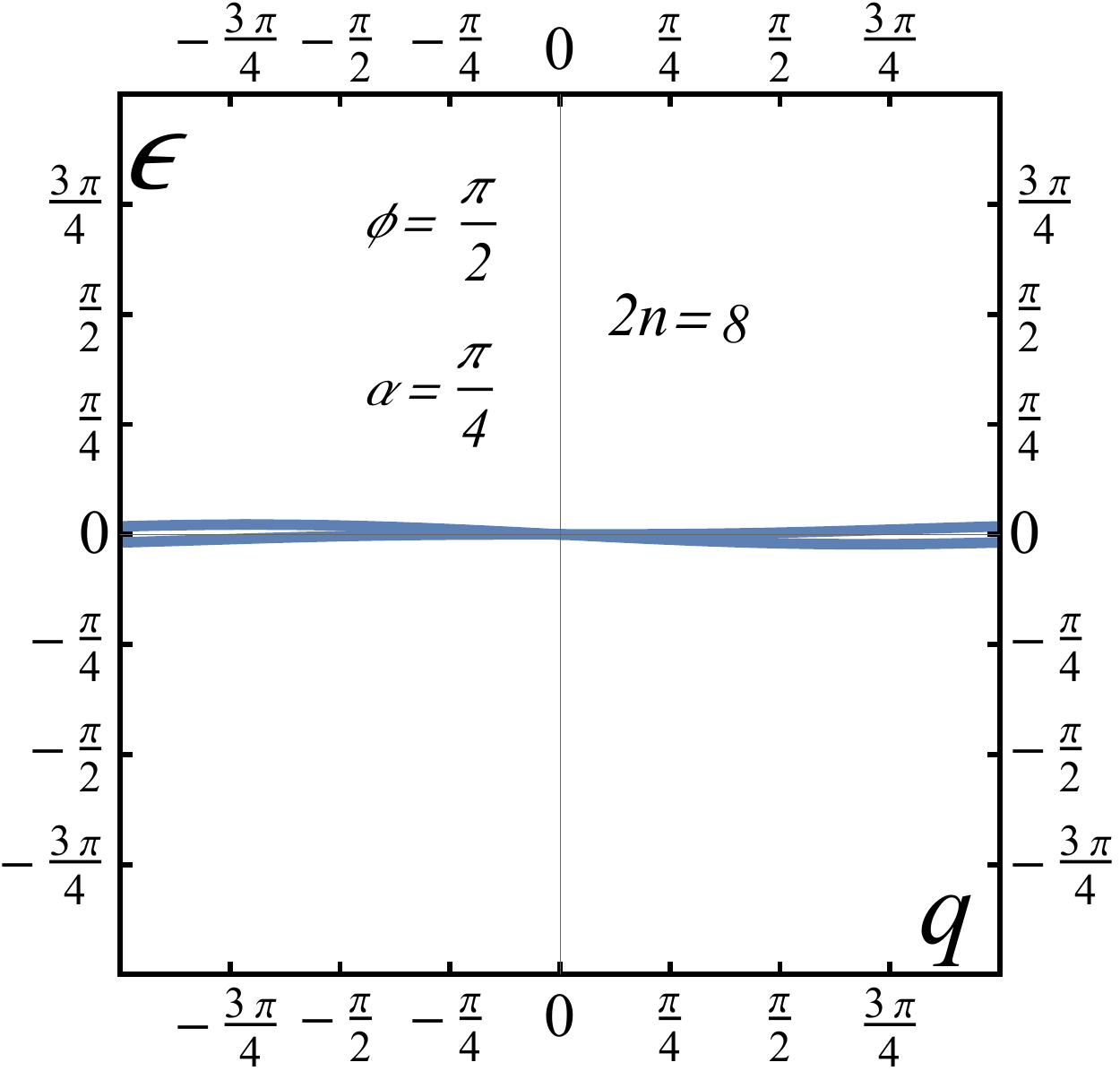}
\label{fig:8bsH-2}
}
  \caption{Spectra of multi-particle bound states for the Chiral Hubbard model found from \req{eq:spectrumnH},
    The swap angle $\alpha$ is fixed, $\alpha=\pi/4$. The phase $\phi$ characterizes the interaction strength, $\phi=\pi/2$ is the
    strongest possible interaction (notice that the multi-particle spectrum becomes practically flat that corresponds to the
    auto-localization by interactions).
  a)-d) Four particle; e)-h) Six particle; and i)-l) eight particle bound states. }
\end{figure}

Unfortunately, the further simplifications of \req{eq:spectrum2H} are not possible for the general case and
we limit ourselves with the plotting of resulting curves for different values of phase $\phi$, see Fig.~\ref{fig:nbsH}


\subsection{Takahashi type equations.}

In this subsection we derive the system of  of non-linear equations describing the multiple string solutions somewhat analogously to
Sec.~\ref{sec:BGD}. We are interested in the limit $L\gg 1$ whereas the excitation number $\mathcal{N}_{1,2}$ remains arbitrary.
The numbers characterizing the quasi-energies and quasi-momenta are the real rapidities of single excitation $\tilde{\varepsilon}_j,
\ j=0,\dots, n_0$,
real rapidities of the $2n$-strings $\lambda^{(2\mu)}_j,\ j=0,\dots, n_\mu$, and the spin rapidities $\lambda_j,\ j=1,\dots, n_\lambda$.
The spin rapidities may be either real or form the string by themselves (so called $\Lambda$-strings). The latter strings become universal only
in the limit $\mathcal{N}_{1,2}\gg 1$, which will be of no particular importance for us. Therefore, the equations for spin rapidities will
be kept in the original Lieb-Wu form allowing for the complex $\lambda_j$ appearing in a complex conjugated pairs. Finally, we do not know
what experiment would allow to separate the $\Lambda$-strings on the background of the continuous spectrum for the Chiral Hubbard model.
The derivation follows mostly Ref.~\cite{Tak72}.

The number of each of $2\mu$-strings is connected to the total number of excitations $\mathcal{N}_{1,2}$ by
\be
\begin{split}
  &\mathcal{N}=\mathcal{N}_{1}+\mathcal{N}_{2}
  =n_0+2\sum_{\mu=1}^{\left[\mathcal{N}/2\right]} \mu n_\mu;
  \\
  &\mathcal{N}_{2}
 =n_\lambda+\sum_{\mu=1}^{\left[\mathcal{N}/2\right]} \mu n_\mu.
\end{split}
\label{eq:stringnumberH}
\ee

The total quasi-energy and the quasi-momentum of the state are
given by addition of that for all the strings,
\be
\begin{split}
  &E=\ldb
\sum_{j=1}^{n_0}\epsilon(\tilde{\varepsilon}_j)+
  \sum_{\mu=1}^{\left[\mathcal{N}/2\right]}
\sum_{j=1}^{n_\mu}
\epsilon^{(2\mu) }\left(\lambda_j^{(2\mu )}\right)\rdb;
\\
&Q=\ldb
\sum_{j=1}^{n_0}q(\tilde{\varepsilon}_j)+
  \sum_{\mu=1}^{\left[\mathcal{N}/2\right]}
\sum_{j=1}^{n_\mu}
q^{(2\mu) }\left(\lambda_j^{(2\mu )}\right)
\rdb.
\end{split}
\label{eq:EQexpressionH}
\ee
where the quasi-energies and quasi-momenta of constituting excitations are given by
\reqs{implicit-H2}
and \rref{eq:spectrumnH}.

\begin{subequations}\label{eq:TakH}
  The equations are derived along the lines of Ref.~\cite{Tak72}.
  The equation for the spin rapidities is decoupled from those for the bound state:
    \be
  \begin{split}
  & (-1)^{\mathcal{N}}
  \prod_{j=1}^{n_0}
   \frac{\sin\tilde{\varepsilon}_j-\lambda_k+iu}{\sin\tilde{\varepsilon}_j-\lambda_k-iu}
  =
  \prod_{\substack{l=1,\\l\neq k}}^{\mathcal{N}_2}
  \frac{\lambda_l-\lambda_k+2iu}{\lambda_l-\lambda_k-2iu},
  \\
  &
  \qquad k=1,\dots n_\lambda.
  \end{split}
  \ee
  As we mentioned before, this equation allows not only for the real roots but also for the $\Lambda$-strings
  with complex $\lambda_j$'s and all real $\tilde{\varepsilon}_j$'s.
  Only in the limit $n_0\gg 1$ the strings become universal but we will not take this limit.

  The single excitations quasi-momenta are quantized including with account
  of the scattering of single excitations on each other (the first factor on the right hand side)
  and on the $2\mu$-strings (the second factor on the right hand side)
  \be
  \begin{split}
  e^{iq(\tilde{\varepsilon}_j)L}
 & =(-1)^{\mathcal{N}_1+1}
  \left\{\prod_{k=1}^{\mathcal{n}_\lambda}
  \frac{\sin\tilde{\varepsilon}_j-\lambda_k-iu}{\sin\tilde{\varepsilon}_j-\lambda_k+iu}
\right\}
\\
&\times \prod_{\mu=1}^{\left[\mathcal{N}/2\right]}
\left\{\prod_{k=1}^{n_\mu}
\frac{\sin\tilde{\varepsilon}_j-\lambda_k^{(2\mu)}-iu\mu}{\sin\tilde{\varepsilon}_j-\lambda_k^{(2\mu)}+iu\mu}
\right\}
;
\\
&
\qquad j=1,\dots, n_0;
\end{split}
\ee
Analogously, the bound states
acquire the scattering phase due to the scatterings on a single particle excitations and on the bound states:
\be
  \begin{split}
 &   e^{
      iq^{(2\mu)}(\lambda_j^{(2\mu)})
      L}
  =(-1)^{\mu\mathcal{N}}
  \left\{\prod_{k=1}^{\mathcal{n}_0}
  \frac{\lambda_k^{(2\mu)}-\sin\tilde{\varepsilon}_j-iu\mu}{\lambda_k^{(2\mu)}-\sin\tilde{\varepsilon}_j+iu\mu}
\right\}
\\
&\qquad \times \prod_{\nu=1}^{\left[\mathcal{N}/2\right]}
\left\{\prod_{k=1}^{n_\nu}
G^{\mu\nu}\left(\lambda^{(2\nu)}_k-\lambda^{(2\mu)}_j\right)
\right\}
;
\\
&
\qquad j=1,\dots, n_\mu;\qquad \mu=1,\dots,\left[\mathcal{N}/2\right].
\end{split}
  \ee
where
\be
  \begin{split}
   &  G^{\mu\nu}\left(v\right)\equiv
     \prod_{m=\left|\mu-\nu\right|/2}^{(\mu+\nu-2)/2}
     \left(\frac{ v+2im\, u }{v-2im\,u}\right)
      \left(\frac{v+2i(m+1)u}{v-2i(m+1)u}\right)
      ;
      \\
      &
      G^{\mu=\nu}\left( v= 0\right)=1.
\end{split}
\label{eq:TakH2}
\ee
Notice that the scattering phase of the bound states on each other is completely analogous to that for the $XXZ$ model, see Sec.~\ref{sec:BGD}.

\end{subequations}

\subsection{What could be observed in modern quantum computer experiments?}

The discussion of Sec.~\ref{sec:QCXXZ} about the observability of the bound states for $XXZ$ Floquet circuit is directly adapted for the Chiral
Hubbard circuit, after we modify \req{eq:Sigma} as
\be
\hat{\Sigma}^{(k)}_x=
\prod_{m=0}^{k-1}\hat{\sigma}_{\ldcb x+m\rdcb,1}^{x}\hat{\sigma}_{\ldcb x+m\rdcb,2}^{x}.
\label{eq:SigmaH}
\ee
The experimental preparation of the initial state is discussed in the end of \ref{sec:ap1}.

\section{Summary.}

In conclusion, I presented complete analytic theory of certain non-trivial quantum circuits.
It is important to emphasize that the spectra of all bound states are expressible in terms of the same limited number of parameters
(i.e. the number of possible experiments significantly exceeds the number of control knobs).
Therefore, analytic results presented here may help not only calibrate the interaction phase $\phi$ in the quantum gates but also put experimental bounds on
the possible effects of excitations on the other qubits on those gates.
What remains interesting is to investigate the effects of static disorder and the noise
\cite{Google2020} on the experimentally observable correlation function as it may help to separate the different mechanisms and the long correlations.
It will be done elsewhere.

\section{Acknowledgements}
I am thankful to V.Cheianov, L. Glazman, L. Ioffe, K. Kechedzhi, T. Ren, V. Smelyanskiy,  and A. Tsvelik for helpful conversations.
I also acknowledge discussions
with C. Neill, P. Roushan, and Z. Jiang about the possible experiments with the bound states performed on the modern Google platform.

\appendix

\section{Observability of the correlation function \rref{eq:function} \cite{IK}}.
\label{sec:ap1}

Prepare the system with the following initial wave-function
(the preparation will be described in the end of this  Appendix.)
\be
\left|\Psi(\chi)\right\rangle=
\frac{1}{\sqrt{2}}
\left(\hat{\openone}+e^{i\chi}\hat{\Sigma}^{(k)}_0\right)
\left|0\right\rangle,
\ee
where we used the property $\left[\hat{\Sigma}^{(k)}\right]^2=\hat{\openone}$
so that the wavefunction is properly normalized.
Let us subject the system to the $n$ steps of the Floquet dynamics
and measure the expectation value of the operator \rref{eq:Sigma}:
\[
  \Sigma_x^{(k)}\left(n;\chi\right)=
\left\langle\Psi(\chi)\right|  \left[\hat{\mathcal{U}}^\dagger\right]^{n}\hat{\Sigma}_x^{(k)}
 \hat{\mathcal{U}}^n
\left|\Psi(\chi)\right\rangle.
\]
We find
\be
\begin{split}
 \Sigma_x^{(k)}\left(n;\chi\right)&=
 \frac{1}{2}
 \left\{
   e^{i\chi}\mathcal{D}^{(k)}\left(n,x\right)
   +e^{-i\chi}\mathcal{D}^{(k)}\left(-n,x\right)
 \right\}
 \\
 &+ \frac{1}{2}
 \left\langle 0 \right|
\hat{\Sigma}_0^{(k)}
 \left[\hat{\mathcal{U}}^\dagger\right]^{n}\hat{\Sigma}_x^{(k)}
 \hat{\mathcal{U}}^n\hat{\Sigma}_0^{(k)}
 \left|0\right\rangle,
\end{split}
\label{ap:obs}
\ee
the last term vanishes for the odd values of $k$.

The correlation function \rref{eq:function} can be ascertained from the measurable quantity \rref{ap:obs} as
\be
\mathcal{D}^{(k)}\left(n,x\right)
=
\frac{2}{q}
\sum_{p=1}^qe^{-\frac{2ip\pi }{q}}\Sigma_x^{(k)}\left(n;\frac{2\pi p}{q}\right),
\ee
where integer $q\geq 3$.

To complete this section we describe the preparation of the initial state $\left|\Psi(\chi)\right\rangle$.
This is achieved by the following set of unitary gates acting on the vacuum state $\left|0\right\rangle$.
\be
\left|\Psi(\chi)\right\rangle
= \left(\prod_{m=2}^{k}\widehat{cX}_{m-3/2}\right)
\hat{H}_0(\chi)
\left|0\right\rangle,
\label{ap:psiXXZ}
\ee
and the gates $\widehat{cX}_{m-3/2}$ in the product are $m$-ordered.

Single qubit unitary
\be
  \hat{H}_x(\chi)=\frac{1}{\sqrt{2}}
  \begin{pmatrix}1 & e^{-i\chi}\\
e^{i\chi}&  -1
  \end{pmatrix}
\ee
is acting in the Hilbert space $\mathcal{h}_x$ and the
two-qubit gate defined as
\be
\widehat{cX}_{x+1/2}
=
  \begin{pmatrix}1 & 0 &0 & 0 \\
    0 & 1 &0 & 0 \\
    0 & 0 &0 & 1 \\
    0 & 0 &1 & 0 \\
  \end{pmatrix}
\ee
acts in the four-dimensional $\mathcal{h}_{\ldcb x\rdcb}\otimes \mathcal{h}_{\ldcb x+1\rdcb}$
Hilbert space.

For the Chiral Hubbard quantum circuit, one should replace \req{ap:psiXXZ} to
\be
\left|\Psi(\chi)\right\rangle
= \prod_{r=1}^2\left(\prod_{m=2}^{k}\widehat{cX}_{m-3/2,r}\right)
\hat{H}_{0,r}(\chi/2)\left|0\right\rangle,
\ee
where the subscript $r$ indicates to which replica the operator is applied.

\end{document}